\documentclass[%
 reprint,superscriptaddress,
 amsmath,amssymb,prb,
]{revtex4-2}

\usepackage{enumerate}
\usepackage{graphicx}
\usepackage{dcolumn}
\usepackage{bm}
\usepackage{color}
\usepackage{siunitx}
\usepackage{tabularx}
\usepackage{booktabs}

\newcommand{\tred}[1]{\textcolor{black}{#1}}
\def \CTS{Co$_{0.325}$TaS$_{2}$}
\def \CTSideal{Co$_{1/3}$TaS$_{2}$}

\def \TN{$T_{\mathrm N}$}
\def \TNo{$T_{\mathrm N1}$}
\def \TNt{$T_{\mathrm N2}$}

\def \Q{$\mathbf{Q}$}

\def \NCTO{Na$_{2}$Co$_{2}$TeO$_{6}$}

\def \Tb{$T<T_{\mathrm{N2}}$}
\def \Ti{$T_{\mathrm{N2}}<T<T_{\mathrm{N1}}$}
\def \Th{$T>T_{\mathrm{N1}}$}

\begin{document}

\preprint{APS/123-QED}

\title{Spin dynamics of triple-Q magnetic orderings in a triangular lattice: Implications for multi-Q orderings in general two-dimensional lattices} \thanks{This manuscript has been authored by UT-Battelle, LLC under Contract No. DE-AC05-00OR22725 with the U.S. Department of Energy.  The United States Government retains and the publisher, by accepting the article for publication, acknowledges that the United States Government retains a non-exclusive, paid-up, irrevocable, world-wide license to publish or reproduce the published form of this manuscript, or allow others to do so, for United States Government purposes.  The Department of Energy will provide public access to these results of federally sponsored research in accordance with the DOE Public Access Plan (http://energy.gov/downloads/doe-public-access-plan).}

\author{Pyeongjae Park}
\email{parkp@ornl.gov}
\affiliation{Materials Science \& Technology Division, Oak Ridge National Laboratory, Oak Ridge, TN 37831, USA}
\affiliation{Center for Quantum Materials, Seoul National University, Seoul 08826, Republic of Korea}
\affiliation{Department of Physics and Astronomy, Seoul National University, Seoul 08826, Republic of Korea}

\author{Woonghee Cho}
\affiliation{Center for Quantum Materials, Seoul National University, Seoul 08826, Republic of Korea}
\affiliation{Department of Physics and Astronomy, Seoul National University, Seoul 08826, Republic of Korea}

\author{Chaebin Kim}
\affiliation{Center for Quantum Materials, Seoul National University, Seoul 08826, Republic of Korea}
\affiliation{Department of Physics and Astronomy, Seoul National University, Seoul 08826, Republic of Korea}

\author{Yeochan An}
\affiliation{Center for Quantum Materials, Seoul National University, Seoul 08826, Republic of Korea}
\affiliation{Department of Physics and Astronomy, Seoul National University, Seoul 08826, Republic of Korea}

\author{Kazuki Iida}
\affiliation{Neutron Science and Technology Center, Comprehensive Research Organization for Science and Society (CROSS), Tokai, Ibaraki 319-1106, Japan}

\author{Ryoichi Kajimoto}
\affiliation{Materials and Life Science Division, J-PARC Center, Japan Atomic Energy Agency, Tokai, Ibaraki 319-1195, Japan}

\author{Sakib Matin}
\affiliation{Center for Nonlinear Studies, Los Alamos National Laboratory, Los Alamos, NM, 87545, USA}
\affiliation{Theoretical Division, Los Alamos National Laboratory, Los Alamos, NM, 87545, USA}

\author{Shang-Shun Zhang}
\affiliation{Department of Physics and Astronomy, The University of Tennessee, Knoxville, TN, 37996, USA}

\author{Cristian D. Batista}
\affiliation{Department of Physics and Astronomy, The University of Tennessee, Knoxville, TN, 37996, USA}
\affiliation{Quantum Condensed Matter Division and Shull-Wollan Center, Oak Ridge National Laboratory, Oak Ridge, TN, 37831, USA}

\author{Je-Geun Park}
\email{jgpark10@snu.ac.kr}
\affiliation{Center for Quantum Materials, Seoul National University, Seoul 08826, Republic of Korea}
\affiliation{Department of Physics and Astronomy, Seoul National University, Seoul 08826, Republic of Korea}
\affiliation{Institute of Applied Physics, Seoul National University, Seoul 08826, Republic of Korea}

\begin{abstract}
Multi-$\mathbf{Q}$ magnetic structures on two-dimensional (2D) lattices provide a key route to realizing topological physics in 2D magnetism. A major experimental challenge is to unambiguously confirm their formation by excluding the possibility of \tred{topologically trivial multi-domain single- or double-$\mathbf{Q}$ magnetic orders}, which cannot be distinguished using conventional diffraction techniques. Here, we propose that long-wavelength spin dynamics offers a \textit{universal} diagnostic for triangular lattices: \tred{triple-$\mathbf{Q}$ orders that preserve rotational symmetry and single- or double-$\mathbf{Q}$ orders that break it} exhibit qualitatively distinct anisotropies in their Goldstone mode velocities, stemming from fundamental differences in their underlying spin configurations. We validate this concept using the metallic triangular lattice antiferromagnet \tred{\CTS{}}, which hosts both a stripe-type single-$\mathbf{Q}$ state and a triple-$\mathbf{Q}$ tetrahedral ordering at different temperatures. Using inelastic neutron scattering (INS) and spin dynamics simulations, we first refine the spin Hamiltonian by fitting the paramagnetic excitation spectra, allowing us to develop an unbiased model independent of magnetic ordering. We then show that the observed velocity profiles of the Goldstone modes agree with the high-temperature model's predictions: markedly anisotropic for the single-$\mathbf{Q}$ phase and near isotropic for the triple-$\mathbf{Q}$ phase. Importantly, this contrast persists across various exchange parameters, highlighting its model-independent nature and suggesting potential applicability to other 2D lattice systems. Beyond the long-wavelength regime, we present a substantial discrepancy between the measured and simulated magnon spectra exclusively in the triple-$\mathbf{Q}$ phase. We attribute this discrepancy to magnon energy renormalization arising from order-of-magnitude enhanced magnon-magnon interactions in the triple-$\mathbf{Q}$ phase, due to its noncollinear configuration. This work provides universal insight into the dynamical properties of \tred{topological} multi-$\mathbf{Q}$ magnetic orderings in 2D lattice structures, offering a broadly applicable diagnostic to distinguishing them from \tred{topologically trivial single- or double-$\mathbf{Q}$ counterparts}. The unequivocal confirmation of the triple-\Q{} structure in \tred{\CTS{}} further establishes it as a prominent material platform for exploring topological spin textures in the genuine 2D limit.

\end{abstract}

\maketitle

\section{Introduction}
Symmetry and topology are central themes in modern magnetism, with antiferromagnetism gaining increasing recognition for its potential in these areas. The diverse configurations of antiferromagnetic spins give rise to various combinations of magnetic symmetry and topological properties, each capable of producing unique phenomena~\cite{Smejkal22,Bonbien22}. Since diffraction techniques are commonly used to reveal the structure of antiferromagnetic textures, these textures are often characterized by their Fourier components ${\mathbf{S}}_{\mathbf{Q}}$, where their magnitudes $|{\mathbf{S}}_{\mathbf{Q}}|^2$ correspond to the intensities of the Bragg peak. Complex spin textures typically involve multiple Bragg peaks located at symmetry-related wave vectors ${\mathbf{Q}}_{\nu}$, generally referred to as multi-${\mathbf{Q}}$ orderings.

Multi-${\mathbf{Q}}$ orderings are attracting growing interest because they are required to generate topologically nontrivial spin textures, such as skyrmion, meron, or vortex crystals~\cite{Okubo12,Leonov2015,Hayami16,Ozawa16,Batista_2016, Ozawa17,Wang2021,Motome21_review,Wang20,wang_skyrmion}. Among them, magnetic skyrmions are a representative example of two-dimensional (2D) topological spin textures, where the spins twist in a manner that wraps around the Bloch sphere~\cite{Skyrmion_1989, skyr_review}. The integer topological invariant, or skyrmion charge $Q_{\rm Skx}$, corresponds to the number of times the spin texture wraps the Bloch sphere. Due to the conservation of this charge under continuous deformations, skyrmions behave as emergent mesoscale particles, potentially playing a crucial role in future spintronic memory devices~\cite{skyr_review, skyr_writing, Skyr_electrody, Skyr_track}.

Hexagonal structures serve as ideal platforms for stabilizing triple-\Q{} skyrmion, meron, or vortex crystals, making them a crucial framework for realizing topological physics in 2D magnetism. Such triple-\Q{} spin textures are more favored in this lattice geometry, as the sum of the three ordering wave vectors ${\mathbf{Q}}_1$, ${\mathbf{Q}}_2$ and ${\mathbf{Q}}_3$ related by the three-fold symmetry equals zero, i.e., ${\mathbf{Q}}_1 + {\mathbf{Q}}_2 + {\mathbf{Q}}_3= {\mathbf{0}}$. In general, the superposition of two spirals with propagation vectors ${\mathbf{Q}}_{\mu}$ and ${\mathbf{Q}}_{\nu}$ requires inclusion of higher harmonics to satisfy the fixed spin-length constraint $|{\mathbf{S}}_j| = S$ at every site $j$ in real space. The zero-sum condition implies that first harmonic ${\mathbf{Q}}_{\mu} + {\mathbf{Q}}_{\nu} = -\epsilon_{\mu \nu \eta} {\mathbf{Q}}_{\eta}$ ($\mu \neq \nu $)  is not  penalized by the exchange interaction because it is symmetry-equivalent to the original wave vectors  ${\mathbf{Q}}_{\mu}$ and ${\mathbf{Q}}_{\nu}$. Among various hexagonal lattices, the triangular lattice is the simplest structure capable of hosting topologically nontrivial spin textures, and has therefore been widely used to study triple-$\mathbf{Q}$ magnetic orderings~\cite{Okubo12,Leonov2015,Hayami16,Batista_2016, vortex_kitaev,Ozawa17,takagi2018,TL_skyrmion2,Wang20,Motome21_review,wang_skyrmion}.

However, it remains challenging to experimentally identify triple-${\mathbf{Q}}$ magnetic ordering in hexagonal lattices, as it is indistinguishable from a superposition of symmetry-related single-${\mathbf{Q}}$ or \tred{double-$\mathbf{Q}$} domains in conventional diffraction experiments. As the three ${\mathbf{Q}}_{\nu}$ vectors ($\nu=1,2,3$) are related by the three-fold rotational symmetry, both cases can produce the same three-fold-symmetric pattern of magnetic Bragg peaks with equal intensities. As a result, identifying triple-${\mathbf{Q}}$ magnetic structures requires advanced experimental tools to distinguish them from alternative scenarios involving three equally-populated
domains of single-${\mathbf{Q}}$ or double-${\mathbf{Q}}$ spin configurations, which spontaneously break the three-fold lattice symmetry.

Although various advanced techniques have successfully confirmed triple-${\mathbf{Q}}$ magnetic structures in several systems~\cite{skyrmion_diffH, skyr_rspace, Skyrmion_STM}, a promising approach to address this challenge is to complement diffraction measurements with inelastic neutron scattering (INS), which probes magnons of each magnetic ordering. Due to their distinct spin configurations, single-${\mathbf{Q}}$, \tred{double-$\mathbf{Q}$}, and triple-${\mathbf{Q}}$ magnetic orderings exhibit different excitation spectra. While this perspective has been used in previous studies~\cite{3Qdynamics1}, a systematic comparison of spin dynamics between these types of orderings is still lacking. In particular, clarifying the characteristic dynamical properties of each phase--independent of specific material conditions or spin models--can greatly aid in establishing a universal framework for distinguishing \tred{topologically non-trivial triple-${\mathbf{Q}}$} orderings from \tred{their single-${\mathbf{Q}}$ or double-$\mathbf{Q}$} counterparts (which are typically topologically trivial) in a broad class of 2D hexagonal magnets, which is the central theme of this study. 

\begin{figure}[t!]
\includegraphics[width=1\columnwidth]{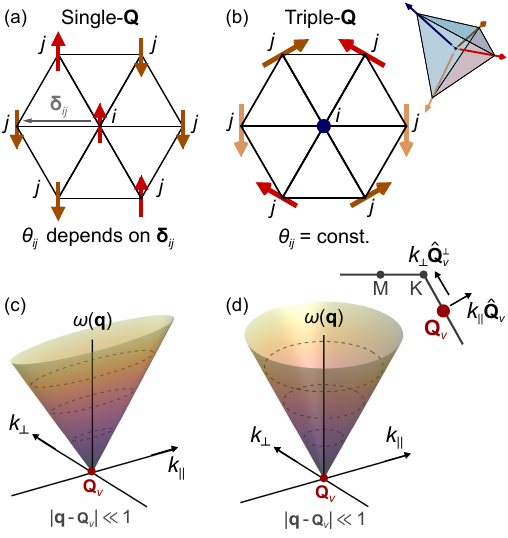} 
\caption{\label{basic} Characteristic real-space spin configurations and long-wavelength magnon profiles of single-$\mathbf{Q}$ and triple-$\mathbf{Q}$ magnetic orderings in a triangular lattice. (a)--(b) Real-space schematics of a single-$\mathbf{Q}$ and triple-$\mathbf{Q}$ magnetic structures, here illustrating the case of $M$-point ordering wave vectors ($\mathbf{Q}_{\nu} = \mathbf{G}_{\nu}/2$). (c)--(d) Schematic illustrations of the corresponding magnon dispersion near the ordering wave vector $\mathbf{Q}_{\nu}$, which is linear for both single-$\mathbf{Q}$ and triple-$\mathbf{Q}$ phases. Dashed lines indicate the constant-energy contour, which visualizes the direction-dependent velocity of the Goldstone mode. The right-top inset of (d) illustrates the relative momentum coordinate system used throughout this work around $\mathbf{Q}_{\nu}$, $(k_{\parallel},k_\perp)$; see section \tred{IV. A.}}
\end{figure}

Unlike high-momentum spin dynamics, which are strongly model-dependent, the long-wavelength limit--where the magnon momentum $\mathbf{q}$ lies near $\mathbf{0}$ or $\mathbf{Q}_{\nu}$--provides a robust foundation for finding a universal framework. In this regime, magnon dispersion in both single-$\mathbf{Q}$ and triple-$\mathbf{Q}$ orderings follows a linear form governed by the Goldstone theorem, independent of microscopic details (Fig.~\ref{basic}). Despite this common behavior, the direction-dependent velocity profile of this linear mode can differ significantly between the \tred{symmetry-preserving triple-$\mathbf{Q}$ ordering and symmetry-breaking single- or double-\Q{} orderings}, even under the same spin Hamiltonian. As demonstrated throughout this work, a triple-$\mathbf{Q}$ order, \tred{, which preserves the underlying rotational symmetry of the hexagonal lattice,} yields a nearly isotropic character, whereas \tred{single- and double-$\mathbf{Q}$ orders that break this symmetry} generally exhibit strong anisotropy in the profile [Fig.~\ref{basic}(c)–(d)]. This concept can be systematically tested by conducting INS measurements with a material that hosts both cases as the external variables (e.g., temperature) are varied. Such a system allows for a direct, ideal comparison of their long-wavelength spin dynamics under the same spin Hamiltonian.

\begin{figure*}[ht]
\includegraphics[width=1\textwidth]{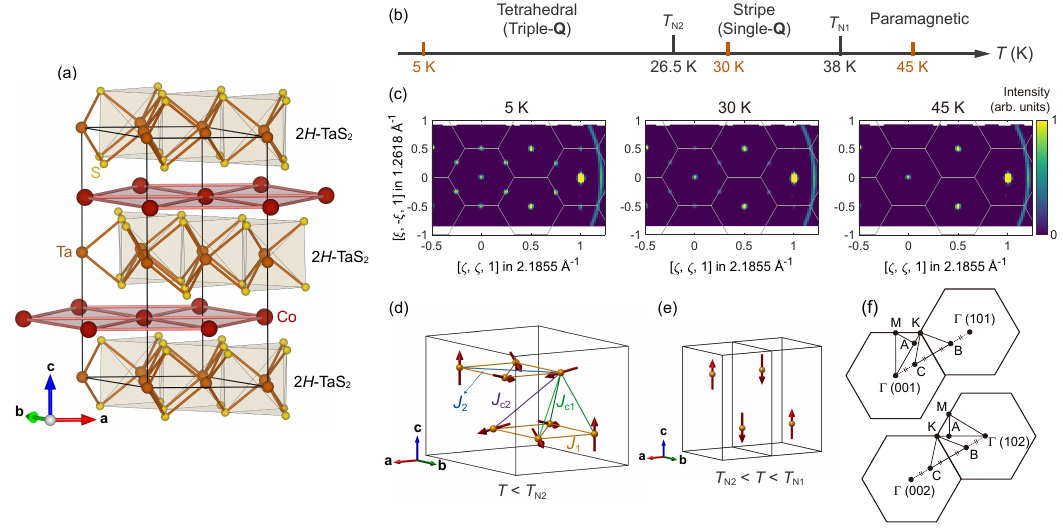} 
\caption{\label{CTS} Basic information on the magnetism in \tred{\CTS{}}. (a) Crystal structure of \CTSideal{}. (b) Schematic temperature-dependent phase diagram of \tred{\CTS{}} as suggested by neutron diffraction and transport measurements~\cite{CTS_tripleQ_natcomm, CTS_tripleQ_nphys}. The three orange vertical ticks indicate the temperatures at which the INS data were collected. (c) Elastic component ($-0.52 \,\mathrm{meV}< E < 0.52 \,\mathrm{meV})$ of the single-crystal INS data ($E_{i} = 13\,\mathrm{meV}$) at 5\,K (\Tb{}), 30\,K (\Ti{}), and 45\,K (\TNo{}\,$<T$). (d)--(e) Tetrahedral triple-${\mathbf{Q}}$ and stripe single-${\mathbf{Q}}$ magnetic ground states at \Tb{} and \Ti{}, respectively~\cite{CTS_tripleQ_natcomm, CTS_tripleQ_nphys}. The exchange interaction paths considered in this work are illustrated in (d). (f) Brillouin zones on [$H$, $K$, 1] and [$H$, $K$, 2] planes. Labels of high-symmetry ${\mathbf{q}}$ points and corresponding high-symmetry contours used throughout this work are also plotted.}
\end{figure*}

The layered triangular-lattice antiferromagnet \tred{\CTS{}} [Fig.~\ref{CTS}(a)] satisfies the above criteria, making it an ideal platform for investigating the contrasting dynamical properties of single-${\mathbf{Q}}$ and triple-${\mathbf{Q}}$ phases. Upon cooling, it undergoes two antiferromagnetic transitions at \TNo{} = 38\,K and \TNt{} = 26.5\,K [Fig.~\ref{CTS}(b)], with neutron diffraction revealing magnetic Bragg peaks at the $M$ points of the Brillouin zone in both phases ($\mathbf{Q}_{\nu} = \mathbf{G}_{\nu}/2$ with $\nu=1,2,3$, where $\mathbf{G}_{\nu}$ are reciprocal lattice vectors related by 120 degree rotations about the $c$-axis)~\cite{CTS_tripleQ_natcomm, CTS_tripleQ_nphys}. As detailed in Section III. B, the triple-${\mathbf{Q}}$ nature of the low-temperature phase has been revealed by a combination of this diffraction result and bulk transport measurements, while the intermediate-temperature phase is believed to exhibit single-${\mathbf{Q}}$ order. This makes \tred{\CTS{}} a rare ideal system for investigating distinct dynamical signatures of single-${\mathbf{Q}}$ and triple-${\mathbf{Q}}$ magnetic orderings under the same spin Hamiltonian, simply by adjusting temperature [Fig.~\ref{CTS}(b)]. Notably, these temperature-dependent spin dynamics can be systematically modeled using Landau-Lifshitz dynamics (LLD) simulations combined with a recently developed technique that accounts for the quantum renormalization of the classical dynamics. This latest approach has proven successful in describing experimental INS data measured at finite temperatures~\cite{rescale_Dahlbom, CoI2_nphys, BLCTO, park_ZVPO}.

In this work, we present a comprehensive theoretical and experimental investigation of the dynamical properties associated with single-\Q{} and triple-\Q{} magnetic orderings in a triangular lattice. The central message of this work is that the contrasting Goldstone mode velocity profiles--strongly anisotropic in the single-$\mathbf{Q}$ \tred{(or double-$\mathbf{Q}$)} phases and nearly isotropic in the triple-$\mathbf{Q}$ phase (Fig.~\ref{basic})--constitute a model-independent hallmark of the two distinct types, rooted in their fundamental symmetry-breaking or symmetry-preserving nature [see Fig.~\ref{basic}(a)--(b)]. As these symmetry properties remain the same for any spin configurations on triangular lattices, the resultant anisotropy contrast can serve as a universal diagnostic for distinguishing topological triple-$\mathbf{Q}$ spin textures from trivial single-$\mathbf{Q}$ \tred{or double-$\mathbf{Q}$} counterparts. We substantiate this framework through a case study of the $M$-ordering wave vector ($\mathbf{Q}_{\nu} = \mathbf{G}_{\nu}/2$) realized in \tred{\CTS{}}. Our analysis shows that the nearly isotropic (strongly anisotropic) velocity profile of the triple-$\mathbf{Q}$ (single-$\mathbf{Q}$) phase remains robust across a wide range of exchange parameters and accurately reproduces the long-wavelength spin dynamics observed in INS measurements of \tred{\CTS{}}. This not only provides \textit{unequivocal} evidence of the single-\Q{} to triple-\Q{} phase transition in \tred{\CTS{}} beyond previous studies, but also strongly supports the proposed dynamical distinction between the \tred{symmetry-preserving} triple-$\mathbf{Q}$ and \tred{symmetry-breaking single- or double-$\mathbf{Q}$} orderings.

The rest of the paper is organized as follows. After the Methods section (Section II), we begin with a qualitative discussion of how the distinct symmetry-breaking or symmetry-preserving nature between single-\tred{/double-$\mathbf{Q}$} and triple-$\mathbf{Q}$ magnetic orderings gives rise to contrasting velocity anisotropy in their long-wavelength spin dynamics (Section III. A). This conceptual insight lays the foundation for identifying universal dynamical signatures of Goldstone modes in general triple-$\mathbf{Q}$ orderings, which can be potentially extended to many more 2D magnets--especially 2D van der Waals magnets--beyond triangular lattice structures. \tred{In Section III. B, we describe the temperature-dependent magnetic phase diagram of \tred{\CTS{}}, highlighting it as an ideal platform for validating the framework proposed in Section III. A}.

\tred{In Section IV. A}, we introduce the effective spin Hamiltonian for \tred{\CTS{}} and its corresponding ground states and long-wavelength excitations, focusing on the case of $M$-point ordering vectors ($\mathbf{Q}_{\nu} = \mathbf{G}_{\nu}/2$). We analytically derive the long-wavelength magnon velocity profiles, which indeed imply the universality and model independence of the contrasting anisotropy. The rest of Section \tred{IV} is devoted to its systematic application to the experimental INS data of single-crystal \tred{\CTS{}} across three distinct phases: the paramagnetic phase (\Th{}) and two ordered phases (\Ti{} and \Tb{}). Section \tred{IV. B} presents our analysis of the paramagnetic spin dynamics. We successfully determined bilinear exchange parameters by fitting \textit{projected} dynamical spin structure factor maps [$S_{\perp}(\mathbf{q},\omega)$, see Appendix C] of the paramagnetic spectra, utilizing our state-of-the-art LLD simulation protocol (see Methods) and Bayesian optimization algorithm. This approach allows us to obtain an optimal exchange parameter set independently of the magnetic ordering information, enabling a fair comparison of single-\Q{} and triple-\Q{} spin dynamics based on the same Hamiltonian, a notable advancement achieved in this work.

In Section \tred{IV. C} we compare the experimental and simulated spin-wave spectra using this Hamiltonian in the low-$T$ (\Tb{}) and intermediate (\Ti{}) phases. A clear contrast in the Goldstone-mode velocity profiles--anisotropic at \Ti{} and nearly isotropic at \Tb{}--emerges, in excellent agreement with our theoretical predictions. LLD simulations based on our optimal spin model confirm that these distinct behaviors can only be explained by the single-${\mathbf{Q}}$ and triple-${\mathbf{Q}}$ nature, respectively. Importantly, we show that the contrast in the Goldstone-mode velocity profiles persists nearly independent of exchange parameters.
 
In Section \tred{IV. D}, we extend the data analysis beyond the long-wavelength regime and reveal the presence of magnon linewidth broadening and energy renormalization in \tred{\CTS{}}. In particular, a comparison between the experimental data and LLD simulations reveals that the latter characteristic is significantly enhanced in the tetrahedral triple-${\mathbf{Q}}$ phase (\Tb{}). We provide an interpretation of this result on the basis of the magnon-magnon interactions, which are substantially enhanced in a noncollinear magnetic ground state. This distinction further contrasts the spin dynamics of the tetrahedral triple-${\mathbf{Q}}$ and stripe single-${\mathbf{Q}}$ phases, as only the former exhibits noncollinearity.

Finally, in Section V, we summarize our key findings and outline a generalized protocol for distinguishing between \tred{symmetry-breaking single- or double-$\mathbf{Q}$ phases and a symmetry-preserving triple-$\mathbf{Q}$ phase} using INS. We discuss the broad applicability and practical limitations of our long-wavelength excitation-based protocol, with implications for other existing materials similar to \tred{\CTS{}}. We also describe significant implications of our finding for the triple-\Q{} magnetism of atomically thin \tred{\CTS{}}, make some remarks about the effect of magnon-magnon interactions in the triple-${\mathbf{Q}}$ phase of \tred{\CTS{}}, and discuss the validity and limitations of our isotropic spin Hamiltonian.

\section{Methods}
Single-crystal \tred{\CTS{}} was synthesized following the recipes described in Refs.~\cite{CTS_npj_2022, CTS_tripleQ_natcomm, CTS_composition_2024}. Since slight over-doping of Co is known to induce qualitatively different magnetic properties~\cite{CTS_composition_2024}, the crystals were meticulously characterized by measuring temperature-dependent magnetization along the $c$-axis. As shown in Fig.~\ref{CTS}(c), our diffraction data confirm accurate composition control we achieved for this work: magnetic Bragg peaks associated with over-doped Co$_x$TaS$_{2}$ ($x > 1/3$), characterized by $\mathbf{Q}_{\nu} = (1/3, 0, 0)$~\cite{CTS_composition_2024}, are absent. Furthermore, the lower magnetic transition temperature \TNt{} is highly sensitive to Co content, reaching a maximum of 26.5\,K at $x \approx 0.325$, corresponding to minimal Co vacancies~\cite{CTS_composition_2024}. Only samples exhibiting \TNt{} = 26.5\,K were used for the study. \tred{Thus,} the term “\tred{\CTS{}}” throughout this work refers specifically to samples that exhibit two magnetic transitions and stabilizes the tetrahedral triple-$\mathbf{Q}$ ground state, \tred{with a minimal level of vacancies} [see Fig.~\ref{CTS}(b)].

Using CYTOP (CTL-809M, Asahi Glass, Japan), a total of 172 single-crystal \tred{\CTS{}} pieces (12.05\,g) were co-aligned on multiple aluminium plates, achieving mosaicity within approximately 2$^{\circ}$. The co-aligned assembly was oriented in the $(HHL)$-horizontal geometry (see Fig.~\ref{align} in Appendix A). A sample holder without \tred{\CTS{}} crystals was also prepared to measure background signals independently.

INS data were collected at the 4SEASONS time-of-flight spectrometer at J-PARC, Japan~\cite{4SEASONS}. Using the repetition-rate-multiplication (RRM) technique~\cite{Jparc_RRM}, we simultaneously collected data from multiple incident neutron energies: 46.7, 22.0, 12.8, 8.3, and 5.8\,meV from a chopper frequency of 200 Hz. Data were acquired at 5, 30, and 45\,K, with azimuthal sample rotation over 160$^{\circ}$, and symmetrized according to the symmetry operations of the \CTSideal{} crystal structure. The symmetrization process does not introduce any artificial distortion to the observed $S_{\perp}(\mathbf{q},\omega)$ profile, as confirmed by a direct comparison with unsymmetrized data (see Appendix G). We used the Horace~\cite{Horace} and Utsusemi~\cite{Utsusemi} software packages to analyze and visualize four-dimensional $S_{\perp}(\mathbf{q},\omega)$ maps. Background estimation was performed by measuring the empty sample holder under identical conditions. Unless otherwise specified, the INS data are shown in this work after background subtraction.

Magnon dispersion [$\omega(\mathbf{q})$] and energy- and momentum-resolved $S_{\perp}(\mathbf{q},\omega)$ without temperature effects were calculated using the linear spin-wave theory (LSWT) within the SpinW~\cite{SpinW} software package. Energy and momentum-resolved $S_{\perp}(\mathbf{q},\omega)$ at finite temperatures were calculated by the LLD simulations of a spin system, using the su(n)ny package~\cite{Sunny,Sunny_ref1}. Renormalization of the scalar bi-quadratic interaction term from higher-order 1/$S$ corrections was applied based on the description in Ref.~\cite{dahlbom_renormalized} (or see Appendix B). In our LLD simulations, a temperature-dependent renormalization scheme for the spin length was used, which allows for accurate simulations of magnetic excitation energies even under sizable thermal fluctuations. Further details on this treatment are provided in Refs.~\cite{rescale_Dahlbom, park_ZVPO}.

For the calculations of $S_{\perp}(\mathbf{q},\omega)$ at 45 K (\Th{}), we simulated the time evolution of a \CTSideal{} supercell of size $24\times24\times16$ (18432 Co sites) using a Langevin time step ($\mathrm{d}t$) and a damping constant of 0.02 meV$^{-1}$ and 0.1, respectively. An initial equilibration phase ($t_{\mathrm{eq}}$) was performed for 4,000 Langevin timesteps. The resulting $S_{\perp}(\mathbf{q},\omega)$ was averaged over 4 supercell replicas. For the simulations at 5 and 30 K, we used a larger supercell size of $30\times30\times24$, with $\mathrm{d}t = 0.025$ meV$^{-1}$ and $t_{\mathrm{eq}} = 10000$ time steps. In this case, $S_{\perp}(\mathbf{q},\omega)$ was averaged over 30 independent replicas to ensure an equal population of multiple magnetic domains: three magnetic domains related by a three-fold rotation about the $c$-axis for the single-${\mathbf{Q}}$ ordering, and two magnetic domains with opposite signs of scalar spin chirality for the triple-${\mathbf{Q}}$ ordering.

The resultant $S_{\perp}(\mathbf{q},\omega)$ was multiplied by the neutron polarization factor and the magnetic form factor of Co$^{2+}$. It was then convolved with the instrumental energy and momentum resolutions, each derived from the geometry of 4SEASONS spectrometer and the full width at half-maximum (FWHM) of the (1/2, 0, 1) magnetic Bragg peak along the [$H$, 0, 0], [$-K$, 2$K$, 0], and [0, 0, $L$] directions, respectively. The effects of finite integration range perpendicular to the plotting axes of $S_{\perp}(\mathbf{q},\omega)$ slices were incorporated into the simulations by accounting for the same pixel histogram as the experimental $S_{\perp}(\mathbf{q},\omega)$ slices. Unless noted otherwise, all simulation results presented in this work include the aforementioned treatments. 

\tred{\section{Conceptual basis and material context}}
\subsection{Qualitative view of contrasting Goldstone mode anisotropy}
Although linear modes in the long-wavelength limit are common for any antiferromagnetic structure that spontaneously breaks a continuous symmetry, their momentum-dependent velocity profiles are shaped by the underlying magnetic structure. An intuitive understanding of the contrasting long-wavelength behaviors in \tred{symmetry-breaking} single- or double-$\mathbf{Q}$ and \tred{symmetry-preserving} triple-$\mathbf{Q}$ phases can be gained by examining their real-space spin configurations. Figures~\ref{basic}(a)--(b) illustrate the generic spin alignment of single-$\mathbf{Q}$ and triple-$\mathbf{Q}$ orderings on a triangular lattice, using the simple case where the ordering wave vectors lie at the $M$ points of the Brillouin zone ($\mathbf{Q}_{\nu} = \mathbf{G}_{\nu}/2$). Their characteristic configurations reveal how the Fourier-transformed interaction matrix $J(\mathbf{q})$--which governs magnon velocities in the long-wavelength limit--differs qualitatively between the two cases. In the triple-$\mathbf{Q}$ phase, the spins form a highly symmetric tetrahedral configuration, preserving the three-fold rotational symmetry about the $c$-axis. This ensures that the six nearest-neighbor exchange paths yield the same relative spin angle ($\theta_{ij} = \cos^{-1}(-1/3)$ for $\mathbf{Q}_{\nu}=\mathbf{G}_{\nu}/2$), leading to a momentum-direction-independent phase factor in $J(\mathbf{q})$. As a result, the linear magnon velocity becomes nearly isotropic around the ordering wave vector. However, single-${\mathbf{Q}}$ ground states do not achieve this highly symmetric configuration because they break the three-fold C$_3$ symmetry. Among the six nearest-neighbor bonds, two exchange paths perpendicular to $\mathbf{Q}_{\nu}$ connect ferromagnetically aligned spins, while the others connect nonparallel spins with their relative angle determined by $|\mathbf{Q}_{\nu}|$. This spin configuration results in distinct phase factors along different directions in $\mathbf{q}$-space, naturally introducing anisotropy in the Goldstone-mode velocity. \tred{Similarly, a double-$\mathbf{Q}$ order also breaks the C$_3$ rotational symmetry, resulting in direction-dependent phase factors in $J(\mathbf{q})$, and is therefore expected to exhibit an anisotropic velocity profile}.

Notably, the above discussion suggests that the distinct long-wavelength velocity profile of \tred{the triple-${\mathbf{Q}}$ ground state, as opposed to other symmetry-breaking cases}, should qualitatively persist even for general incommensurate ordering wave vectors of the form $\mathbf{Q}_{\nu} = (q, 0, 0)$ with $0 < q < 1/2$. Regardless of the modulation period, a single-${\mathbf{Q}}$ spiral phase always exhibits ferromagnetic spin alignment along bond vectors ($\boldsymbol{\delta}$) perpendicular to $\mathbf{Q}_{\nu}$. In contrast, the three-fold rotational symmetry ($C_{3z}$) of the triple-${\mathbf{Q}}$ structure ensures a uniform relative spin angle across all six $J_{n}$ bonds. These features imply that the long-wavelength velocity $\mathbf{v}_{\mathrm{L}}({\mathbf{k}})$ can serve as an effective diagnostic tool for distinguishing triple-${\mathbf{Q}}$ orders from \tred{symmetry-breaking single-${\mathbf{Q}}$ or double-$\mathbf{Q}$} structures on the triangular lattice.

More intriguingly, the distinctive pattern of single-\Q{} and triple-\Q{} magnetic orderings highlighted in Fig.~\ref{basic}(a)–-(b) is intrinsic to any 2D lattice types, suggesting that the aforementioned unique characteristics of the symmetry-breaking and symmetry-preserving multi-$\mathbf{Q}$ orderings may be universally applicable to general 2D magnetic systems beyond the triangular lattice structure. A notable example is the comparison between a single-\Q{} spiral and a double-\Q{} skyrmion lattice on a square lattice magnet. Unlike the single-${\mathbf{Q}}$ spiral phase, the four-fold symmetry ($C_{4z}$) of the double-${\mathbf{Q}}$ structure ensures uniform relative spin alignment, naturally leading to (nearly-) isotropic Goldstone mode velocity profile.

Based on the above arguments, we propose that the nearly isotropic velocity of the Goldstone mode is a universal feature of triple-${\mathbf{Q}}$ spin dynamics on hexagonal lattices in sharp contrast to the strongly anisotropic velocity characteristic of single-${\mathbf{Q}}$ \tred{or double-$\mathbf{Q}$} order. \tred{Importantly, while this long-wavelength characteristic may not distinguish between single-${\mathbf{Q}}$ and double-$\mathbf{Q}$ scenarios, it remains a deterministic and broadly applicable approach for identifying rotationally invariant multi-${\mathbf{Q}}$ spin textures with nontrivial topology (e.g., Skyrmion crystals), by systematically excluding their symmetry-breaking counterparts, which are typically topologically trivial}. The following sections present a comprehensive validation of this proposal for the specific case $\mathbf{Q}_{\nu} = \mathbf{G}_{\nu}/2$ and the real material example \tred{\CTS{}}. Further case-specific studies involving different ordering wave vectors and lattice geometries will be essential to test the broader universality of this behavior. However, such investigations require carefully tailored microscopic spin Hamiltonians for each case, which we leave for future work.

\tred{\subsection{Experimental validation using \tred{\CTS{}}}}
\tred{\CTS{}} belongs to a broad family of 3d transition-metal-intercalated van der Waals (vdW) magentic materials~\cite{Parkin_80_v1}. Co atoms are intercalated between the vdW gaps of 2$H$-TaS$_{2}$--a well-known transition metal dichalcogenide--and crystallize in the $P6_{3}22$ space group [Fig.~\ref{CTS}(a)]. The intercalated Co atoms adopt a divalent (Co$^{2+}$) state and carry localized magnetic moments, realizing a layered triangular-lattice magnetic system. Notably, this crystal structure remains unchanged across both the magnetically ordered and paramagnetic phases, as confirmed by the absence of any variation in the nuclear Bragg peak profiles in neutron diffraction measurements~\cite{CTS_tripleQ_natcomm, CTS_tripleQ_nphys}.

\tred{\CTS{}} has recently gained considerable attention for its noncoplanar triple-${\mathbf{Q}}$ magnetic ground state~\cite{CTS_tripleQ_natcomm, CTS_tripleQ_nphys}. Figure~\ref{CTS}(b) summarizes the temperature-dependent magnetic ground states of \tred{\CTS{}}. Neutron diffraction measurements~\cite{CTS_tripleQ_natcomm, CTS_tripleQ_nphys}--including those presented in this work [Fig.~\Ref{CTS}(c)]--reveal magnetic Bragg peaks at the M points of the Brillouin zone in both phases ($\mathbf{Q}_{\nu} = \mathbf{G}_{\nu}/2$).  Below \TNt{}, \tred{\CTS{}} exhibits a large spontaneous Hall conductivity $\sigma_{xy}(\mathbf{H}=0)$~\cite{CTS_npj_2022, CTS_tripleQ_natcomm, CTS_tripleQ_nphys}, which rules out the multi-domain single-${\mathbf{Q}}$ scenario. This is because $\sigma_{xy}(\mathbf{H}=0)$ is strictly forbidden in a single-${\mathbf{Q}}$ long-range order with $\mathbf{Q}_{\nu} = \mathbf{G}_{\nu}/2$, due to the symmetry of time reversal combined with the translation of a lattice vector ($\tau_{1a}T$)~\cite{CTS_tripleQ_natcomm, CTS_tripleQ_nphys}.

The fundamental importance of this commensurate triple-${\mathbf{Q}}$ ordering merits more explanation. This ordering consists only of four sublattices pointing along the principal directions of a regular tetrahedron [Fig.~\ref{CTS}(d)], thereby referred to as the \textit{tetrahedral} triple-${\mathbf{Q}}$ ordering~\cite{Batista_3q_08, CTS_tripleQ_natcomm}. In particular, along with two-sublattice stripe and three-sublattice 120$^{\circ}$ magnetic orderings, it is one of the three fundamental antiferromagnetic configurations in a triangular lattice system~\cite{CTS_tripleQ_natcomm}. Moreover, this spin configuration is the highest density limit of a skyrmion lattice, sharing the same topological characteristics as skyrmions despite lacking a continuous real-space texture~\cite{wang_skyrmion, CTS_tripleQ_natcomm}. The dense real-space Berry curvature due to its small Skyrmion radius can indeed result in a substantial topological Hall effect (THE), explaining the observed spontaneous Hall effect in \tred{\CTS{}} ($\rho_{xy}\sim4\,\mu\Omega\, \mathrm{cm}$, or $\sigma_{xy}\sim70\,\Omega^{-1} \mathrm{cm}^{-1}$)~\cite{CTS_npj_2022, CTS_tripleQ_nphys} that is a few orders of magnitude larger than that observed in typical Skyrmion crystals, such as FeGe~\cite{FeGe_THE} and MnSi~\cite{MnSi_THE}.

In contrast, the intermediate-temperature phase at \Ti{} is suggested to be a single-$\mathbf{Q}$ ordering. This phase is characterized by zero $\sigma_{xy}(\mathbf{H}=0)$ and $M_{z}(\mathbf{H}=0)$, and Co$^{2+}$ magnetic moments are aligned along the out-of-plane direction according to neutron diffraction results~\cite{CTS_tripleQ_natcomm, CTS_tripleQ_nphys}. Based on these findings, previous studies have suggested that the intermediate phase should be a stripe single-${\mathbf{Q}}$ ordering [Fig.~\ref{CTS}(b) and \ref{CTS}(e)]~\cite{CTS_tripleQ_natcomm, CTS_tripleQ_nphys}.

\tred{Given this background, \tred{\CTS{}} provides an ideal platform for experimentally validating the proposed contrasting dynamical properties of single-${\mathbf{Q}}$ and triple-${\mathbf{Q}}$ magnetic orderings within the same spin Hamiltonian, simply by tuning temperature across the two phases.}

\section{Results}
\subsection{Spin model and low-energy dynamics of \tred{\CTS{}}}
The two-step phase transition of the $M$-ordering in \tred{\CTS{}} [Fig.~\ref{CTS}(b)] can be modeled using the phenomenological Hamiltonian including Heisenberg and scalar biquadratic interactions, where the latter term breaks the accidental degeneracy between the stripe single-${\mathbf{Q}}$ and tetrahedral triple-${\mathbf{Q}}$ orderings present in the pure Heisenberg model~\cite{CTS_tripleQ_natcomm}:
\begin{eqnarray}
\mathcal{\hat H} = \mathcal{\hat H}_{\rm Heis} + \mathcal{\hat H}_{\rm Bq}
\label{Hamiltonian}
\end{eqnarray}
with 
\begin{eqnarray}
\mathcal{\hat H}_{\rm Heis} &=&  \frac{1}{2} \sum_{ \substack{{\mathbf{r}}, {\bm{\delta}} \\ a, b}} J^{a b}_{\bm{\delta}} \hat {\mathbf{S}}^{a}_{\mathbf{r}} \cdot \hat {\mathbf{S}}^{b}_{{\mathbf{r}}+{\bm{\delta}}},
\nonumber \\
\mathcal{\hat H}_{\rm Bq} &=& \frac{K}{2} \sum_{{\mathbf{r}}, {\bm{\delta}}_1, a} (\hat {\mathbf{S}}^a_{\mathbf{r}} \cdot \hat {\mathbf{S}}^a_{{\mathbf{r}}+{\bm{\delta}}_1})^{2},
\label{Hamiltonian-terms}
\end{eqnarray}
where ${\bm{\delta}}$ runs over the position vectors of each unit cell, expressed in the basis of primitive vectors $\{{\mathbf {a}}, {\mathbf{b}}, {\mathbf{c}} \}$ shown in Fig.~\ref{CTS}(d), when the origin is at the unit cell ${\mathbf{r}}$ and $a , b \in \{o,e \}$ run over the two Co sublattices corresponding to even and odd Co-layers. The factor of $1/2$ is included to avoid double-counting of each exchange interaction (each bond is shared between two sites). Finally, ${\bm{\delta}}_1$ runs only over nearest-neighbor sites on the same layer.

First, to describe magnetic orderings with wave vectors $\mathbf{Q}_{\nu} = \mathbf{G}_{\nu}/2$ [see Fig.~\ref{CTS}(c)], it is convenient to Fourier transform the Heisenberg term: 
\begin{equation}
\mathcal{\hat H}_{\rm Heis} = \sum_{ \mathbf{Q},a, b} \tilde{J}^{a b}_{\mathbf{q}} \tilde{\mathbf{S}}^{a}_{\mathbf{Q}} \cdot  \tilde{\mathbf{S}}^{b}_{\bar {\mathbf{Q}}},
\end{equation}
with $\bar {\mathbf{Q}} \equiv -\mathbf{Q} $,
\begin{equation}
\tilde{\mathbf{S}}^{a}_\mathbf{Q} = \frac{1}{\sqrt{N}} \sum_{\mathbf{r}} e^{-i \mathbf{Q} \cdot {\mathbf{r}}} \mathbf{S}^{a}_{\mathbf{r}},
\end{equation}
$N$ is the number of unit cells and the Fourier-transformed interaction matrix is given by
\begin{equation}
\tilde{J}^{a b}_{\mathbf{q}} \equiv \frac{1}{2} \sum_{\boldsymbol{\delta}} J^{a b}_{\boldsymbol{\delta}} e^{-i\mathbf{q}\cdot\boldsymbol{\delta}}.
\end{equation}
In particular, to ensure $\mathbf{Q}_{\nu} = \mathbf{G}_{\nu}/2$, $\tilde{J}^{a b}_{\mathbf{q}}$ should have its global minimum in the $\mathbf{q}$-space at $\mathbf{q}=\mathbf{G}_{\nu}/2$ (at the M points). The Fourier components obey the sum rule 
\begin{equation} \label{eq:gc}
    \sum_{\mathbf{q}} \tilde{\mathbf{S}}^{a}_{\mathbf{q}} \cdot \tilde{\mathbf{S}}^{a}_{-\mathbf{q}} = N{\mathbf{S}} \cdot {\mathbf{S}},
\end{equation}
that arises from the real space Casimir invariant ${\mathbf{S}}_{\mathbf{r}} \cdot {\mathbf{S}}_{\mathbf{r}} = S(S+1)$, which becomes $S^2$ in the classical limit ($S\to \infty$). 

For the two Co sublattices with a hexagonal close-packed stacking [see Fig.~\ref{CTS}(a)], antiferromagnetic exchange interactions between even and odd layers~\cite{CTS_tripleQ_natcomm} guarantee that a spin configuration of each layer conincides with the other after being translated along the vector $\mathbf{t} = (1, 1, 1/2)$. Thus, the vector amplitudes on even and odd sublattices are related in the following simple expression:
\begin{equation}
\tilde{\mathbf{S}}^{o}_{\mathbf{Q}_{\nu}} = e^{i {\bf Q}_{\nu} \cdot {\mathbf{t}}}\, \tilde{\mathbf{S}}^{e}_{\mathbf{Q}_{\nu}}.
\end{equation}

However, when only Heisenberg interactions are present, the stripe single-${\mathbf{Q}}$ and tetrahedral triple-${\mathbf{Q}}$ orderings remain accidentally degenerate. More generally, the single-$\mathbf{Q}$ ordering has exactly the same energy as any multi-$\mathbf{Q}$ ordering of the form:
\begin{equation}
    \tilde{\mathbf{S}}^{a}_{\mathbf{q}} =  \tilde{\mathbf{S}}^{a}_{\mathbf{Q}_1} \delta_{\mathbf{q},\mathbf{Q}_1} + \tilde{\mathbf{S}}^{a}_{\mathbf{Q}_2}\delta_{\mathbf{q},\mathbf{Q}_2} +
    \tilde{\mathbf{S}}^{a}_{\mathbf{Q}_3}\delta_{\mathbf{q},\mathbf{Q}_3},
    \label{eq:multi-q}
\end{equation}
where $\mathbf{Q}_{\nu}=\mathbf{G}_{\nu}/2$ ($\nu = 1,2,3$). The three vector amplitudes $\mathbf{S}_{\mathbf{Q}_{\nu}}$ ($\nu = 1,2,3$) are mutually orthogonal and obey the normalization condition~\eqref{eq:gc}
\begin{equation}
|\tilde{\mathbf{S}}^{a}_{\mathbf{Q}_1}|^2 + |\tilde{\mathbf{S}}^{a}_{\mathbf{Q}_2}|^2 + |\tilde{\mathbf{S}}^{a}_{\mathbf{Q}_3}|^2 = NS^2.
\label{eq:norm}
\end{equation}

Notably, despite this degeneracy, the single-${\mathbf{Q}}$ phase becomes the true ground state at any $T>0$ since thermal fluctuations favor a collinear magnetic order~\cite{od_by_disod1,od_by_disod2}. Thus, it is necessary to consider four-spin interactions to realize the tetrahedral triple-${\mathbf{Q}}$ ground state within the spin Hamiltonian framework. The scalar biquadratic interaction in Eq. \ref{Hamiltonian} with $K>$\,0 is the simplest example of such. Yet, it is important to note that other forms of four-spin interactions (e.g. see Ref.~\cite{4spin_general}) should also be considered to develop a complete spin model, although they are often omitted in experimental studies due to the extensive number of interaction coefficients that largely complicates the analysis.

Although $K>0$ indeed favors the noncollinear tetrahedral ordering at $T=0$, the collinear single-${\mathbf{Q}}$ ordering can still emerge as a ground state at finite temperatures due to an order-by-thermal-disorder mechanism~\cite{od_by_disod1, od_by_disod2}. Thus, tuning the magnitude of $K$ controls the presence and position of the single-${\mathbf{Q}}$ to triple-${\mathbf{Q}}$ transition at $T=$\,\TNt{} [Fig.~\ref{CTS}(b)]~\cite{CTS_tripleQ_natcomm}. Our choice of $K \sim 0.06J_{1}$ reproduces \TNt{}/\TNo{} $\sim 0.7$ observed in \tred{\CTS{}}, where the higher-order renormalization for the biquadratic term is considered (see Appendix B).

For the long-wavelength limit, we introduce the relative coordinate ${\mathbf{k}} = \mathbf{q} - \mathbf{Q}_{\nu}$ with $|{\mathbf{k}}| \ll 1$, which measures the deviation from the ordering wave vector. In this limit, both the three-domain single-$\mathbf{Q}$ and single-domain triple-$\mathbf{Q}$ phase result in the universal profile of magnons consisting of linear and quadratic dispersion. We will use subscripts ``$s$'' and ``$t$'' to indicate the low-energy dispersions of the single domain single-$\mathbf{Q}$ and triple-$\mathbf{Q}$ orderings, respectively. Note that, unlike the universally present gapless linear Goldstone mode, the quadratic mode arises only for the $M$-ordering wave vectors ($ \mathbf{Q}_{\nu} =  \mathbf{G}_{\nu}/2$), as a consequence of the accidental degeneracy described above. For generic wave vectors with $0<|\mathbf{Q}_{\nu}|<|\mathbf{G}_{\nu}|/2$, this degeneracy is typically lifted and the gapless quadratic mode is no longer present.

For a mono-domain single-${\mathbf Q}_{\nu}$ ordering, there is a Goldstone mode centered at $\mathbf{Q}_{\nu}$ with the linear dispersion~\cite{CTS_tripleQ_natcomm} 
\begin{eqnarray}
    \omega_{\mathrm{L,s}}({\mathbf{k}}) = 
 \sqrt{v_{\parallel,s}^2 k_{\parallel}^2 + v_{\perp,s}^2 k_{\perp}^2},
\label{eq_magvel}    
\end{eqnarray}
where ${\mathbf k} = k_{\parallel} \hat{{\mathbf Q}}_{\nu} + k_{\perp} \hat{{\mathbf Q}}^{\perp}_{\nu}$, 
\begin{eqnarray}
v_{\parallel,s}  &=& \sqrt{c_{\parallel} (\tilde{J}_{\mathbf 0}^{aa}-\tilde{J}_{\mathbf{Q}_{\nu}}^{aa})}, 
\nonumber \\
v_{\perp,s} &=& \sqrt{c_{\perp}(\tilde{J}_{\mathbf 0}^{aa}-\tilde{J}_{\mathbf{Q}_{\nu}}^{aa})},
\label{eq_1q_vl}    
\end{eqnarray}
with the constants defined through the expansion 
\begin{equation}
\tilde{J}^{aa}_{{\mathbf Q}_{\nu}+ {\mathbf k}} \simeq  \tilde{J}^{aa}_{{\mathbf Q}_{\nu}} + c_{\parallel} {k_{\parallel}}^2 + c_{\perp} {k_{\perp}}^2.
\end{equation}
There are two branches of quadratic modes centered at $\mathbf{Q}_{\nu'}$ ($\nu'\neq\nu$) with anisotropic dispersion~\cite{CTS_tripleQ_natcomm}
\begin{eqnarray}
    \omega_{\mathrm{Q,s}}(\mathbf{k}) = \sqrt{(\alpha-\beta) {k_\parallel}^4 +(\alpha+\beta) {k_\perp}^4},
\end{eqnarray}
where $(k_{\parallel},k_{\perp})$ are coordinates of ${\mathbf k}$ along the two principal axes of $\mathbf{Q}_{\nu'}$ ($\nu'\neq \nu$), and
\begin{eqnarray}
    \alpha &=& 5 c_{\parallel}^2+6c_\parallel c_{\perp} + 5 c_{\perp}^2, \nonumber \\
    \beta &=& \sqrt{25 (c_{\parallel}^4+c_{\perp}^4) - 132 (c_{\parallel}^3 c_{\perp} + c_{\parallel} c_\perp^3) + 470 c_\parallel^2 c_\perp^2 }. \nonumber \\
\end{eqnarray}
However, with three equally populated magnetic domains, $\mathbf{Q}_{1}$, $\mathbf{Q}_{2}$, and $\mathbf{Q}_{3}$ (i.e., all M points) exhibit the same long-wavelength excitation spectrum with both the linear and quadratic magnon modes. 

For the triple-$\mathbf{Q}$ ordering, there is one Goldstone mode around each ordering wave vector, whose velocities along the local principal axes are given by 
\begin{eqnarray}
v_{\parallel,t} &=& \sqrt{(\tilde{J}_{\mathbf 0}^{aa}-\tilde{J}_{\mathbf{Q}_{\nu}}^{aa})(3c_\parallel+c_\perp)/6}, 
\nonumber \\
v_{\perp,t} &=& \sqrt{(\tilde{J}_{\mathbf 0}^{aa}-\tilde{J}_{\mathbf{Q}_{\nu}}^{aa})(c_\parallel +3c_\perp)/6}.
\label{eq_3q_vl}    
\end{eqnarray}
There is also a quadratic mode, 
\begin{equation}
\omega_{\mathrm{Q},t}(\mathbf{k})\simeq \frac{1}{4} \sqrt{(3c_\parallel+c_\perp)(c_\parallel+3c_\perp)} k^2. 
\end{equation}
with $k \equiv (k^2_{\parallel} + k^2_{\perp})^{1/2}$, which results from the accidental degeneracy of multi-$\mathbf{Q}$ orderings defined by Eqs.~\eqref{eq:multi-q} and \eqref{eq:norm}. 

The general trend that $v_{\perp,t}/v_{\parallel,t}$ is nearly unity (isotropic) while $v_{\perp,s}/v_{\parallel,s}$ strongly deviates from unity (anisotropic) is already apparent from the analytic expressions. According to Eqs.~\eqref{eq_1q_vl} and \eqref{eq_3q_vl}, $v_{\perp,s}/v_{\parallel,s}$ can, in principle, vary from zero to infinity, indicating unbounded anisotropy. In contrast, $v_{\perp,t}/v_{\parallel,t}$ is restricted within the range $\sqrt{1/3} \approx 0.58$ to $\sqrt{3} \approx 1.73$. This already highlights, to some extent, the model-independent nature of the contrasting velocity profiles associated with \tred{symmetry-breaking} single-${\mathbf{Q}}$ and \tred{symmetry-preserving} triple-${\mathbf{Q}}$ orderings in the long-wavelength limit. Moreover, as we will demonstrate later, the actual deviation of $v_{\perp,t}/v_{\parallel,t}$ from unity is much smaller than what is implied by the upper ($\sqrt{3}$) and lower ($\sqrt{1/3}$) bounds across the entire parameter space, even reinforcing the robustness of the contrast.

To describe \tred{\CTS{}} using Eq.~\eqref{Hamiltonian}, we incorporated intra-layer exchange interactions up to third nearest neighbors ($J_{n}\equiv J^{aa,bb}_{\bm{\delta}}$ where $\boldsymbol{\delta}$ connects $n^{\mathrm{th}}$ intralayer nearest neighbors) and inter-layer exchange interactions up to second nearest neighbors (NNs)($J_{\mathrm{c}m}\equiv J^{ab}_{\bm{\delta}}$ where $\boldsymbol{\delta}$ connects $m^{\mathrm{th}}$ interlayer NNs), as illustrated in Fig.~\ref{CTS}(d). This inclusion of multiple interactions accounts for the long-ranged nature of magnetic interactions mediated by conduction electrons, such as the Ruderman–Kittel–Kasuya–Yosida (RKKY) mechanism, which plays a key role in the collective behavior of localized Co$^{2+}$ moments in \tred{\CTS{}}~\cite{CTS_tripleQ_natcomm}. These interactions cover all possible paths up to a bond length of approximately $11.5$\,\AA{} (see Table \ref{table:opt}). 

As demonstrated above, the velocity of the linear mode [$\mathbf{v}_{\mathrm{L}}({\mathbf{k}})\equiv (v_{\parallel},v_{\perp})$] is always direction-dependent in momentum space ($v_{\parallel} \neq v_{\perp}$), with its quantitative profile determined by the relative ratios between multiple $J_{n}$ and $J_{\mathrm{c}m}$ parameters. However, as we will show in the following sections, single-${\mathbf{Q}}$ and triple-${\mathbf{Q}}$ magnetic orderings yield qualitatively different ${\mathbf{k}}$-dependence of $\mathbf{v}_{\mathrm{L}}({\mathbf{k}})$ even under the same set of $J_{n}$ and $J_{\mathrm{c}m}$. Thus, once the bilinear exchange parameters are known, comparing the experimental $\mathbf{v}_{\mathrm{L}}({\mathbf{k}})$ with its theoretical expectation from each phase serves as an effective method for unambiguously distinguishing between the single-${\mathbf{Q}}$/triple-${\mathbf{Q}}$ phases, which is the central idea of our approach.

\begin{figure*}[ht]
\includegraphics[width=1\textwidth]{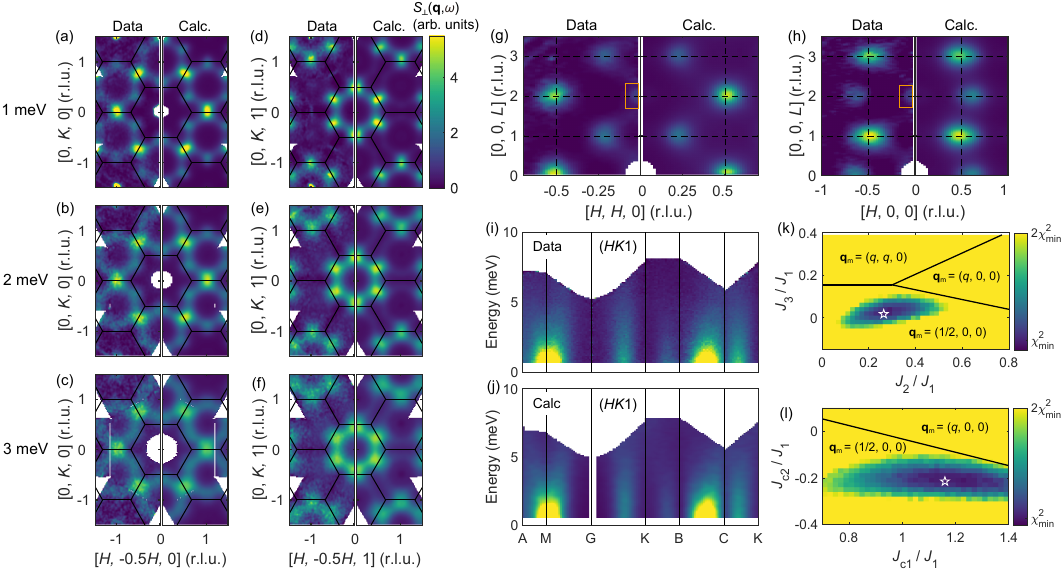} 
\caption{\label{highT} Least-squares fitting of the four-dimensional $S_{\perp}(\mathbf{q},\omega)$ maps from the paramagnetic phase of \tred{\CTS{}} (\Th{}) using the LLD technique. (a)--(f) The left side of each panel shows the measured constant-$\omega$ cuts on the [$H$, $K$, 0] and [$H$, $K$, 1] planes with $\hbar\omega=1, 2,$ and 3\,meV. The right side shows the corresponding LLD simulation results from the optimal exchange parameter set suggested by our Bayesian optimization algorithm (Table \ref{table:opt}). For better presentation, overall scaling factors of 2 and 3 are multiplied to $S_{\perp}(\mathbf{q},\omega)$ for $\hbar\omega= 2,$ and 3\,meV, respectively. (g)--(h) Similar cuts to (a)--(f) but on the [$H$, $H$, $L$] and [$H$, $0$, $L$] planes and at $\hbar\omega= 1.5$\,meV. Orange rectangular boxes in (g) and (h) denote the acoustic phonon signals, which were masked during the optimization process. All constant-$\omega$ cuts are based on data measured with $E_{i} = $\,8.3\,meV and include energy and momentum integrations of $\pm0.5$\,meV and $\pm0.08L$ (r.l.u.), respectively. (i)--(j) Measured and simulated energy-momentum slices along the high-symmetry lines in momentum space [see Fig.~\ref{CTS}(f)]. Data from different $E_{i}$ values are overlaid. The masked low-energy region ($E<0.5\,$meV), dominated by quasi-elastic background signals, was not used for the fitting. (k)--(l) Goodness-of-fit (i.e., $\chi^{2}$-metric) maps around the optimal solution calculated by brute-force scans of the parameter space. White stars denote the best parameter set found by the Bayesian optimization algorithm, listed in Table \ref{table:opt} with uncertainty.}
\end{figure*}

\subsection{Analysis of paramagnetic excitation spectra}
Rather than using the conventional method of spin-wave fitting in magnetically ordered states, we determined the exchange parameters $J_{n}$ and $J_{\mathrm{c}m}$ in \tred{\CTS{}} by analyzing its energy-resolved paramagnetic excitation spectra through semi-classical LLD simulations. This approach offers the following two key advantages for studying \tred{\CTS{}}. 

First, it does not rely on a predefined magnetic ground state, allowing for the determination of optimal exchange parameters independent of the magnetic structure. This flexibility enables a systematic comparison between experimental data and theoretical spin-wave spectra for both single-${\mathbf{Q}}$ and triple-${\mathbf{Q}}$ magnetic structures using a consistent set of exchange parameters. Such consistency is crucial for accurately identifying the correct ground state from spin-wave analysis. Notably, this approach was not employed in Ref.~\cite{CTS_tripleQ_natcomm}, which limited the ability of the previous work in discerning critical differences between the spin dynamics of single-${\mathbf{Q}}$ and triple-${\mathbf{Q}}$ magnetic structures below \TN{}.

Second, analyzing the paramagnetic phase provides a more reliable estimate of the spin Hamiltonian, particularly when significant quantum effects beyond the LSWT are expected in the excitation spectrum below \TN{}. These quantum effects, such as magnon decay, are generally pronounced in $S=1/2$ systems. Analyzing the paramagnetic phase using LLD has recently been recognized as a highly effective method to estimate the spin Hamiltonian in such cases~\cite{park_ZVPO, BLCTO, CoI2_nphys}. Notably, as we discuss in Section \tred{IV. D}, the INS data of \tred{\CTS{}} indicates the presence of nonlinear effects beyond LSWT in the triple-${\mathbf{Q}}$ phase.

The left panels of Figs. \ref{highT}(a)-(h) and Fig.~\ref{highT}(i) display nine slices from a four-dimensional $S_{\perp}(\mathbf{q},\omega)$ map measured at 45 K ($T = 1.18$\TNo{}), covering all principal directions in the $\mathbf{q}-\omega$ space. Despite broadening due to large thermal fluctuations, each slice shows a distinct distribution of $S_{\perp}(\mathbf{q},\omega)$ along both the ${\mathbf{q}}$ and $\omega$ axes. For example, the strongest diffuse scattering signal at the M points of the Brillouin zones exhibits an elongated shape towards the [\textit{H}, 0, 0] or its symmetry-equivalent directions in constant-$\omega$ cuts [e.g., Fig.~\ref{highT}(a) and \ref{highT}(d)]. These patterns across the nine slices put sufficient constraints on estimating multiple bilinear exchange parameters with high accuracy (see Appendix E and Fig.~\ref{wrong}). The exchange parameters of \tred{\CTS{}} were refined through least-squares fitting of our LLD simulations to the nine measured $S_{\perp}(\mathbf{q},\omega)$ slices in Fig.~\ref{highT}. To efficiently search for a global minimum of the goodness-of-fit in a reasonable time frame, we adopted an advanced optimization algorithm, specifically Bayesian optimization, detailed in Appendix D.

The right panels of Figs. \ref{highT}(a)-(h) and Fig.~\ref{highT}(j) show the LLD simulation results obtained using the best-fit parameter set ($J_{1},J_{2},J_{3},J_{c1},J_{c2}$) suggested by the Bayesian optimization algorithm. These results demonstrate remarkable agreement with the observed $S_{\perp}(\mathbf{q},\omega)$, indicating that these five exchange interactions effectively capture the spin Hamiltonian of \tred{\CTS{}}. The set of optimal parameters and their uncertainties are summarized in Table \ref{table:opt}. In particular, the nearest-neighbor interlayer exchange $J_{c1}$ is larger than the nearest-neighbor intralayer exchange $J_{1}$, reflecting the 3D nature of the spin Hamiltonian. The solution suggested by our optimization algorithm has been further validated by examining the $\chi^{2}$ metric — the measure of goodness-of-fit — around the optimal solution in the ($J_{1},J_{2},J_{3},J_{c1},J_{c2}$) parameter space [Figs. \ref{highT}(k)--(l)]. A well-defined minimum of $\chi^{2}$ ($\chi^{2}_{\mathrm{min}}$) is found at the position indicated by the optimization algorithm (white stars). Additional diagnostic analyses described in Appendix E further corroborate the solution.

\renewcommand{\arraystretch}{1.1}
\begin{table}[!h]

 \caption{Optimal parameter set for the five exchange interactions ($J$) and individual standard deviations ($\sigma_{J}$). We used $S=3/2$ for the spin length. The relative magnitudes to $J_{1}$ are also listed for easier comparison with theoretical magnetic phase diagrams in Fig.~\ref{highT}(k)--(l).}
 \begin{center}
 \centering
 \begin{tabularx}{\columnwidth}{c c c c c c} 
 \midrule \midrule
  & $J_1$ & $J_2$ & $J_3$ & $J_{c1}$ & $J_{c2}$ \\
 \midrule
 \ \ $J$ (meV) \ \ & 1.212 & 0.320 & 0.022 & 1.406 & -0.260 \\
 \ \ $\sigma_{J}$ (meV) \ \ & $\pm$0.104 & $\pm$0.061 & $\pm$0.036 & $\pm$0.097 & $\pm$0.036 \\ 
 \ \ $J/J_{1}$ \ \ & 1 & 0.264 & 0.018 & 1.160 & -0.215 \\  
 \ \ Bond length (\AA) \ \ & 5.75 & 9.96 & 11.5 & 6.80 & 8.91 \\
 \midrule \midrule
 \end{tabularx}
 \end{center}
\label{table:opt}
\end{table}

\begin{figure*}[hbtp]
\includegraphics[width=1\textwidth]{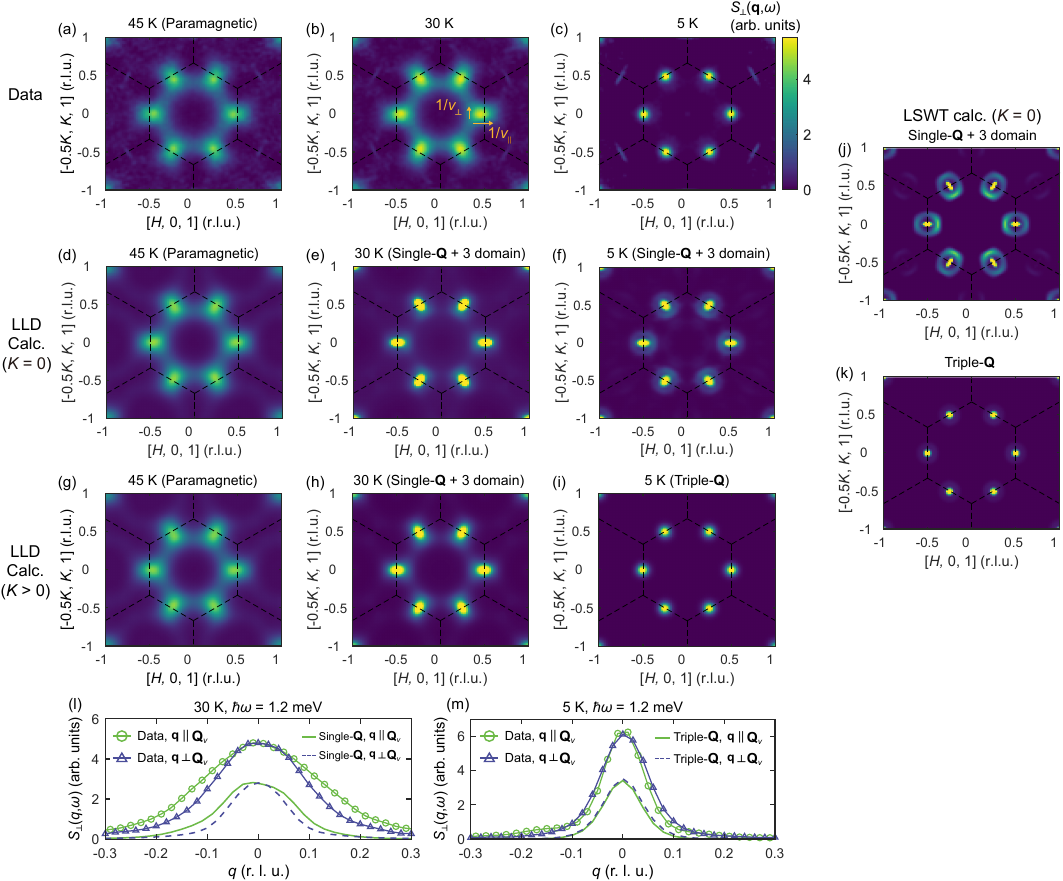} 
\caption{\label{longw} Distinct long-wavelength magnetic excitation spectra of the paramagnetic, stripe single-${\mathbf{Q}}$ and tetrahedral triple-${\mathbf{Q}}$ states. (a)--(c) Constant-$\omega$ slices at $\hbar\omega=1.2$\,meV measured at (a) 45\,K (\TNo{}\,$<T$), (b) 30\,K (\Ti{}), and (c) 5\,K (\Tb{}). $v_{\parallel}$ and $v_{\perp}$ in (b) denote $v_{1}(\mathbf{q}\,\parallel\,\mathbf{Q}_{\nu})$ and $v_{1}(\mathbf{q} \perp \mathbf{Q}_{\nu})$, respectively; see Section IV. A. (d)--(f) LLD simulation results corresponding to (a)--(c) obtained from the Hamiltonian determined above \TNo{}, but without finite $K$. This results in single-${\mathbf{Q}}$ spin dynamics at both (e) 30\,K and (f) 5\,K. (g)--(i) Same as (d)--(f) but with finite $K>0$, resulting in single-${\mathbf{Q}}$ and triple-${\mathbf{Q}}$ spin dynamics at 30 and 5\,K, respectively (see Fig.~\ref{MC} in Appendix F). All constant-$\omega$ cuts are based on the data measured with $E_{i} = $\,5.8\,meV and include energy and momentum integrations of $\pm0.3$\,meV and $\pm0.08L$ (r.l.u.), respectively. (j)--(k) LSWT simulation results of the INS spectra in (b)--(c), which, unlike the LLD simulations, do not include any thermal fluctuation effects. (l) Line cuts of (b) and (h) along directions parallel and perpendicular to $\mathbf{Q}_{\nu}$ at $\hbar\omega = 1.2$\,meV, centered at the M point ($q = 0$). (m) Same as (l), but for the 5\,K data. Symbols represent experimental data, while solid and dashed curves without symbols are LLD simulations with $K>0$. For effective comparison of velocity profiles (i.e., inverse linewidths), $q$-axes were shifted and intensities rescaled to align and normalize peak positions. Error bars are smaller than the symbol size.}
\end{figure*}

Overlaying the $\chi^{2}(J_{1},J_{2},J_{3},J_{c1},J_{c2})$ map on a theoretical magnetic phase diagram at $T=0$ elucidates the magnetic order suggested by our spin model. The phase boundaries, calculated from classical Monte-Carlo simulations, are shown on the $\chi^{2}$ map in Figs. \ref{highT}(k)--(l). The optimal parameter set indeed stabilizes a magnetic order with $\mathbf{Q}_{\nu}$ = $\mathbf{G}_{\nu}/2$ or its symmetry-equivalent vectors [i.e., $\tilde{J}_{\mathbf{q}}$ has a global minimum at ${\mathbf{Q}}$ = $\mathbf{G}_{\nu}/2$], in accordance with observations in \tred{\CTS{}} [Fig.~\ref{CTS}(c)]~\cite{CTS_tripleQ_natcomm, CTS_tripleQ_nphys}.

However, this does not reveal whether the ground state is triple-${\mathbf{Q}}$ or single-${\mathbf{Q}}$, as they are degenerate under isotropic bilinear exchange interactions. It should be noted that estimating $K$ from the fit of the high-temperature spectrum is subject to substantial uncertainty due to its smaller magnitude relative to $|J_{1}|$. For instance, Figs. \ref{longw}(d) and \ref{longw}(g), which show constant-$\omega$ slices without and with finite $K$, are almost identical. Nevertheless, as we will demonstrate in the next section, successfully determining the bilinear interaction coefficients is sufficient to distinguish single-${\mathbf{Q}}$ and triple-${\mathbf{Q}}$ magnetic ground states from spin-wave spectra.

\subsection{Long-wavelength magnon spectra}
With the bilinear exchange interactions determined at $T>$\,\TN{}, we show that analyzing in-plane profile of $\mathbf{v}_{\mathrm{L}}(\mathbf{k})$ in an ordered phase can unambiguously differentiate between a single-domain triple-${\mathbf{Q}}$ phase and a triple-domain single-${\mathbf{Q}}$ phase. This distinction arises from the inherent contrast in their velocity anisotropy, as established in previous sections. Crucially, we demonstrate that this contrast is robust against variations in the exchange parameters, indicating that it is not specific to \tred{\CTS{}} but broadly applicable to systems with arbitrary \tred{interaction strengths}.

The in-plane profile of $\mathbf{v}_{\mathrm{L}}(\mathbf{k})$ can be visualized by plotting a constant-$\omega$ slice of $S(\mathbf{q}, \omega)$ with $\hbar \omega$ set sufficiently below the overall energy bandwidth of the magnetic excitations [see dashed lines in Fig.~\ref{basic}(c)--(d)]. For example, an isotropic $\mathbf{v}_{\mathrm{L}}({\mathbf{k}})$ will produce a circular pattern centered at $\mathbf{q}=\mathbf{Q}_\nu$ in the constant-$\omega$ slice, while a higher velocity along $\mathbf{k}\,//\,\mathbf{Q}_{\nu}$ ($v_{\parallel}>v_{\perp}$) will result in an ellipsoidal pattern elongated in the direction perpendicular to $\mathbf{Q}_\nu$. See the orange text and arrows in Fig.~\ref{longw}(b) or Fig.~\ref{magvel_detail} in Appendix I.

Figs. \ref{longw}(a)-(c) show the constant-$\omega$ slices of $S_{\perp}(\mathbf{q},\omega)$ at $\hbar\omega = 1.2$\,meV, each measured from the paramagnetic (45\,K, \Th{}), the intermediate (30\, K, \Ti{}), and the low-$T$ (5\,K, \Tb{}) phases, respectively. While, as shown in Fig.~\ref{longw}(b), the intermediate phase possesses strong in-plane anisotropy of $\mathbf{v}_{\mathrm{L}}({\mathbf{k}})$, the low-$T$ phase exhibits nearly isotropic $\mathbf{v}_{\mathrm{L}}({\mathbf{k}})$, as shown in Fig.~\ref{longw}(c). This contrast suggests distinct underlying magnetic structures at \Ti{} and \Tb{}, likely due to the single-$\mathbf{Q}$ and triple-$\mathbf{Q}$ nature of each phase, as anticipated from the general considerations discussed in Sections \tred{III. A and IV. A}.

As expected, comparing the observed spectra with the corresponding LLD simulation results with and without finite $K>0$ demonstrates that the intermediate and low-$T$ phases are single-${\mathbf{Q}}$ and triple-${\mathbf{Q}}$, respectively. It is important to note that for these simulations, we consistently use the optimal set of bilinear exchange parameters determined at \Th{}. First, as shown in Fig.~\ref{MC}(b) of Appendix F, the triple-${\mathbf{Q}}$ ordering does not appear in the classical thermodynamic phase diagram for $K=0$, thereby yielding a single-${\mathbf{Q}}$ magnon spectrum at both 30 and 5\,K, as shown in Figs.~\ref{longw}(e) and \ref{longw}(f) respectively. Contrary to the experimental observations, the simulated spectra at 30 and 5\,K display nearly the same $\mathbf{v}_{\mathrm{L}}({\mathbf{k}})$, except that the 30\,K spectrum is broader due to enhanced thermal fluctuations. The calculation result remains similar for negative $K$, as the system still retains the stripe single-${\mathbf{Q}}$ ground state.

However, the simulation results with $K=0.06J_{1}$ successfully capture the measured spectra. LLD simulations of this model reproduce the two-step phase transition process depicted in Fig.~\ref{CTS}(b) [see also Fig.~\ref{MC}(a) in Appendix F] and consequently provide triple-${\mathbf{Q}}$ magnon spectra at 5\,K [Fig.~\ref{longw}(i)] and single-${\mathbf{Q}}$ spectra at 30\,K [Fig.~\ref{longw}(h)]. The LLD simulations reproduce both the anisotropic $\mathbf{v}_{\mathrm{L}}({\mathbf{k}})$ observed at 30\,K [Fig.~\ref{longw}(b)] and the isotropic $\mathbf{v}_{\mathrm{L}}({\mathbf{k}})$ at 5\,K [Fig.~\ref{longw}(c)], as further illustrated by the line cuts around the M points [Fig.~\ref{longw}(l)--(m)]. In other words, the combination of the optimal exchange parameter set determined at \Th{} and $K>0$ describes the spin dynamics observed in all three phases simultaneously. This result not only supports the parameters presented in Table \ref{table:opt}, but also provides evidence that the phases at \Tb{} and \Ti{} correspond to the triple-${\mathbf{Q}}$ and single-${\mathbf{Q}}$ orderings, respectively.
 
Another noticeable contrast between these two magnetic orderings is the intensity of quadratic magnon modes. This contrast is more clearly illustrated in Figs. \ref{longw}(j)--(k), showing the single-${\mathbf{Q}}$ and triple-${\mathbf{Q}}$ magnon spectra calculated at 0~K using LSWT, which does not include thermal fluctuation effects. Although a quadratic mode signal is still present in both the single-${\mathbf{Q}}$ and triple-${\mathbf{Q}}$ calculations, its relative spectral weight compared to the linear mode is extremely weak in the triple-${\mathbf{Q}}$ phase. This observation is consistent with our data at 30 and 5\,K, further supporting that \TNt{} marks the transition between the tetrahedral triple-${\mathbf{Q}}$ and stripe single-${\mathbf{Q}}$ orderings. For completeness, we also considered the case of a double-${\mathbf{Q}}$ magnetic ordering, which, although not realized in \tred{\CTS{}}, provides an important point of comparison \tred{for firmly isolating the triple-$\mathbf{Q}$ nature of the ground state}. As elaborated in Appendix H, the double-${\mathbf{Q}}$ state exhibits a strongly anisotropic Goldstone-mode velocity profile, closely resembling the single-${\mathbf{Q}}$ case. This indicates that the nearly isotropic dispersion is a distinctive signature of the triple-${\mathbf{Q}}$ phase, \tred{consistent with our proposal in Section III. A, which is based on the qualitative distinction between triple-${\mathbf{Q}}$ and other symmetry-breaking orders}.

\begin{figure}[ht]
\includegraphics[width=1.0\columnwidth]{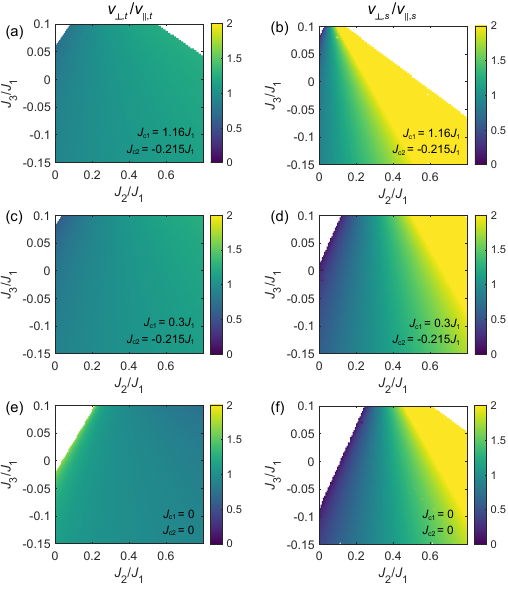} 
\caption{\label{magvel} Distinct anisotropy in the linear magnon's velocity [$v_{\mathrm{L}}(\mathbf{k})$] between the tetrahedral triple-${\mathbf{Q}}$ and stripe single-${\mathbf{Q}}$ magnetic structures. (a) $J_{2}/J_{1}$ and $J_{3}/J_{1}$ dependence of $v_{\perp}/v_{\parallel}$ for the triple-${\mathbf{Q}}$ phase calculated by LSWT, where $v_{\parallel}$ and $v_{\perp}$ are $v_{\mathrm{L}}(\mathbf{k}\,//\,\mathbf{Q}_{\nu})$ and $v_{\mathrm{L}}(\mathbf{k}\perp \mathbf{Q}_{\nu})$, respectively; see Fig.~\ref{longw}(b). (b) Same as (a), but for the single-${\mathbf{Q}}$ phase. For the calculations in (a) and (b), the optimal values of $J_{\mathrm{c1}}$ and $J_{\mathrm{c2}}$ were used (Table \ref{table:opt}). (c)--(f) Same as (a)-(b), but calculated with (c)--(d) reduced and (e)--(f) zero interlayer interactions $J_{\mathrm{c1}}$ and $J_{\mathrm{c2}}$. Empty regions in the color plots indicate instability of the stripe single-${\mathbf{Q}}$ or tetrahedral triple-${\mathbf{Q}}$ magnetic orderings for given $J_{2}/J_{1}$ and $J_{3}/J_{1}$.}
\end{figure}

Next, we investigate whether the observed anisotropic (nearly isotropic) $\mathbf{v}_{\mathrm{L}}({\mathbf{k}})$ of the single-${\mathbf{Q}}$ (triple-${\mathbf{Q}}$) phase is a characteristic dynamical feature that persists independently of the specific exchange parameters. Fig.~\ref{magvel} shows the degree of anisotropy in $\mathbf{v}_{\mathrm{L}}({\mathbf{k}})$ across a wide parameter space. This anisotropy can be quantified by the ratio $v_{\perp}/v_{\parallel}$, where $v_{\parallel}$ and $v_{\perp}$ are $\mathbf{v}_{\mathrm{L}}({\mathbf{k}})$ for $\mathbf{k}\parallel \mathbf{Q}_{\nu}$ and $\mathbf{k} \perp \mathbf{Q}_{\nu}$, respectively [see Section \tred{IV. A} and orange texts in Fig.~\ref{longw}(b)]. Figs. \ref{magvel}(a)--(b) show $v_{\perp}/v_{\parallel}$ for the triple-${\mathbf{Q}}$ and single-${\mathbf{Q}}$ orderings as a function of $J_{3}/J_{1}$ and $J_{2}/J_{1}$, using the optimal interlayer exchange parameters: $J_{c1} = 1.16J_{1}$ and $J_{c2}=-0.22 J_{1}$. Interestingly, $v_{\perp,t}/v_{\parallel,t}$ (anisotropy ratio for the triple-${\mathbf{Q}}$ phase) remains close to 1 across the wide parameter space, whereas $v_{\perp}/v_{\parallel}$ of the single-${\mathbf{Q}}$ phase ($v_{\perp,s}/v_{\parallel,s}$) generally deviates significantly from 1. 

Furthermore, this contrast remains qualitatively intact even with reduced or zero interlayer interactions (i.e., 2D spin Hamiltonian). This is illustrated in Fig.~\ref{magvel}(c)--(f), which display the $v_{\perp,t}/v_{\parallel,t}$ and $v_{\perp,s}/v_{\parallel,s}$ maps resulting from much weaker or zero $J_{c1}$ and $J_{c2}$. Thus, the distinction in $\mathbf{v}_{\mathrm{L}}({\mathbf{k}})$ between the single-${\mathbf{Q}}$ and triple-${\mathbf{Q}}$ phases persists across a broad range of exchange parameters. See Fig.~\ref{magvel_detail} in Appendix I for a more explicit presentation of this contrast.

A quantitative assessment of $v_{\perp,t}/v_{\parallel,t}$ and $v_{\perp,s}/v_{\parallel,s}$ further highlights their stark contrast. As described in Section \tred{IV. A}, $v_{\perp,s}/v_{\parallel,s}$ can vary from 0 to infinity, whereas $v_{\perp,t}/v_{\parallel,t}$ is limited from $\sqrt{1/3} = 0.58$ to $\sqrt{3} = 1.732$. However, the variation of $v_{\perp,t}/v_{\parallel,t}$ within the parameter space is not uniform; it changes rapidly around the phase boundary of $\mathbf{Q}_{\nu}$\,=\,$1/2\mathbf{G}_{\nu}$ and converges toward its upper or lower bound at the boundary. As a result, the expected deviation of $v_{\perp,t}/v_{\parallel,t}$ from 1 across the entire parameter space is much smaller than what is implied by the upper ($\sqrt{3}$) and lower ($\sqrt{1/3}$) bounds. Indeed, average and standard deviation values of the $v_{\perp,t}/v_{\parallel,t}$ map in Fig.~\ref{magvel}(a) and~\ref{magvel}(e) are $1.07\pm0.02$ and $1.00\pm0.04$, demonstrating its narrow distribution centered around 1. On the other hand, the variation of $v_{\perp,s}/v_{\parallel,s}$ in the parameter space is much more pronounced, meaning that $v_{\perp,s}/v_{\parallel,s}\sim1$ is realized only in a very confined region of the parameter space.

These observations consistently reveal an intrinsic distinction in $\mathbf{v}_{\mathrm{L}}({\mathbf{k}})$ between single-${\mathbf{Q}}$ and triple-${\mathbf{Q}}$ orderings, clearly demonstrated for the $M$-ordering case ($\mathbf{Q}_{\nu} = \mathbf{G}_{\nu}/2$) and expected to apply more broadly based on the microscopic arguments presented in Section~III.A. This distinction is particularly significant in the context of two-dimensional spin Hamiltonians with negligible interlayer coupling; that is, in genuine 2D magnets. We further note that the biquadratic interaction $K$ has minimal influence on these results, primarily due to its magnitude being much smaller than that of the bilinear exchange terms, as is typically observed in real materials (see Appendix~I). The sole role of $K$ is to select the triple-${\mathbf{Q}}$ ordering and, consequently, to gap out the otherwise quadratic mode associated with the accidental degeneracy.

\begin{figure*}[ht]
\includegraphics[width=1\textwidth]{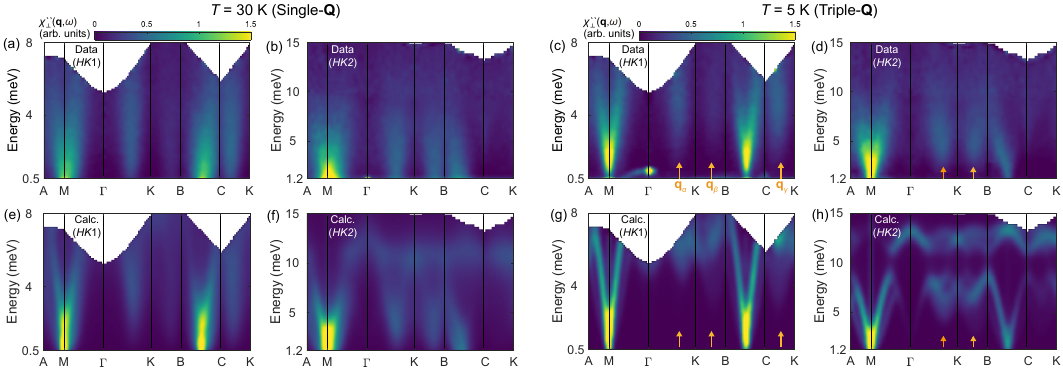} 
\caption{\label{dispersion} Full magnon spectra of the stripe single-${\mathbf{Q}}$ and tetrahedral triple-${\mathbf{Q}}$ magnetic orderings. (a)--(b) Magnetic spectrum measured at 30\,K along the high-symmetry lines on the [$H$, $K$, 1] and [$H$, $K$, 2] planes [see Fig.~\ref{CTS}(f)]. (c)--(d) Same as (a)--(b), but measured at 5\,K. The energy range not shown in these plots is dominated by quasi-elastic background signals. (e)--(h) LLD simulation results corresponding to (a)--(d), respectively. Note that the colorplots in this figure shows dynamical susceptibility [$\chi''_{\perp}(\mathbf{q},\omega)$]. Orange arrows in (c)--(d) and (g)--(h) indicate apparent discrepancies between the data and simulation results for the triple-${\mathbf{Q}}$ phase, suggesting the presence of additional effects beyond the noninteracting magnon picture.}
\end{figure*}

\begin{figure}[ht]
\includegraphics[width=1\columnwidth]{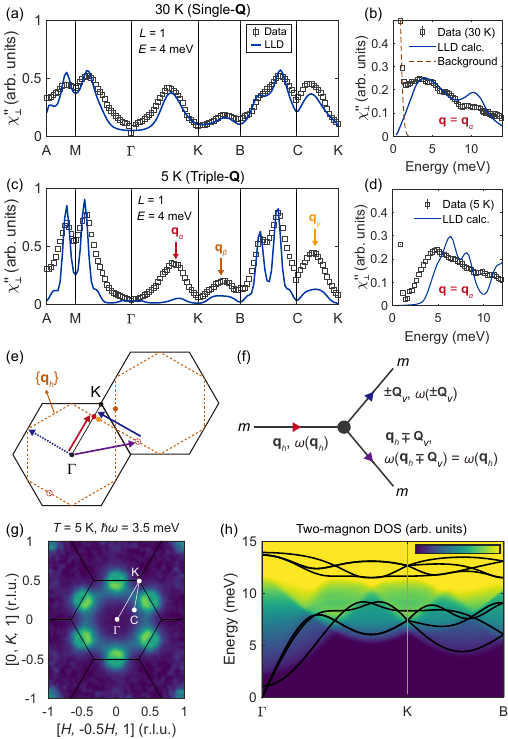} 
\caption{\label{2mag} Magnon energy renormalization in the tetrahedral triple-${\mathbf{Q}}$ phase. (a)--(b) Comparison between the INS data and LLD simulation results at 30\,K through constant-$\omega$ (4\,meV) and ${\mathbf{Q}}$ ($\mathbf{q}_{\alpha}$) cuts from Figs. \ref{dispersion}(a) and \ref{dispersion}(e). The dashed brown line in (b) represents the estimated quasielastic background signal based on a single Gaussian peak fitting. (c)--(d) Same as (a)--(b), but for 5\,K [Figs. \ref{dispersion}(c) and \ref{dispersion}(g)]. (e) Momentum positions $\{\mathbf{q}_{h}\}$ (dashed orange lines) where sizable magnon energy renormalization is observed at 5\,K. Red-, orange-, and yellow-filled circles indicate $\mathbf{q}_{\alpha}$, $\mathbf{q}_{\beta}$, and $\mathbf{q}_{\gamma}$ in (c), respectively. The three solid arrows illustrate a decay process of a single magnon on $\{\mathbf{q}_{h}\}$ (red) into two magnons. (f) Diagrammatic representation of the magnon decay process depicted in (e). $\mathbf{Q}_{\nu}$ denotes three ordering wave vectors ($\nu=1,2,3$) corresponding to the three M points of the Brillouin zone. (g) A constant-$\omega$ slice at 3.5\,meV on the [$H$, $K$, 1] plane, demonstrating the magnon signal along the line $\{\mathbf{q}_{h}\}$. (h) Noninteracting two-magnon density of states (DOS) calculated from bare one-magnon dispersion. Black solid lines show the one-magnon dispersion obtained from LSWT.}
\end{figure}

\subsection{Failure of noninteracting magnon picture in the triple-${\mathbf{Q}}$ phase}
The full spin-wave spectrum of \tred{\CTS{}} beyond the long-wavelength regime provides more insights into the nature of the triple-${\mathbf{Q}}$ and single-${\mathbf{Q}}$ spin dynamics. Figs. \ref{dispersion}(a)--(b) and \ref{dispersion}(c)--(d) show the magnon spectra over a wide energy-momentum space at 30\,K (single-${\mathbf{Q}}$) and 5\,K (triple-${\mathbf{Q}}$), respectively. The most conspicuous feature of these spectra is their broadness. For the 5\,K spectra, where thermal fluctuations are minimal, this broadness is undoubtedly an intrinsic feature rather than an artifact of sample mosaicity or instrumental resolution effects. This is evidenced by the fact that the phonon branches observed in the high-${\mathbf{Q}}$ region of the same dataset exhibit much sharper spectra (see Fig.~\ref{phonon_com} in Appendix J). This observation, at least for the triple-${\mathbf{Q}}$ phase, indicates a clear breakdown of the noninteracting magnon picture. On the other hand, sizable thermal fluctuations contribute significantly to the broadness for the single-${\mathbf{Q}}$ spectra measured at 30\,K, (=\,0.79\TNo{}), which is difficult to disentangle from the intrinsic linewidth broadening relevant to finite magnon lifetime. Thus, although some intrinsic broadening likely exists at 30\,K as well, it is less clear than at 5\,K.

Figs. \ref{dispersion}(e)--(h) show the magnon spectra simulated by LLD using the parameters in Table \ref{table:opt} and $K=0.06J_{1}$. For $\hbar\omega>10\,$meV, where the optical branches appear in our LLD simulation [Figs. \ref{dispersion}(f) and \ref{dispersion}(h)], the INS spectra are heavily damped for both the single-${\mathbf{Q}}$ and triple-${\mathbf{Q}}$ phases compared to the simulation results. While this makes a more detailed comparison with the LLD simulations challenging, such deviations from the spin-wave theory's predictions, which are based on a fully localized picture of the magnetic moments, are commonly observed for high-energy excitations in metallic antiferromagnets~\cite{AFM_metal_1, AFM_metal_2, AFM_metal_3, CaFe2As2_nphys, FeSn_INS}. This is due to the prevalence of the Stoner continuum in their energy-momentum space, whose influence generally increases with energy transfer~\cite{AFM_metal_1, AFM_metal_2, AFM_metal_3, CrB2}. This metallic character may also be responsible for the reduced ordered moment of \tred{\CTS{}} in the triple-${\mathbf{Q}}$ phase: $\mu_{s}=1.3\mu_{\mathrm{B}}$~\cite{CTS_tripleQ_natcomm}, which is less than half of the value expected for a fully localized scenario ($gS = 3\mu_{\mathrm{B}}$)~\cite{moriya}.

However, analysis of the low-energy region ($\hbar\omega<10$\,meV) suggests that the Stoner continuum may not fully explain the spin dynamics beyond LSWT in \tred{\CTS{}}. A clear magnon dispersion observed in this low-energy region still offers insights when comparing the data with the results of LLD simulations. At 30\,K, LLD shows overall satisfactory agreement with the data regarding magnon dispersion [Figs. \ref{dispersion}(a)--(b) and \ref{dispersion}(e)--(f)], which is further demonstrated by the constant-${\mathbf{Q}}$ and constant-$\omega$ cuts in \ref{2mag}(a)--(b). While LLD overestimates the intensity at low energy transfer values around the M point [see Fig.~\ref{dispersion}(a) and \ref{dispersion}(e)], this is partially attributed to a calculation artifact: the gapless linear mode at the M point possesses a diverging structure factor in the calculation, which, due to resolution convolution, generates sizable intensity that extends up to finite energy transfer values in the simulation.

Despite exhibiting a magnon spectrum similar to that of the single-${\mathbf{Q}}$ phase, the spectrum of the triple-${\mathbf{Q}}$ phase measured at 5\,K shows an apparent inconsistency with the LLD simulations. The orange arrows indicate this discrepancy in Figs. \ref{dispersion}(c)--(d) and \ref{dispersion}(g)--(h): although our calculation reproduces a local minimum of the magnon dispersion at these ${\mathbf{Q}}$ points, it predicts a much shallower dip into lower energy [see also Fig.~\ref{2mag}(c)--(d)]. Importantly, such steep downward dispersion cannot be reproduced by any reasonable set of exchange parameters in LLD; see Appendix J and Fig.~\ref{mageig} therein. This suggests the presence of substantial magnon energy renormalization beyond LSWT in the triple-${\mathbf{Q}}$ phase. This contrasts with the single-${\mathbf{Q}}$ phase at 30\,K, where the LLD calculation captures the deep downward dispersion. Note that the deep downward dispersion observed at 30\,K is attributed to both a steeper magnon dispersion in the single-\Q{} phase and the sizable thermal fluctuations at 30\,K, further depressing the dispersion minimum in the spectrum.

The magnon energy renormalization pronounced only in the triple-${\mathbf{Q}}$ phase hints at its microscopic origin. In \tred{\CTS{}}, possible origins for this renormalization include: (i) interactions between magnons and conduction electrons (i.e., renormalization by the Stoner continuum), and (ii) magnon-magnon interactions (i.e., renormalization by the multi-magnon continuum). However, these two factors depend differently on the detailed spin configuration. The second mechanism is significantly enhanced when a magnetic structure becomes noncollinear due to the generation of cubic vertices~\cite{Noncollinear_SWT1, Noncollinear_SWT2, RMP_Mdecay}. This leads to significantly larger magnon decay and renormalization in the noncollinear triple-${\mathbf{Q}}$ ordering compared to the collinear single-${\mathbf{Q}}$ ordering. On the other hand, the effect of the Stoner continuum is anticipated to be similar in both magnetic structures, or even smaller in a noncollinear magnet since a Stoner excitation process requires a full spin flip. See Appendix K for further explanation. Thus, our observation agrees better with the magnon-magnon interaction mechanism, suggesting that it is the primary origin of energy renormalization.

Interestingly, the ${\mathbf{q}}$ positions where pronounced energy renormalization occurs provide unique insights into the three-magnon process of the tetrahedral triple-${\mathbf{Q}}$ ordering. These ${\mathbf{q}}$ positions, $\mathbf{q}_{\alpha,\beta,\gamma}$ in Figs. \ref{dispersion}(c) and \ref{2mag}(c)--(d), lie on the hexagon that connects the six M points of a Brillouin zone [denoted as $\{\mathbf{q}_{h}\}$ in Fig.~\ref{2mag}(e)]. The intuition behind why this specific ${\mathbf{q}}$-path might be significant for the three-magnon process of the triple-$\mathbf{Q}$ phase relates to its stabilization mechanism, which involves Fermi surface nesting (or quasi-nesting). The $M$--ordering wave vectors extensively connect different positions on $\{\mathbf{q}_{h}\}$, thereby acting as nesting wave vectors when $\{\mathbf{q}_{h}\}$ corresponds to the Fermi surface and thus stabilizing the tetrahedral triple-${\mathbf{Q}}$ ordering~\cite{Batista_3q_08, Batista_2016_review}. An analogous scattering process can occur for magnons: a magnon with $\mathbf{q}_{1}\in\{\mathbf{q}_{h}\}$ can decay into two magnons with $\mathbf{q}_{2}=\mathbf{Q}_{\nu}$ and $\mathbf{q}_{3}\in\{\mathbf{q}_{h}\}$, where $\mathbf{q}_{3}$ is equivalent to $\mathbf{q}_{1}$. An example of such a process is illustrated in Fig.~\ref{2mag}(e)--(f). Importantly, the kinematic conditions for these three-magnon processes can always be satisfied since Goldstone modes arising from Eq. \ref{Hamiltonian} have zero energy at $\mathbf{q}=\mathbf{Q}_{\nu}$. This implies that observing the pronounced effects of the magnon-magnon interaction specifically at $\mathbf{q}\in\{\mathbf{q}_{h}\}$ is feasible. In fact, a signal from the renormalized magnon mode is observed consistently at all positions on $\{\mathbf{q}_{h}\}$ [Fig.~\ref{2mag}(g)].

For a deeper understanding, we calculated the energy and momentum-dependent two-magnon density-of-states (DOS), $D(\mathbf{q},\omega_{\mathbf{q}})$. Although accurate magnon energy renormalization can only be determined through nonlinear spin-wave theory (NLSWT)~\cite{Noncollinear_SWT2, TLAF_NLSWT}, applying NLSWT to the triple-\Q{} ordering in \tred{\CTS{}} is quite cumbersome due to its multiple magnetic sublattices. Instead, $D(\mathbf{q},\omega_{\mathbf{q}})$ can serve as a simple yet effective quantity to qualitatively examine the extent of the three-magnon process~\cite{CrB2,tmdos_ref2, CoI2_nphys}. $D(\mathbf{q},\omega_{\mathbf{q}})$ is calculated by counting the number of three-magnon channels at $(\mathbf{q},\omega_{\mathbf{q}})$ that satisfy the kinematic condition:

\begin{align}\label{eq_DOS}
D(\mathbf{q},\omega_{\mathbf{q}}) = \, \frac{1}{N} \sum_{i,j}{\sum_{\mathbf{k}}{\delta(\omega_{\mathbf{q}}-(\omega_{\mathbf{k}}+\omega_{\mathbf{q-k}}))}},
\end{align}

where $N$ is a normalization factor, $\mathbf{k}$ runs over the set of $\mathbf{q}$ points in the first Brillouin zone, and $i$ and $j$ are the magnon band indices. Fig.~\ref{2mag}(h) shows the calculated $D(\mathbf{q},\omega_{\mathbf{q}})$ on the $\Gamma-K-B$ lines, which encompasses $\mathbf{q}_{\alpha}$ and $\mathbf{q}_{\beta}$. Indeed, all magnon branches at around $\mathbf{q}_{\alpha}$ and $\mathbf{q}_{\beta}$ are embedded in the two-magnon continuum with sizable $D(\mathbf{q},\omega_{\mathbf{q}})$, supporting our interpretation of magnon decay and renormalization due to magnon-magnon interactions. Nonetheless, a more detailed analysis based on NLSWT is necessary to rigorously confirm our scenario, and we leave this as a future theoretical challenge.

\section{Summary and Discussion}
Through theoretical analysis and experimental validation on \tred{\CTS{}}, which exhibits both single-\Q{} and triple-\Q{} ground states, we have highlighted that a nearly isotropic (strongly anisotropic) Goldstone-mode velocity serves as a distinctive hallmark of a \tred{symmetry-preserving} triple-\Q{} (a \tred{symmetry-breaking single- or double-$\mathbf{Q}$}) phase in triangular lattices. This indeed suggests that long-wavelength (= low-energy) magnetic excitation spectra can significantly aid in identifying the triple-\Q{} nature\tred{--typically accompanied by non-trivial topology--}of long-range order. To this end, we propose the following generalized protocol:

\begin{enumerate}
    \item Measure the magnetic excitation spectra of the paramagnetic phase ($T>T_{\mathrm{N}}$) and refine the spin Hamiltonian by fitting these spectra.
    
    \item Measure long-wavelength excitation spectra of the ordered phase of interest ($T<T_{\mathrm{N}}$) and compare the results with the theoretical spectra of the single-${\mathbf{Q}}$ (\tred{or double-$\mathbf{Q}$ if feasible}) and triple-${\mathbf{Q}}$ phases. The theoretical calculations should use the Hamiltonian obtained in Step 1, which will likely yield different profiles of Goldstone mode dispersion for the triple-\Q{} and \tred{the other two symmetry-breaking} orderings.
\end{enumerate}

A crucial step in this approach is to obtain unbiased optimal exchange parameters from the paramagnetic phase through high-temperature simulation techniques based on LLD~\cite{rescale_Dahlbom}. Yet a more practical protocol is also available based on the model-independent universality proposed in this work: if the observed Goldstone-mode velocity profile is strongly anisotropic (nearly isotropic), the ground state is likely single \tred{or double-$\mathbf{Q}$} (triple-\Q{}). This qualitative distinction can be made directly from the experimental data without requiring Step 1.

Identifying triple-${\mathbf{Q}}$ order through long-wavelength excitations offers significant advantages. For instance, magnons in metallic systems usually experience substantial decay or renormalization, especially at higher energies, due to interactions with conduction electrons. Moreover, in highly itinerant magnets, the Heisenberg model may not even be a valid description of magnetic dynamics~\cite{moriya}. Nevertheless, the Goldstone theorem guarantees well-defined collective excitations in the long-wavelength limit, ensuring that our analysis based on their velocity remains valid even under such situations. From an experimental standpoint, long-wavelength excitations around ${\mathbf{q}}$=$\mathbf{Q}_{\nu}$ typically exhibit the strongest $S_{\perp}(\mathbf{q},\omega)$, allowing for high-quality measurements even in systems with weak magnetic signals. These considerations suggest that identifying the triple-${\mathbf{Q}}$ phases \tred{(whose textures are typically topologically nontrivial)} based on $\mathbf{v}_{\mathrm{L}}({\mathbf{k}})$ would be applicable to a broad range of magnetic materials, regardless of their specific characteristics.

This approach should also be applicable to systems with significant quantum fluctuations (e.g., $S=1/2$ systems). While the noninteracting magnon picture may fail across a wide energy-momentum space, it still remains robust in the long-wavelength limit. Even when decay is kinematically allowed, Goldstone modes retain an infinite lifetime near the ordering wave vector on symmetry grounds, making their velocities well-defined. Notably, even in the extreme quantum limit of $S = 1/2$, 1/$S$ corrections to the Goldstone-mode velocity in noncollinear magnets (e.g., the 120$^\circ$ order) are known to be modest--on the order of 10\,\%~\cite{Chubukov1994}. These corrections further diminish in multi-$\mathbf{Q}$ orderings with larger spin or longer modulation wavelengths, reinforcing the robustness of the key contrast in velocity anisotropy between a triple-$\mathbf{Q}$ phase \tred{and the other two symmetry-breaking phases} against quantum corrections.

The successful demonstration of triple-\Q{} ordering through its characteristic $\mathbf{v}_{\mathrm{L}}({\mathbf{k}})$ profile possesses a fundamentally different level of significance compared to the approach of two recent studies on \tred{\CTS{}}, which confirmed the triple-\Q{} nature through neutron diffraction and bulk transport measurements~\cite{CTS_tripleQ_natcomm, CTS_tripleQ_nphys}. While these studies successfully excluded the single-\Q{} scenario based on $\mathbf{Q}_{\nu} =\mathbf{G}_{\nu}/2$ and nonzero $\sigma_{xy}(\mathbf{H}=0)$, the underlying symmetry argument ($\tau_{1a}T$, see Section III. B) remains indirect and could leave room for alternative interpretations. Providing more definitive evidence for triple-\Q{} ordering is therefore crucial, which is precisely what this study achieves through spin dynamics analysis.

More importantly, the symmetry argument used in Refs.~\cite{CTS_tripleQ_natcomm, CTS_tripleQ_nphys} is not applicable to other ordering wave vectors where $0 < |\mathbf{Q}_{\nu}| < |\mathbf{G}_{\nu}|/2$, as the $\tau_{1a}T$ symmetry is present only when $\mathbf{Q}_{\nu} = \mathbf{G}_{\nu}/2$. As a result, the identification method used in these studies is not generalizable and provides very limited insight into the broader challenge of unambiguously identifying topologically nontrivial multi-\Q{} magnetic ground states. Furthermore, relying on the $\tau_{1a}T$ symmetry argument becomes more challenging in insulating systems, where transverse conductivity can no longer be easily measured through routine bulk transport experiments. In contrast, our distinction protocol based on the contrasting long-wavelength dynamical characteristics is universally valid, regardless of the magnitude of $\mathbf{Q}_{\nu}$, specific material properties, and potentially lattice type. Indeed, this approach provides broad and general insights into the multi-domain single-\Q{} (\tred{or double-\Q{}}) versus single-domain multi-\Q{} problem, one of the key bottlenecks in the experimental study of 2D topological spin textures.

One practical limitation of our approach arises when the magnitude of the ordering wave vector $\mathbf{Q}_{\nu}$ becomes very small, corresponding to long-period modulations in real space. Since neutron spectroscopy is a momentum-space probe, resolving the Goldstone-mode dispersion with small $|\mathbf{Q}_{\nu}|$ would become challenging. This stands in contrast to real-space imaging techniques~\cite{skyr_rspace, Skyrmion_STM}, which excel at visualizing long-period multi-$\mathbf{Q}$ textures (e.g. Skyrmion lattices with $\sim$10\,nm periodicity or longer) but lack the spatial resolution to capture atomic-scale multi-$\mathbf{Q}$ configurations like that found in \tred{\CTS{}} [Fig.~\ref{basic}(a)]. In this sense, our momentum-space approach and real-space probes serve complementary roles: our protocol is best suited for multi-$\mathbf{Q}$ structures with relatively short spatial periodicities, while real-space techniques are better suited to mesoscopic spin textures.

Several existing materials are suitable for direct application of our approach. First, Co$_{1/3}$NbS$_{2}$, a metallic antiferromagnet isostructural to \tred{\CTS{}}, also exhibits magnetic Bragg peaks at $\mathbf{Q}_{\nu}=\mathbf{G}_{\nu}/2$ or its symmetry-related positions~\cite{Parkin_1983_magstr} and a sizable $\sigma_{xy}(\mathbf{H}=0)$~\cite{CNS_ncomm}. However, it has an additional incommensurate order wave vector perpendicular to $\mathbf{Q}_{\nu}=\mathbf{G}_{\nu}/2$~\cite{CNS_RXS}, which compromises the symmetry argument associated with the $\tau_{1a}T$ operation used in \tred{\CTS{}}. For the spin configuration of $\mathbf{Q}_{\nu}=\mathbf{G}_{\nu}/2$, both tetrahedral triple-${\mathbf{Q}}$ and stripe single-${\mathbf{Q}}$ orderings have been suggested as candidate ground states~\cite{CTS_tripleQ_nphys, CNS_RXS, Parkin_1983_magstr, Co1/3NbS2_ND}. Although the origin of the coexisting commensurate and incommensurate modulations is unclear, measuring its long-wave excitation spectra will tell which ground state is correct for $\mathbf{Q}_{\nu}=\mathbf{G}_{\nu}/2$.

Another promising candidate for the application is \NCTO{}, which has recently garnered significant interest as a candidate material for the Kitaev honeycomb model with a putative proximate spin-liquid phase~\cite{NCTO_field, NCTO_INS1, NCTO_INS2}. Notably, \NCTO{} shares the same space group and $\mathbf{Q}_{\nu}$ as \tred{\CTS{}}. In this compound, both collinear single-${\mathbf{Q}}$~\cite{Ma_NCTO_1q, NCTO_1q_2, NCTO_1q_3} and noncoplanar triple-${\mathbf{Q}}$~\cite{Li_NCTO_3q_1, Li_NCTO_3q2, Li_NCTO_3q_3} magnetic structures have been suggested as potential ground states. While the latter scenario has been supported convincingly, applying the approach introduced in this work could help undoubtedly confirm the true magnetic ground state of this intriguing compound. In addition, successful application to \NCTO{} could be a meaningful milestone in extending this approach to a more 2D magnetism: unlike \tred{\CTS{}}, \NCTO{} has negligible interlayer interactions and can be considered a 2D hexagonal antiferromagnet~\cite{Li_NCTO_3q_1}. While our analysis of $\mathbf{v}_{\mathrm{L}}({\mathbf{k}})$ already implies the feasibility of our approach even in the case of negligible inter-layer interactions, an experimental demonstration would further promote the investigation of triple-${\mathbf{Q}}$ spin textures in 2D systems.

\begin{figure}[t]
\includegraphics[width=1\columnwidth]{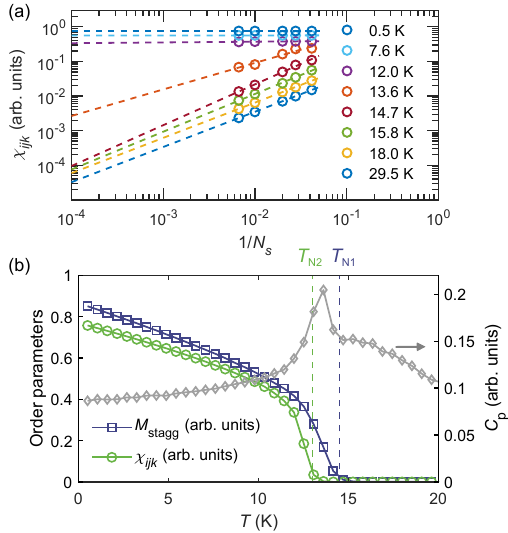} 
\caption{\label{2D_MC} Staggered magnetization ($M_{\mathrm{stagg}}$) and scalar spin chirality ($\chi_{ijk}$) of one-unit-cell-thick (= bilayer) \CTSideal{} calculated by classical Monte-Carlo simulations (see Appendix F). (a) $\chi_{ijk}$ calculated for various in-plane lattice
sizes ($N_{s}$) plotted on a logarithmic scale. Dashed lines represent linear fits of the logarithmic values at each temperature, from which the results with minimal finite-size effects ($N_{s}=10,000$, approximately a $6\,\mu \mathrm{m}$-sized nanoflake) were obtained through extrapolation. (b) Temperature dependence of $M_{\mathrm{stagg}}$ and $\chi_{ijk}$ for one-unit-cell-thick \CTSideal{} with $N_{s}=10,000$. \tred{Grey data points represent the calculated heat capacity of bilayer \CTSideal{} with $N_{s}=150$ (not extrapolated to $N_{s}=10,000$), indicating the presence of two phase transitions.} The simulations were performed using the bilinear exchange parameters listed in Table \ref{table:opt} and $K=0.06J_{1}$.}
\end{figure}

In the context of 2D magnetism above, we emphasize that \tred{\CTS{}} itself also represents an excellent material platform for studying topological triple-\Q{} spin textures in the genuine 2D limit. Notably, it is extremely important to identify topologically nontrivial phases among 2D van der Waals magnets. Despite its three-dimensional crystal structure resulting from Co intercalation, recent studies on the isostructural TM$_{1/3}$TaS$_{2}$ family have demonstrated the feasibility of obtaining atomically thin flakes of \tred{\CTS{}} and exploring its 2D magnetic behavior~\cite{CTS_npj_2022, TTS_chem_inter, TTS_2D_limit, TTS_chem_inter2}. Notably, Refs.~\cite{TTS_chem_inter, TTS_2D_limit} successfully prepared one-unit-cell-thick (or bilayer) Fe$_{1/3}$TaS$_{2}$--the thinnest stoichiometric form of TM$_{1/3}$TaS$_{2}$--and observed ferromagnetism similar to its bulk properties. Consequently, with the definitive confirmation of the triple-\Q{} magnetic ground state provided by this study, \tred{\CTS{}} emerges as a promising candidate for investigating the genuine 2D limit of topologically nontrivial triple-${\mathbf{Q}}$ magnetism, a critical missing element in the current field of 2D magnetism~\cite{Nature_review}.

The exchange parameters of \tred{\CTS{}} reported in this work provide valuable input for future experimental studies on triple-\Q{} magnetism in its atomically thin limit. As a useful preliminary prediction, we performed classical Monte Carlo simulations for one-unit-cell-thick \CTSideal{} (see Appendix F for the methodology), assuming that the interaction strengths remain intact in the atomically thin limit. Using the set of exchange parameters determined at \Th{} (Table~\ref{table:opt}) with $K=0.06J_{1}$ (see Appendix B), we calculated the temperature-dependent staggered magnetization ($M_{\mathrm{stagg}}$) and scalar spin chirality ($\chi_{ijk}$)--each representing an order parameter of the stripe single-\Q{} and tetraherdal triple-\Q{} orderings. As shown in Fig.~\ref{2D_MC}(b), the results reveal the onset of chiral long-range order below 13\,K. Additionally, we note that the weak but finite easy-axis magnetic anisotropy present in \CTS--discussed in detail at the end of this section--could further stabilize the long-range order in the atomically thin limit. Based on these arguments, we speculate that the tetrahedral triple-\Q{} state could persist in atomically thin \tred{\CTS{}}, highlighting the experimental exploration of reduced material thickness as a promising direction for future research.

Concerning the full spin dynamics beyond the long-wavelength limit (Section \tred{IV. D}), the suggested significance of the magnon-magnon interactions in \tred{\CTS{}} also warrants further discussion. Although the magnon-magnon interaction is the most plausible explanation, to the best of our knowledge, for the enhanced renormalization in the triple-\Q{} phase, it is rather unusual to observe such a significant influence in a classical spin system ($S>1/2$). Previous studies consistently suggested that Co$^{2+}$ ions in \tred{\CTS{}} develop localized magnetic moments of $S=3/2$ through a high-spin $d^{7}$ configuration~\cite{Parkin_80_v1, Parkin_80_v2, Parkin_1983_magstr, CTS_npj_2022, CTS_tripleQ_natcomm, CTS_tripleQ_nphys}. In a triangular lattice antiferromagnet that develops coplanar 120$^{\circ}$ ordering, quantum effects beyond LSWT are expected to be marginal for $S=3/2$~\cite{TLAF_NLSWT}. One possibility is that the noncoplanar triple-${\mathbf{Q}}$ phase exhibits much stronger magnon-magnon interactions than those in the 120$^{\circ}$ phase. To confirm this, the analytic $1/S$ expansion calculation for the tetrahedral triple-${\mathbf{Q}}$ magnetic structure should be performed, which, to the best of our knowledge, has yet to be done.

Although it does not align well with our observation of temperature-dependent renormalization at $\{\mathbf{q}_{h}\}$, the Stoner continuum should not be excluded from a source of magnon decay and renormalization in the overall low-energy spectra ($<10$\,meV) of \tred{\CTS{}}. While the two-magnon continuum can dominate over that of the Stoner continuum at specific momentum positions in a noncollinear magnet~\cite{CrB2}, the Stoner continuum would still play a role in the observed spectrum to some extent, considering that broad magnetic spectra are consistently observed in any metallic antiferromagnets~\cite{moriya, AFM_metal_1, AFM_metal_2, AFM_metal_3, CaFe2As2_nphys, Mn3Sn_INS}. Possible Stoner excitation processes at $\mathbf{q}\in\{\mathbf{q}_{h}\}$ are qualitatively discussed in Appendix K. Generally, disentangling these two factors in an INS spectrum is very challenging and thus requires a specific situation where these two contributions behave differently~\cite{CrB2}. We note that the contrasting magnitude of the magnon-magnon interactions between triple-${\mathbf{Q}}$ and single-${\mathbf{Q}}$ orderings--the key feature that led us to interpret our observation as being due to the two-magnon continuum rather than the Stoner continuum--applies only when $\mathbf{Q}_{\nu}$ = (1/2, 0, 0). For $\mathbf{Q}_{\nu}$ = ($q$, 0, 0) with $0<q<1/2$, the single-${\mathbf{Q}}$ phase is also noncollinear and can manifest nonzero three-magnon terms (except in the case of spin density wave).

We finally acknowledge one limitation of the spin Hamiltonian suggested in this work (Eq. \ref{Hamiltonian}): omission of magnetic anisotropy. Although the isotropic spin model in Eq. \ref{Hamiltonian} captures the phase diagram and key spin dynamics we observed, the presence of a few small anisotropy terms is implied in its static and dynamic magnetic properties. First, as reported in Ref.~\cite{CTS_tripleQ_natcomm}, the triple-${\mathbf{Q}}$ phase develops a magnon energy gap (approximately 0.5\,meV or smaller) at the M points. Interestingly, any symmetry-allowed single-ion anisotropy terms in \tred{\CTS{}} cannot open this energy gap, indicating the presence of slight exchange anisotropy. Bond-dependent exchange anisotropy $J_{\pm\pm}$ is a rare term that can open this gap in the tetrahedral triple-\Q{} phase; see Ref.~\cite{TLAF_generalized} for its definition. Another component is the easy-axis anisotropy along the $c$-axis, which is necessary to describe the out-of-plane spin configuration of the single-${\mathbf{Q}}$ phase at \Ti{}. This can be included in Eq. \ref{Hamiltonian} as either the $XXZ$-type exchange anisotropy or single-ion anisotropy [$\hat{H}_{i} = K_{z}(\mathbf{S}^{z}_{i})^{2}$ where $K<0$ and $i$ indices triangular sites].

\tred{That said}, we emphasize that the observation of the tetrahedral triple-${\mathbf{Q}}$ ground state limits the size of these anisotropy terms to be smaller than $K$, as they would incur energy costs to its spatially uniform tetrahedral configuration. The observed energy gap of 0.5\,meV at 5\,K and the absence of any energy gap at 30\,K (down to the precision of $0.3$\,meV) further support their small magnitude. As the four-spin interaction term with the coefficient $K$ is shown to have marginal effects on the observed spin dynamics (Appendix I), these anisotropy terms with coefficients smaller than $K$ are unlikely to affect the spin dynamics analysis presented in this study.

\begin{acknowledgments}
 We acknowledge H.-J. Noh, I. Martin, H. Park, and M. J. Han for their helpful discussions. The Samsung Science \& Technology Foundation supported this work at Seoul National University (Grant No. SSTF-BA2101-05). P.P. acknowledges support from the U.S. Department of Energy, Office of Science, Basic Energy Sciences, Materials Science and Engineering Division. The neutron scattering experiment at the Japan Proton Accelerator Research Complex (J-PARC) was performed under the user program (Proposal No. 2022B0075). One of the authors (J.-G.P.) is partly funded by the Leading Researcher Program of the National Research Foundation of Korea (Grant No. 2020R1A3B2079375). 
 This research was partially supported by the National Science Foundation Materials Research Science and Engineering Center program through the UT Knoxville Center for Advanced Materials and Manufacturing (DMR-2309083). C.D.B. acknowledges support from the U.S. Department of Energy, Office of
Science, Office of Basic Energy Sciences, under Award Number DE-SC0022311. S. Matin acknowledges the Center for Nonlinear Studies at Los Alamos National Laboratory.
\end{acknowledgments}

\providecommand{\noopsort}[1]{}\providecommand{\singleletter}[1]{#1}%

\clearpage

\appendix
\section{Co-aligned single crystals}
\begin{figure}[h]
\includegraphics[width=1\columnwidth]{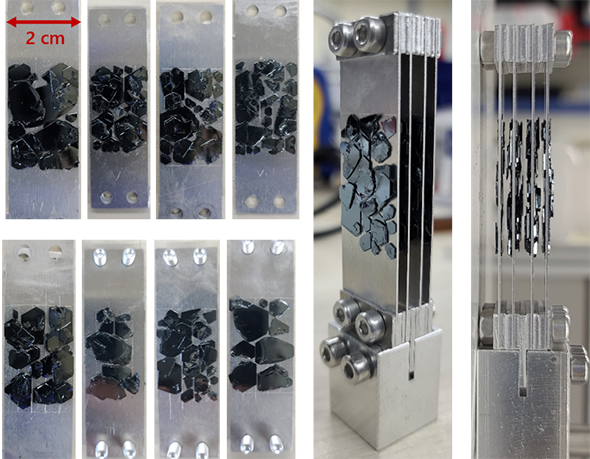} 
\caption{\label{align} The co-aligned \tred{\CTS{}} single crystals used for the INS experiment. More details can be found in the Methods section of the main text.}
\end{figure}

Fig.~\ref{align} shows photos of the co-aligned \tred{\CTS{}} single crystals used in this study. Their good alignment is evident in the diffraction spectra presented in Fig.~\ref{CTS}(c).

\section{Higher-order renormalization of scalar bi-quadratic interactions}
If a Hamiltonian term involves a nonlinear order of spin operators at a single site (i.e., higher than the first order), its amplitude becomes renormalized in higher-order spin wave theory~\cite{dahlbom_renormalized}. The scalar biquadratic term used in Eq. \ref{Hamiltonian} falls into this category. Based on the 1/$S$ expansion, the renormalized magnitude of $K$ ($\Tilde{K}$) is derived as follows~\cite{dahlbom_renormalized}:

\begin{align}\label{eq_renor}
\Tilde{K} = (1-\frac{1}{S}+\frac{1}{4S^{2}})K
\end{align}

In other words, in \tred{\CTS{}} ($S=3/2$), the actual magnitude of biquadratic interactions in the simulation ($=\Tilde{K}$) is 4/9 of the input value ($K$). Since $\Tilde{K}\sim0.027J_{1}$ is needed to reproduce \TNt{}/\TNo{} = 0.7, the input value of $K$ should be $\sim0.06J_{1}$.

\section{Dynamical susceptibility and dynamical structure factor}
In INS with unpolarized neutrons, the measured intensity is proportional to the \textit{projected} dynamical structure factor, which includes only spin fluctuations perpendicular to the momentum transfer vector $\mathbf{q}$. This quantity, denoted as $S_{\perp}(\mathbf{q}, \omega)$ throughout this work, differs from the conventional dynamical structure factor $S(\mathbf{q},\omega)$, which simply sums over all components of the spin correlation tensor, i.e., $\sum_{\alpha,\beta}{}S^{\alpha\beta}(\mathbf{q}, \omega)$ ($\alpha, \beta = x, y, z$). Indeed, all experimental and simulated spectra presented in this work correspond to $S_{\perp}(\mathbf{q}, \omega)$.

When analyzing a spin-wave spectrum with sizable thermal fluctuations, it can sometimes be useful to visualize the dynamical susceptibility [$\chi''_{\perp}(\mathbf{q},\omega)$] instead of the dynamical structure factor [$S_{\perp}(\mathbf{q},\omega)$]~\cite{park_ZVPO}. $\chi''_{\perp}(\mathbf{q},\omega)$ can be easily calculated from $S_{\perp}(\mathbf{q},\omega)$ through the following equation:

\begin{align}\label{eq:chi2p}
\chi''_{\perp}(\mathbf{q},\omega) = \pi(1-e^{-\hbar\omega/k_{\mathrm{B}}T})\times S_{\perp}(\mathbf{q},\omega),
\end{align}

where $T$ is the measurement or LLD simulation temperatures. This conversion effectively deconvolutes the Bose factor $\frac{1}{\pi(1-e^{-\hbar\omega/k_{\mathrm{B}}T})}$ and removes an increased spectral weight in the low-$E$ region due to thermal excitations, which often conceals the low-energy structure of the magnetic excitations in the color plot of $S_{\perp}(\mathbf{q},\omega)$. However, it is important to note that the INS data measured at 5\,K (i.e., the triple-${\mathbf{Q}}$ phase) show no noticeable different between $\chi''_{\perp}(\mathbf{q},\omega)$ and $S_{\perp}(\mathbf{q},\omega)$, as the thermal energy at 5\,K is marginal compared to the energy of the magnon spectrum (see Fig.~\ref{dispersion}). 

\section{Procedure of fitting paramagnetic excitation spectra}
In this section, we describe the optimization process used to fit the paramagnetic excitation spectra using LLD. The optimal solution was found by performing the least-squares fitting between the nine measured and simulated $S_{\perp}(\mathbf{q},\omega)$ slices in Fig.~\ref{highT}. Completing this optimization job within a reasonable time frame requires a good understanding of its characteristics. A LLD simulation to calculate spin dynamics at finite temperatures is a forward simulation. Since this is fairly time-consuming (a few minutes for each slice of $S_{\perp}(\mathbf{q},\omega)$ in Fig.~\ref{highT}), it is crucial to minimize the number of times the optimization process runs this simulation. For this reason, using common gradient-based optimization methods, which require repeated evaluations of the simulation to calculate gradients, would be computationally expensive. This is especially true for our problem, as it deals with a high-dimensional space of variables due to multiple exchange interactions for both intralayer and interlayer bonds, and should fit a four-dimensional profile of $S_{\perp}(\mathbf{q},\omega)$. Moreover, forward simulations contain noise in their results, which further compromises the accuracy of gradient calculations.

To address this challenge, we adopted a Bayesian optimization algorithm, which is gradient-free and very effective for problems where the fitting object is difficult to evaluate due to computational costs. This approach reaches an optimal solution with relatively fewer iterations by performing intelligent parameter space searches based on the surrogate modeling and the acquisition function~\cite{Ref_Bopt}. We used the Bayesian optimization package implemented in Python~\cite{Bopt_python}, with the Gaussian process for the surrogate model~\cite{Bopt_Matin,Bopt_python}. The optimization algorithm searched a wide 5D parameter space of $J_{1}$, $J_{2}/J_{1}$, $J_{3}/J_{1}$, $J_{\mathrm{c1}}/J_{1}$, and $J_{\mathrm{c2}}/J_{1}$, with each parameter allowed to range as follows: $0.8\,\mathrm{meV}<J_{1}<1.4\,\mathrm{meV}$, $0<J_{2}/J_{1}<0.8$,  $-0.15<J_{3}/J_{1}<0.4$, $0.7<J_{\mathrm{c1}}/J_{1}<1.4$, and $-0.4<J_{\mathrm{c2}}/J_{1}<0.1$.

The algorithm converged on the optimal parameter set shown in Table \ref{table:opt}, or an equivalent set within the uncertainty range, after 150--200 Bayesian optimization steps. To ensure the credibility of the suggested solution, optimization was performed starting from several different initial parameter sets, and all trials reached the same solution. Additionally, to check if other nontrivial solutions exist or if the suggested solution is merely a local minimum of $\chi^{2}$, we conducted a brute-force exploration of the $\chi^{2}$ map around the optimal parameter set, as shown in Figs. \ref{highT}(k)--(l). As mentioned in the main text, a unique minimum of $\chi^{2}$ ($\chi^{2}_{\mathrm{min}}$) was found at the solution's position suggested by the optimization algorithm. Further assessment of this optimal solution is described in Appendix E.

\begin{figure}[ht]
\includegraphics[width=1\columnwidth]{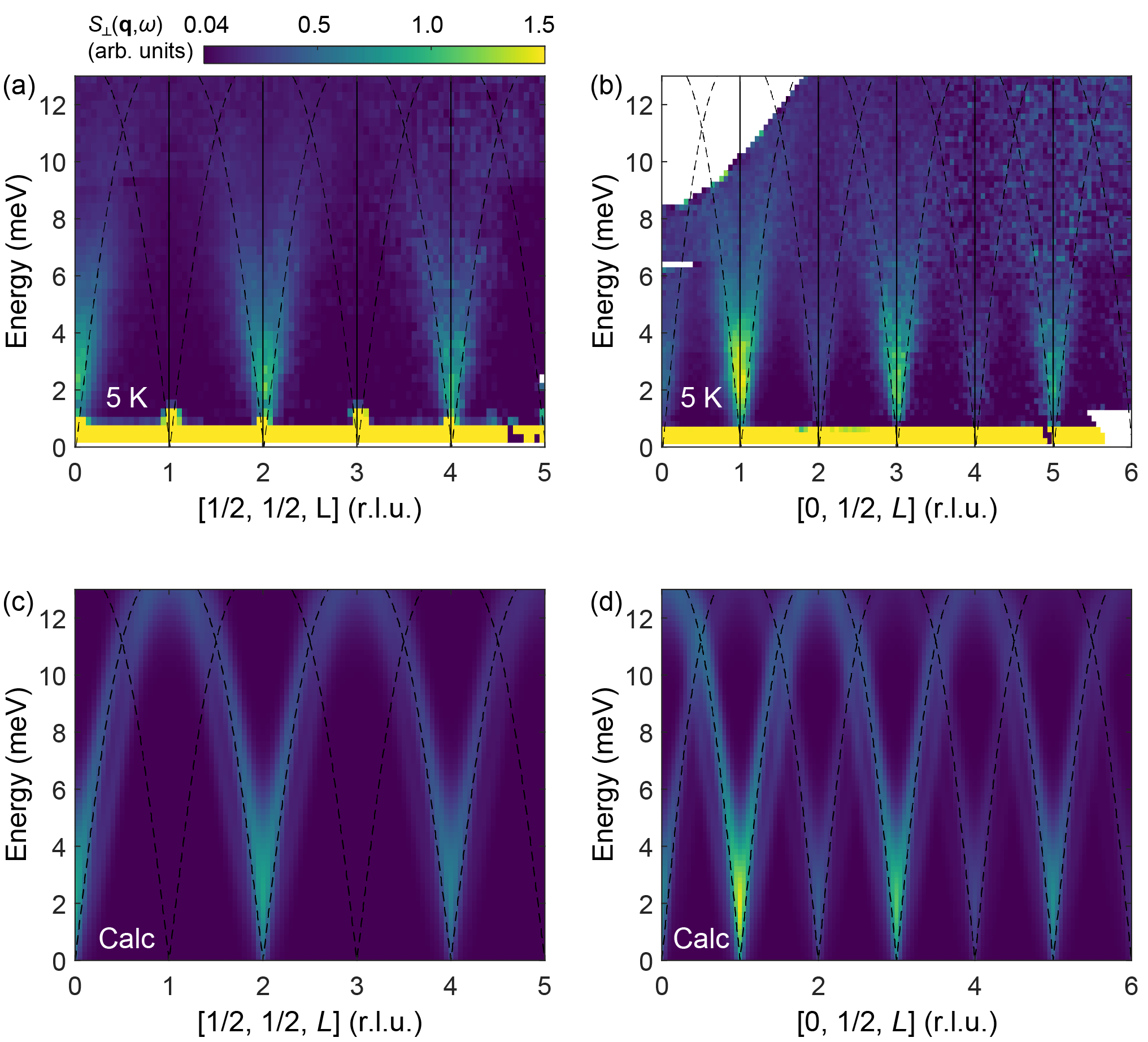} 
\caption{\label{00L} (a)--(b) Measured and (c)--(d) simulated magnon spectra along the [0, 0, $L$] direction at 5\,K (i.e., the triple-${\mathbf{Q}}$ phase). The data collected from different $E_{i}$ are overlaid. Dashed black lines are the dispersion of linear magnon modes. The calculation results in this figure was obtained using LSWT.}
\end{figure}

\begin{figure}[h]
\includegraphics[width=0.95\columnwidth]{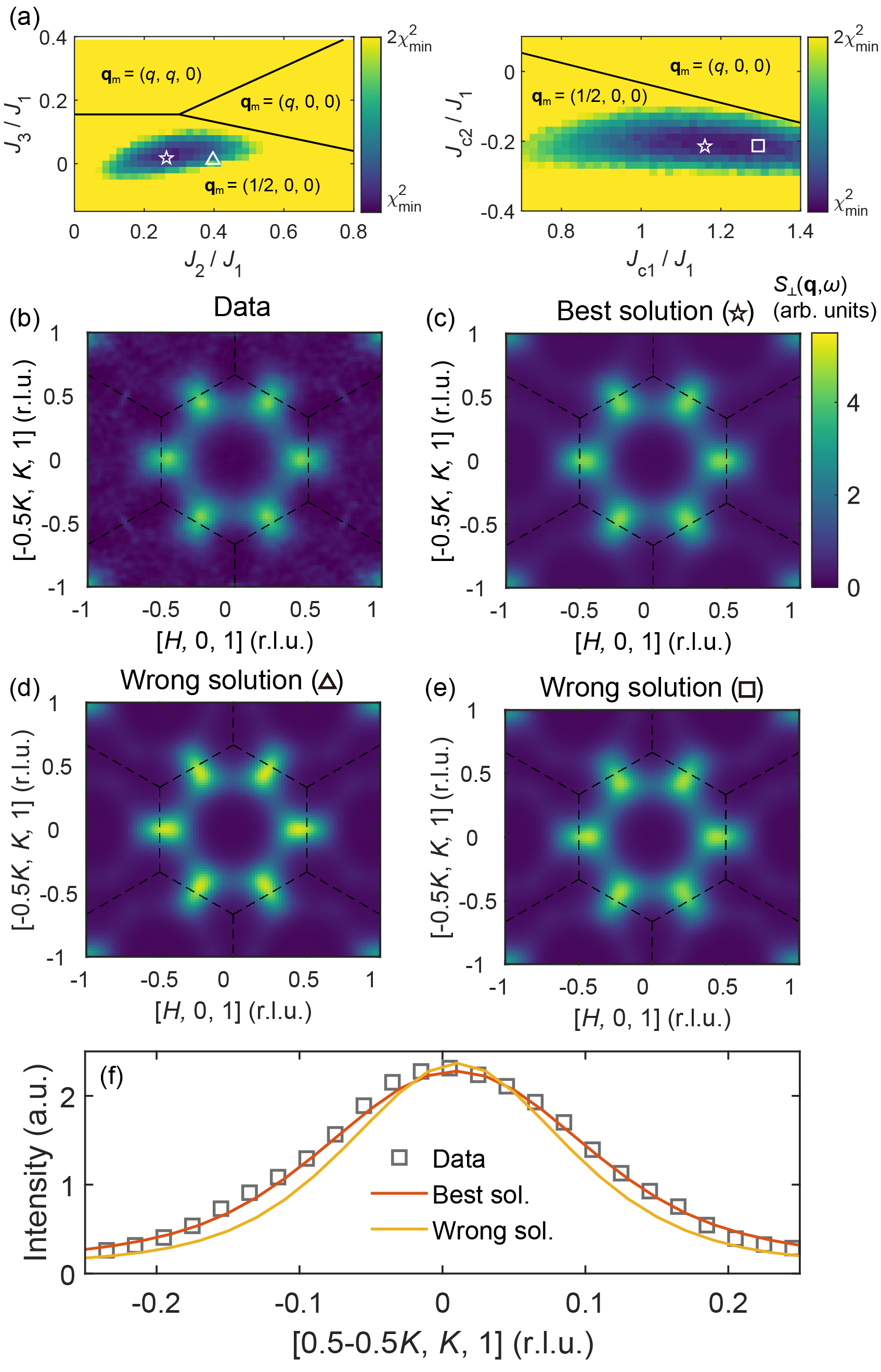} 
\caption{\label{wrong} Comparing paramagnetic excitation spectra obtained from multiple exchange parameter sets that yield different level of goodness-of-fit (i.e., $\chi^{2}$ metric). (a) The same color plots as Figs. \ref{highT}(k)--(l), which additionally marks two parameter sets (a white square and triangle) possessing worse $\chi^{2}$ than the optimal solution (a white star). (b) The same color plot as Fig.~\ref{longw}(a). (c)--(e) Corresponding LLD spectra obtained from the three different parameter sets indicated in (a). (f) A vertical cut of the constant-$\omega$ spectra in (b)--(d). Error bars are much smaller than the data symbol.}
\end{figure}

\section{Reliability of the optimal parameter set found by Bayesian optimization}
In addition to the nice agreement between the data and the simulations as shown in Figs. \ref{highT}(a)--(j) and the well-identified $\chi^{2}$ minimum in Figs. \ref{highT}(k)--(l), we further assess the reliability of the bilinear exchange parameters suggested by our Bayesian optimization algorithm. 

First, the magnitudes of the interlayer exchange parameters $J_{c1}$ and $J_{c2}$ in Table \ref{table:opt} can be validated by examining the magnon spectrum along the [0, 0, $L$] direction, which exhibits simple V-shaped magnon branches [Figs. \ref{00L}(a)--(b)]. In Figs. \ref{00L}(a)--(b), we overlay the LSWT magnon dispersion calculated with the optimal exchange parameters and the triple-${\mathbf{Q}}$ magnetic ordering on the data. Also, the corresponding $S_{\perp}(\mathbf{q},\omega)$ maps calculated by LSWT, convoluted with instrumental resolution and momentum integration effects, are plotted in Figs. \ref{00L}(c)--(d). Indeed, the optimal $J_{c1}$ and $J_{c2}$ parameters in Table \ref{table:opt} accurately capture the measured bandwidth along the [0, 0, $L$] direction.

Furthermore, we directly compared the paramagnetic excitation spectra of the optimal parameter set with those of surrounding parameter sets to evaluate the degradation in fit quality for these alternative solutions. Fig.~\ref{wrong} shows constant-$\omega$ slices of $S_{\perp}(\mathbf{q},\omega)$ simulated from the optimal parameter set [panel (c)] and from suboptimal sets [panels (d)--(e)] near the optimal solution. The $\chi^{2}$ metric of the chosen suboptimal solutions (white triangular and square symbols) is 1.25$\chi^{2}_{min}$, which is not far from $\chi^{2}_{\mathrm{min}}$. However, their $S_{\perp}(\mathbf{q},\omega)$ spectra exhibit clear discrepancies with the data; the signal's pattern around the M point is more elongated to the [$H$, 0, 0] direction than observed while being narrower along the [$-0.5K$, $K$, 0] direction. Fig.~\ref{wrong}(f) demonstrates this more explicitly. This comparison further confirms the reliability of the fitted exchange parameters.

\section{Temperature-dependent magnetic ground state revealed by classical Monte Carlo simulations}
Finite-temperature magnetic ground states of our optimal spin model (see Table \ref{table:opt}) with finite $K>0$ were investigated using classical Monte Carlo simulations, combined with the LLD equation and simulated annealing technique. We created a $30\times30\times8$ sized \tred{\CTS{}} supercell (14,400 Co sites) and sampled its time evolution at each temperature point while cooling down the system from 50 K to 5 K. The Langevin time step ($\mathrm{d}t$) and damping constant were set to 0.02\,meV$^{-1}$ and 0.1, respectively. Each temperature was sampled over 512,000 Langevin time steps. From the collected samples, we calculated staggered magnetization and scalar spin chirality using the equations below, which represent an order parameter for the stripe single-${\mathbf{Q}}$ and tetrahedral triple-${\mathbf{Q}}$ magnetic orderings, respectively:

\begin{align}\label{eq_Mstagg}
M_{\mathrm{stagg}} = \frac{1}{N} \sum_{i}{(-1)^{2\pi(\mathbf{q}^{\nu}_{\mathrm{m}}\cdot\mathbf{r}_{i})} \langle\mathbf{S}_{i} \rangle},
\end{align}

\begin{align}\label{eq_chi}
\chi_{ijk} = \frac{\sum_{\Delta}{\langle\mathbf{S}_{\Delta1}\cdot(\mathbf{S}_{\Delta2}\times\mathbf{S}_{\Delta3}) \rangle}}{N_{t}},
\end{align}

where $i$ is a Co site index, $\Delta$ indexes a single triangular plaquette on a Co triangular lattice consisting of three sites ($\Delta1$, $\Delta2$, $\Delta3$), $\mathbf{q}^{\nu}_{\mathrm{m}}$ ($\nu=1,2,3$) is three possible ordering wave vectors for the stripe single-${\mathbf{Q}}$ order, and $N$ ($N_{t}$) is the total number of spins (triangular plaquettes). We used S = 3/2 for the simulations.

\begin{figure}[h]
\includegraphics[width=0.95\columnwidth]{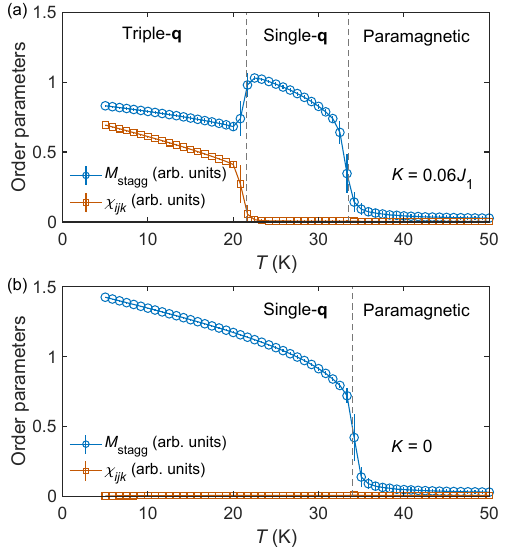} 
\caption{\label{MC} Temperature-dependent staggered magnetization ($M_{\mathrm{stagg}}$) and scalar spin chirality calculated by classical Monte-Carlo simulations. The bilinear exchange parameters in Table \ref{table:opt} are used for the calculation. Panel (a) [panel (b)] shows the result with $K=0.06J_{1}$ ($K=0)$.}
\end{figure}

Fig.~\ref{MC}(a) shows the resultant temperature-dependent order parameters obtained with finite $K$. Our optimal exchange parameter set yields \TNo{} = 34\,K very close to the experimental observation \TNo{} = 38\,K, again supporting our solution in addition to other features described Appendix E. Moreover, $K = 0.06J_{1} = 0.07$\,meV reproduces not only the two-step transition process illustrated in Fig.~\ref{CTS}(b) but also its quantitative interval \TNo{}/\TNt{} = 0.7. Thus, the LLD simulations with $K = 0.06J_{1}$ successfully provide the triple-${\mathbf{Q}}$ and single-${\mathbf{Q}}$ spin dynamics at 5\,K and 30\,K, enabling a direct comparison with our experimental data. 

For comparison, we also plotted the results without $K$ in Fig.~\ref{MC}(b). As expected, the single-${\mathbf{Q}}$ stripe ordering becomes the magnetic ground state. Notably, this outcome is not captured in a simple classical energy comparison, as the single-${\mathbf{Q}}$ and triple-${\mathbf{Q}}$ states are simply degenerate. However, thermal fluctuations break this degeneracy in both the simulation and reality.

Simulations for one-unit-cell-thick \tred{\CTS{}} (comprising two Co triangular layers), representing the thinnest stoichiometric form, were performed using the same approach as described above, but with open boundary conditions along the c-axis. Finite-size effects were carefully addressed by conducting simulations across various lattice sizes along the $a$ and $b$ directions $N_{s}$~\cite{NiPS3_ncomm, CrPS4}; we derived the sampled values of $M_{\mathrm{stagg}}$ and $\chi_{ijk}$ for a 10,000 $\times$ 10,000 supercell (i.e., $N_{s}=10,000$) through extrapolation based on linear fitting. Note that $N_{s}=10,000$ approximately corresponds to a $6\,\mu \mathrm{m}$-sized nanoflake, which is comparable to the typical size of real atomically thin nanoflakes in 2D van der Waals materials~\cite{NiPS3_ncomm, CrPS4, CTS_npj_2022}. For these simulations, each temperature was sampled over 2,048,000 Langevin time steps, four times longer than the length used for the bulk simulations.

As shown in Fig.~\ref{2D_MC}, the resultant temperature-dependent $M_{\mathrm{stagg}}$ and $\chi_{ijk}$ reveals the emergence of chiral triple-\Q{} magnetic ordering below $T=13$\,K. Notably, a temperature range of the intermediate single-\Q{} ordering (\Ti{}) is markedly suppressed in the atomically thin limit.

\section{Robustness of velocity profile $v_{\mathrm{L}}(\mathbf{k})$ against data processing effects}

To confirm the robustness of the observed spin-wave anisotropy, we tested the impacts of the data processing procedures used in this study. As described below, none were found to artificially alter the $v_{\mathrm{L}}(\mathbf{k})$ profiles observed in constant-energy slices.

\begin{figure}[ht]
\includegraphics[width=1.0\columnwidth]{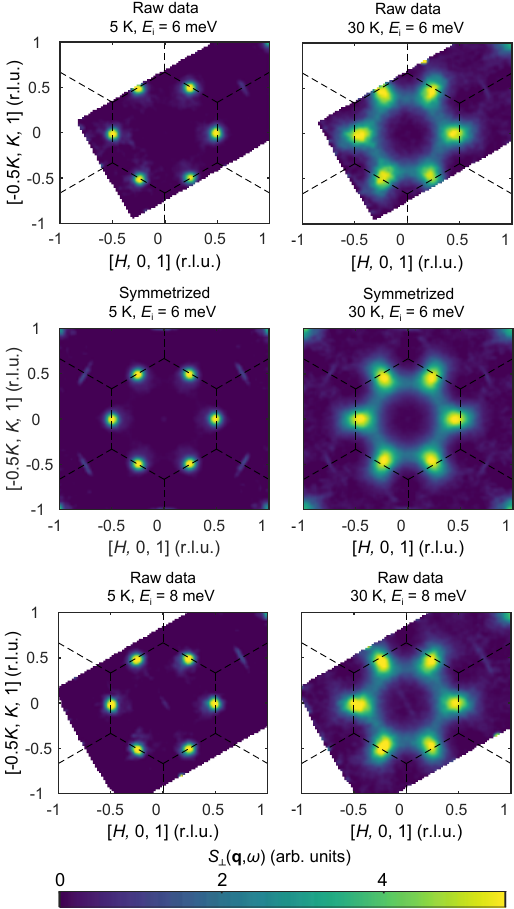} 
\caption{\label{symeffect} Comparison of INS spectra before and after applying the symmetrization procedure.}
\end{figure}

Figure~\ref{symeffect} compares constant-energy cuts from symmetrized (identical to Fig.~\ref{longw}(b)–(c)) and unsymmetrized datasets. The nearly isotropic (triple-$\mathbf{Q}$) and anisotropic (single-$\mathbf{Q}$) spin-wave cones remain clearly distinguishable in both cases, confirming that symmetrization does not distort the intrinsic features of $S_{\perp}(\mathbf{q}, \omega)$ but merely enhances statistical quality. Furthermore, independent measurements performed with a different incident energy ($E_i = 8$\,meV) yield consistent results, indicating that the $v_{\mathrm{L}}(\mathbf{k})$ contrast is unaffected by instrumental resolution.

\begin{figure*}[ht]
\includegraphics[width=1\textwidth]{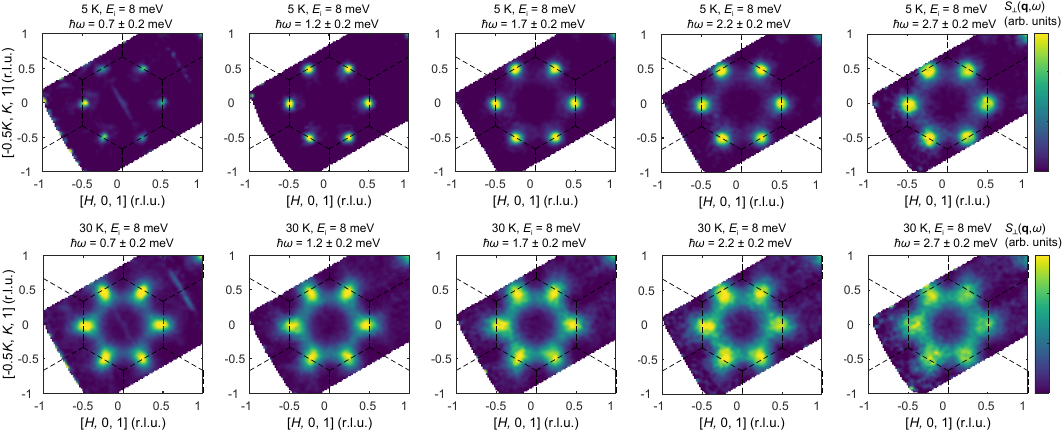} 
\caption{\label{eposeffect} Constant-energy slices of the measured magnon spectra at various energy transfers, extending beyond the long-wavelength regime.}
\end{figure*}

We also examined how the visibility of velocity anisotropy depends on the energy transfer. As shown in Fig.~\ref{eposeffect}, the distinction between single-$\mathbf{Q}$ and triple-$\mathbf{Q}$ cones remains pronounced up to $\sim$1.7,meV, which is 14\,\% of the one-magnon bandwidth of \tred{\CTS{}}. Beyond this point, the linear mode's momentum moves too far from the M points, leading to blurred spectral features--particularly in the single-$\mathbf{Q}$ case--that hinder reliable assessment of directional velocity. Thus, the contrast in $v_{\mathrm{L}}(\mathbf{k})$ remains intact for any energy transfers lower than 14\,\% of the magnon bandwidth, rather than strictly limited to a specific energy window.

\begin{figure}[ht]
\includegraphics[width=1\columnwidth]{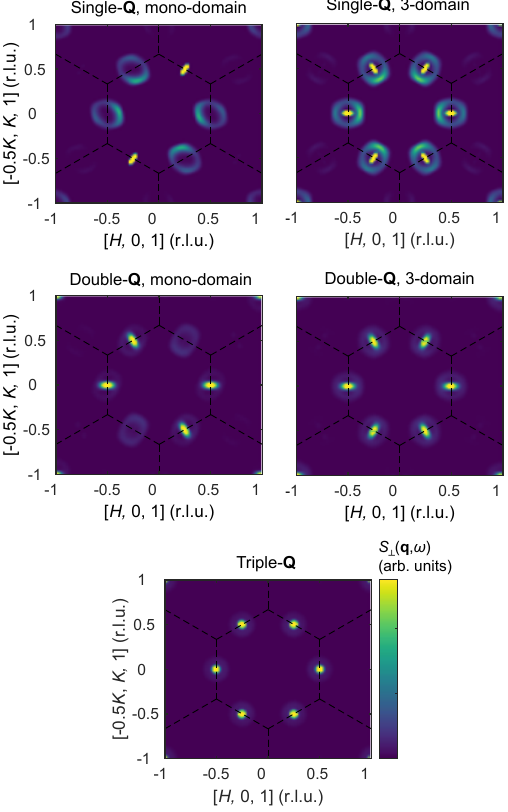} 
\caption{\label{2Q} Comparison of long-wavelength spin dynamics for single-, double-, and triple-$\mathbf{Q}$ magnetic orderings with $\mathbf{Q}_{\nu} = \mathbf{G}_{\nu}/2$. All calculations (LSWT) are based on the exchange parameters in Table~\ref{table:opt} with $K = 0$, under which the three states are energetically degenerate.}
\end{figure}

\newpage
\section{Extension to the case of double-$\mathbf{Q}$ ordering}

While \tred{\CTS{}} does not host a double-$\mathbf{Q}$ phase, it is instructive to consider this case for completeness, as identifying a triple-$\mathbf{Q}$ ground state requires distinguishing it from both single-$\mathbf{Q}$ and double-$\mathbf{Q}$ orderings. The double-$\mathbf{Q}$ configuration with $\mathbf{Q}_{\nu} = \mathbf{G}_{\nu}/2$ is co-planar (but noncollinear) and exhibits nonuniform relative spin angles among the six nearest-neighbor spin pairs, similar to the single-$\mathbf{Q}$ stripe phase. \tred{Therefore, following our universal prediction based on the symmetry-preserving versus symmetry-breaking nature of the magnetic orders in Section III. A, the double-$\mathbf{Q}$ state is likewise expected to show a strongly anisotropic Goldstone-mode velocity profile, similar to the single-$\mathbf{Q}$ case.}

Figure~\ref{2Q} shows that the resulting Goldstone-mode velocity profile for the double-$\mathbf{Q}$ state is strongly anisotropic, closely resembling that of the single-$\mathbf{Q}$ case. However, it is clearly distinguished from the single-$\mathbf{Q}$ case by its weak quadratic-mode intensity, which is rather close to that seen in the triple-$\mathbf{Q}$ spectrum. This comparison supports two key conclusions: (i) the two magnetic phases observed in \tred{\CTS{}} correspond to single-$\mathbf{Q}$ and triple-$\mathbf{Q}$ orderings; and (ii) nearly isotropic long-wavelength dispersion is a unique and robust hallmark of the triple-$\mathbf{Q}$ phase. Consequently, long-wavelength velocity anisotropy serves as a powerful diagnostic for identifying topologically nontrivial triple-$\mathbf{Q}$ states in triangular lattice systems. It also reinforces the central insight of our work: the uniformity of relative spin angles among nearest-neighbor sites \tred{(i.e., whether the magnetic order breaks the three-fold rotational symmetry or not)} is the key determinant of the isotropic $v_{\mathrm{L}}(\mathbf{k})$ profile, supporting the broader applicability of our finding to generic multi-$\mathbf{Q}$ orderings, as described in the main text.

\section{Parameter dependence of $v_{\mathrm{L}}(\mathbf{k})$}

\begin{figure*}[ht]
\includegraphics[width=1\textwidth]{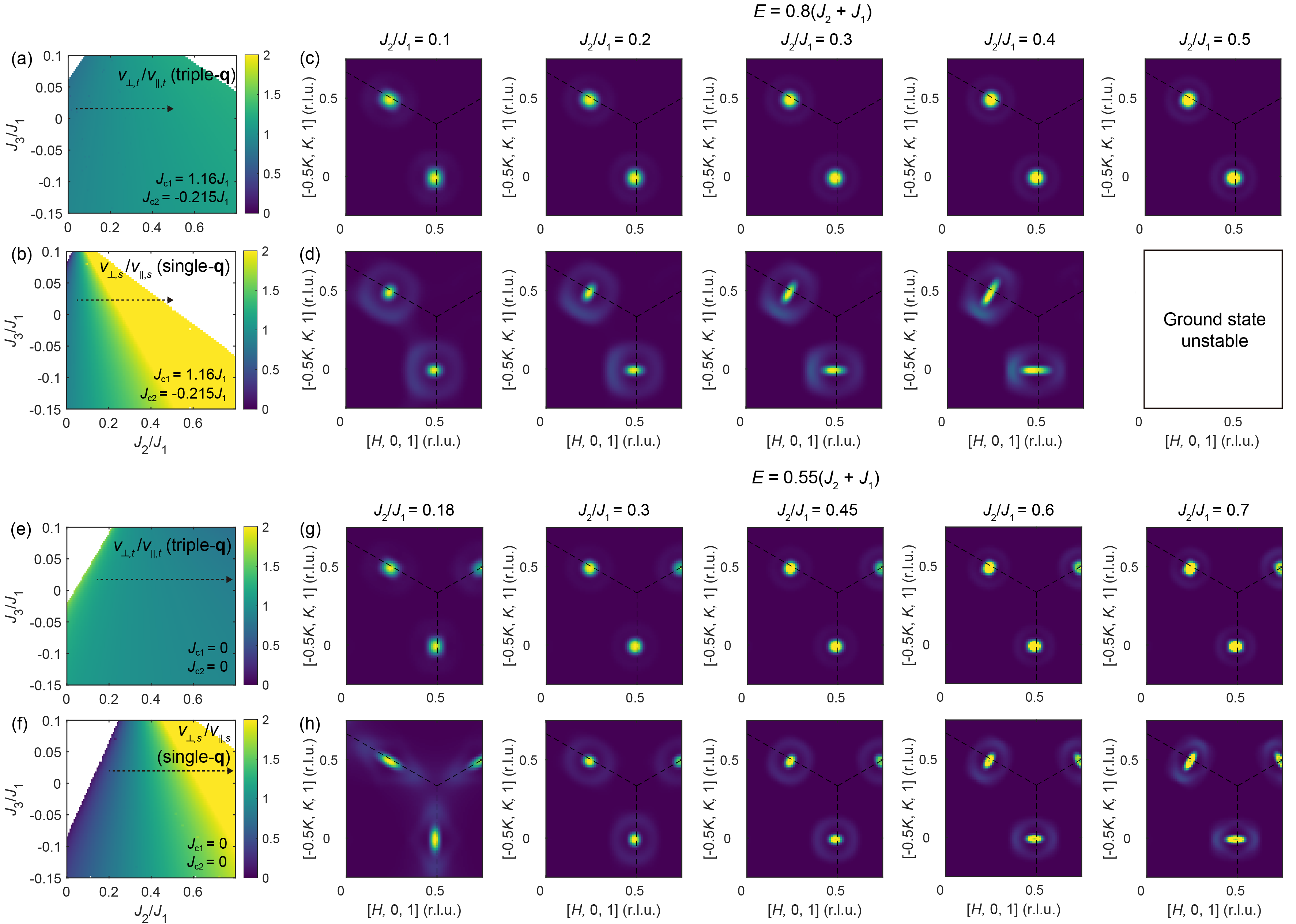} 
\caption{\label{magvel_detail} Visualizing characteristic patterns of the constant-$\omega$ cut stemming from the anisotropic $v_{\mathrm{L}}(\mathbf{k})$. (a)--(b) Parameter-dependent anisotropy of $v_{\mathrm{L}}(\mathbf{k})$ for the triple-${\mathbf{Q}}$ and single-${\mathbf{Q}}$ phases, quantified by $v_{\perp}/v_{\parallel}$ [same as the plots shown in Fig.~\ref{magvel}(a)--(b)]. (c)--(d) Constant-$\omega$ cuts of the triple-${\mathbf{Q}}$ and single-${\mathbf{Q}}$ magnon spectrum at $\hbar\omega = 0.8(J_{2}+J_{1})$ from several different $J_{2}/J_{1}$. They visualize individual $v_{\perp}/v_{\parallel}$ as an eccentricity of the linear magnon's signal at the M Points. (e)--(h) Same as (a)--(d), but based on the spin model without any interlayer interactions $J_{\mathrm{c1}}$ and $J_{\mathrm{c2}}$.}
\end{figure*}

\begin{figure}[ht]
\includegraphics[width=1\columnwidth]{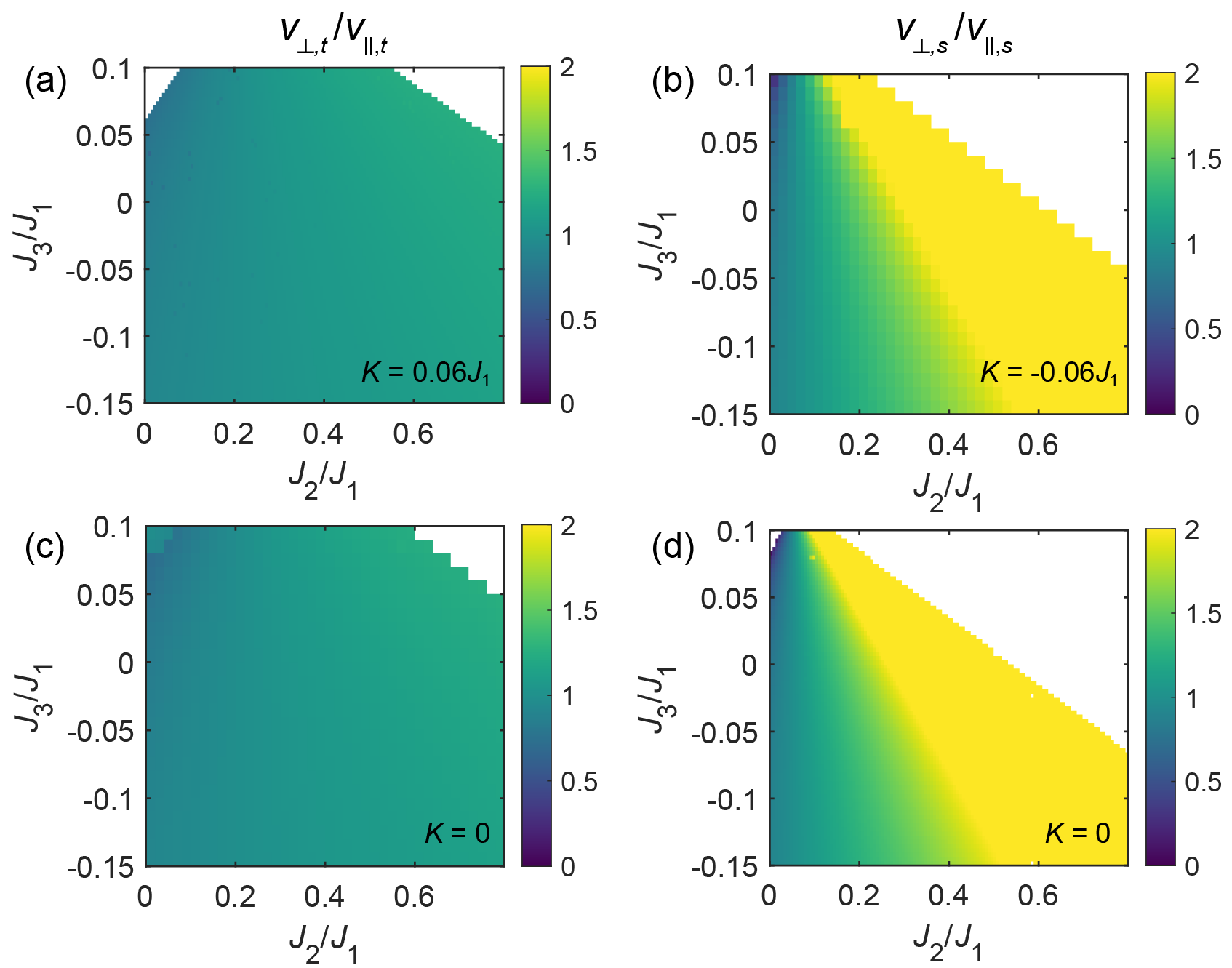} 
\caption{\label{vel_Kbq} A marginal effect of $K$ on theoretical $v_{\perp}/v_{\parallel}$. (a)--(b) $J_{2}$ and $J_{3}$ dependence of $v_{\perp}/v_{\parallel}$ for the triple-${\mathbf{Q}}$ and single-${\mathbf{Q}}$ phase with finite $K$, respectively. Note that we used an opposite sign of $K$ for each calculation to stabilize the magnetic ground state. (c)--(d) The same calculation result as (a)--(b), but without finite $K$.}
\end{figure}

In this section, we provide a more detailed analysis of the parameter dependence of $v_{\mathrm{L}}(\mathbf{k})$. First, Fig.~\ref{magvel_detail} visualizes the consequence of contrasting $v_{\mathrm{L}}(\mathbf{k})$ between the stripe single-${\mathbf{Q}}$ and tetrahedral triple-${\mathbf{Q}}$ orderings through constant-$\omega$ cuts at low energies. Indeed, different profiles of $v_{\mathrm{L}}(\mathbf{k})$ across most of the parameter space generally result in distinct constant-$\omega$ spectra for these two magnetic ground states [Fig.~\ref{magvel_detail}(a)--(b)]. This contrast remains clear even without interlayer interactions [i.e. 2D limit, Fig, \ref{magvel_detail}(e)--(f)]. However, $v_{\mathrm{L}}(\mathbf{k})$ of these two magnetic orderings can coincidentally be very similar for specific parameter sets, e.g., $J_{2}/J_{1}=0.3$ in Fig.~\ref{magvel_detail}(g)--(h). Thus, even though the likelihood is low, caution is advised when analyzing $v_{\mathrm{L}}(\mathbf{k})$ to distinguish single-${\mathbf{Q}}$ and triple-${\mathbf{Q}}$ magnetic structures if the exchange parameter set of a system, derived independently from the spin-wave analysis, happens to lie in such a region.

We also examine the effect of $K$ on $v_{\mathrm{L}}(\mathbf{k})$, in addition to the effect of interlayer interactions shown in Fig.~\ref{magvel}. Fig.~\ref{vel_Kbq} compares $v_{\perp,t}/v_{\parallel,t}$ and $v_{\perp,s}/v_{\parallel,s}$ calculated with and without finite $K$. For the the single-${\mathbf{Q}}$ phase calculation, we used $K<0$ to ensure its stabilization. The magnitude of $K$ was set to 0.06$J_{1}$, as determined from reproducing \TNt{}/\TNo{} = 0.6 in classical Monte Carlo simulations (see Appendix F). 

The results show that $K$ has a very marginal effect on the momentum dependence of $v_{\mathrm{L}}(\mathbf{k})$. Unlike $J_{\mathrm{c1}}$ and $J_{\mathrm{c2}}$, $K$ hardly changes the parameter dependence of $v_{\perp,t}/v_{\parallel,t}$ [Fig.~\ref{vel_Kbq}(a) and \ref{vel_Kbq}(c)]. Also, the variation of $v_{\perp,s}/v_{\parallel,s}$ due to $J_{2}$ and $J_{3}$ remains qualitatively the same, except for a uniform shift toward the right when adding $K=-0.06J_{1}$. We attribute this outcome primarily to the smaller magnitude of $K$ compared to bilinear exchange parameters. Such an order-of-magnitude smaller $K$ is expected to arise in most materials, considering its higher-order nature. Thus, $K$ or other four-spin interactions generally would not play a significant role in determining the momentum dependence of $v_{\mathrm{L}}(\mathbf{k})$ for both single-${\mathbf{Q}}$ and triple-${\mathbf{Q}}$ orderings. Consequently, refining only bilinear exchange interactions in the paramagnetic phase still leads to the successful identification of triple-${\mathbf{Q}}$ magnetic orderings by comparing the resultant $v_{\mathrm{L}}(\mathbf{k})$.

\newpage
\section{Linewidth broadening and renormalization of magnon branches}
\begin{figure}[h]
\includegraphics[width=1\columnwidth]{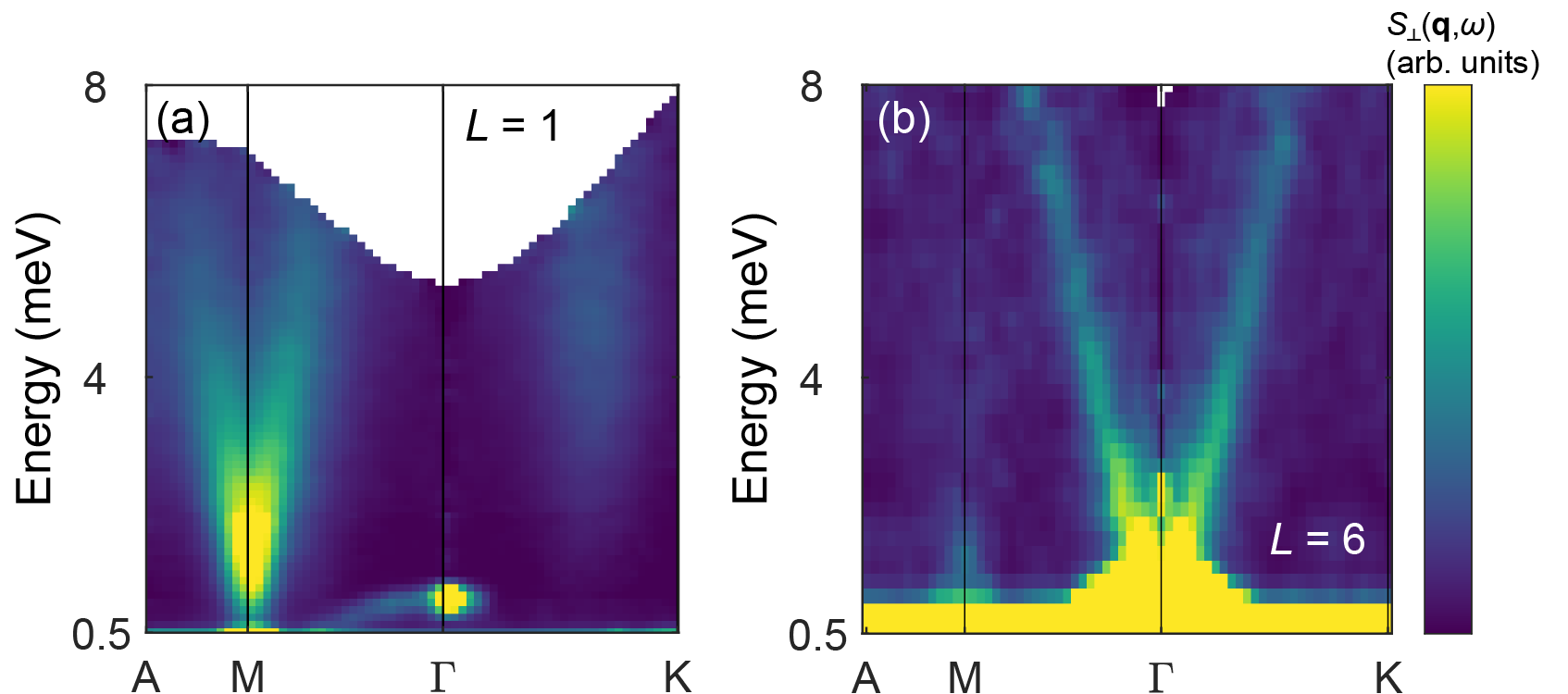} 
\caption{\label{phonon_com} Two energy-momentum slices of the INS data ($T=5$\,K) from different Brillouin zones. (a) A data slice on the [HK1] plane (low-${\mathbf{q}}$), showing a magnon spectrum. (b) A data slice on the [HK6] plane (high-${\mathbf{q}}$), showing a phonon spectrum. The phonon spectrum is much sharper than the magnon spectrum, indicating the presence of intrinsic linewidth broadening for the magnon modes in \tred{\CTS{}}.}
\end{figure}

We describe in more detail the evidence suggesting intrinsic magnon linewidth broadening (i.e., magnon decay) and renormalization beyond LSWT in \tred{\CTS{}}. The most unambiguous way to demonstrate intrinsic magnon linewidth broadening is to compare a magnon spectrum to a phonon spectrum measured in the same experiment. Except for a few special cases, phonons typically have a fairly long lifetime and thus represent a reference spectrum broadened only by instrumental resolution. Thus, a magnon spectrum that is significantly broader than the phonon signal directly indicates a finite magnon lifetime. Indeed, Fig.~\ref{phonon_com} shows that the magnon spectrum of \tred{\CTS{}} is much broader than its phonon spectrum, indicating the prevalence of magnon decay channels over a wide energy-momentum space. Meanwhile, analyzing the intrinsic linewidth broadening quantitatively (e.g., its momentum dependence) is very challenging due to the complex resolution ellipsoid of the time-of-flight spectrometer~\cite{CrB2}.

\begin{figure}[h]
\includegraphics[width=1\columnwidth]{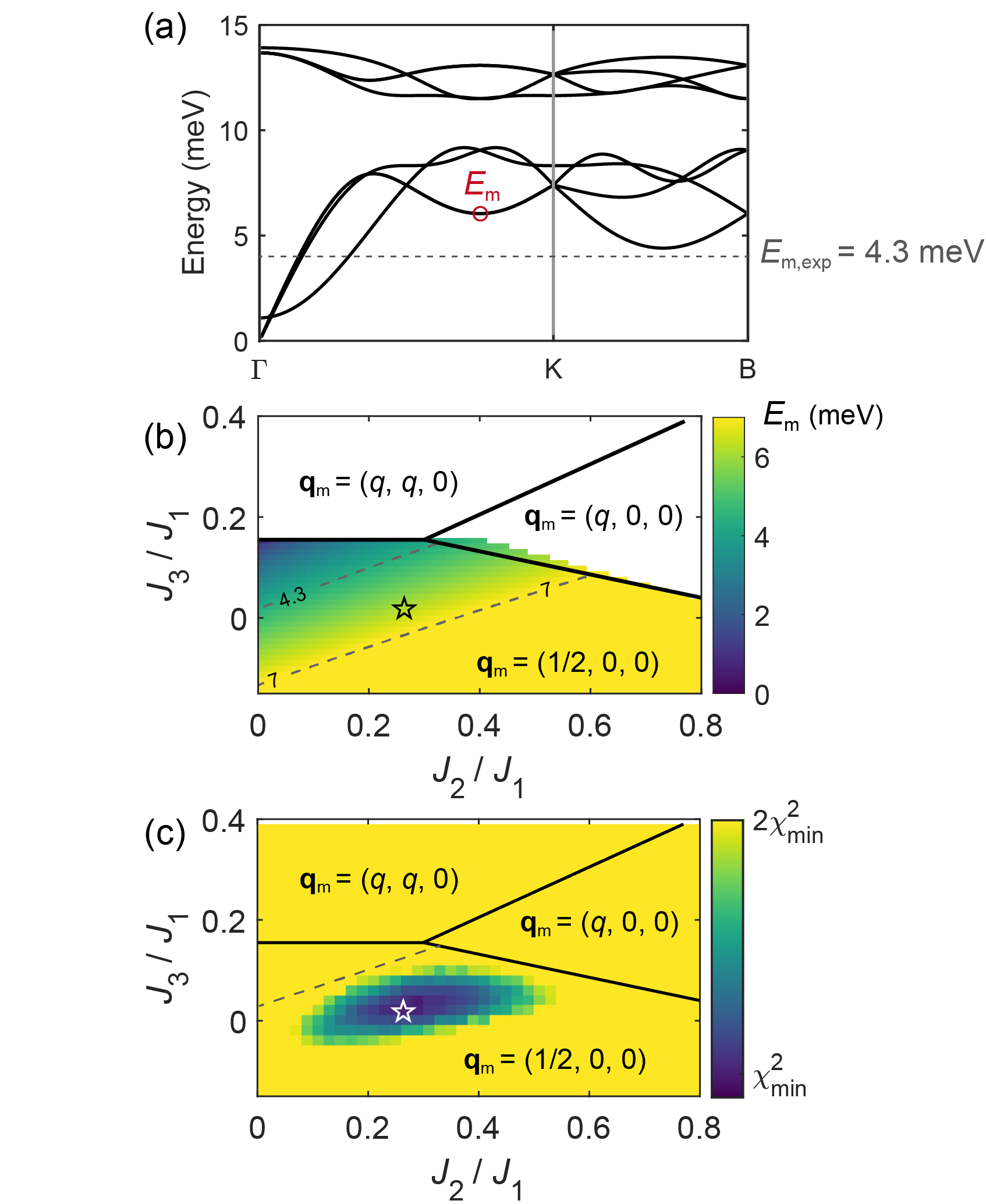} 
\caption{\label{mageig} Failure of LSWT in describing the observed magnon spectrum of \tred{\CTS{}} in the triple-${\mathbf{Q}}$ phase. (a) The magnon dispersion of the tetrahedral triple-${\mathbf{Q}}$ phase calculated using LSWT with the parameters in Table \ref{table:opt} and $K=0.06J_{1}$. The red circle in (a) denotes the lowest magnon eigenvalue ($E_{\mathrm{m}}$) at ${\mathbf{Q}}$\,=\,$\mathbf{q}_{\alpha}$ [see Figs. \ref{2mag}(c)] where the magnon dispersion forms a local minimum. The dashed grey line in (a) indicates the experimental  $E_{\mathrm{m}}$ [Fig.~\ref{2mag}(d)]. (b) $J_{2}/J_{1}$ and $J_{3}/J_{1}$ dependence of $E_{\mathrm{m}}$. The dashed grey lines in (b) are the trajectory of $E_{\mathrm{m}}= 4.3$ and 7\,meV. (c) The goodness-of-fit map ($\chi^{2}$ metric) from the paramagnetic phase analysis (Fig.~\ref{highT}), plotted in the same axes ranges as (b). The dashed grey line in (c) is the trajectory of $E_{\mathrm{m}}= 4.3$\,meV.}
\end{figure}

The magnon energy renormalization suggested at $\mathbf{q}_{\alpha,\beta,\gamma}$ in the triple-${\mathbf{Q}}$ phase (see the main text) is further supported by our advanced analysis illustrated in Fig.~\ref{mageig}. Fig.~\ref{mageig}(b) shows $J_{2}/J_{1}$ and $J_{3}/J_{1}$ dependence of the lowest magnon eigenvalue at ${\mathbf{q}}$\,$=\mathbf{q}_{\alpha}$ calculated by LSWT for the triple-${\mathbf{Q}}$ ordering [$E_{\mathrm{m}}$, see Fig.~\ref{mageig}(a)], where we used $J_{1}, J_{\mathrm{c1}},$ and $J_{\mathrm{c2}}$ shown in Table \ref{table:opt}. The optimal parameter set is different from those yielding $E_{\mathrm{m}}=4.3$\,meV observed experimentally [Fig.~\ref{2mag}((d))]. Notably, the parameter sets that produce $E_{\mathrm{m}}=4.3$\,meV [the dashed lines with the number 4.3 in Fig.~\ref{mageig}(b)] exhibits a very poor $\chi^{2}$-metric (=$2.8\chi^{2}_{\mathrm{min}}$) in the paramagnetic excitation analysis [Fig.~\ref{mageig}(c)]. This metric is much worse than that of the parameter sets that already show apparent disagreement with the measured $S_{\perp}(\mathbf{q},\omega)$ (see Fig.~\ref{wrong} in Appendix E). Moreover, using the parameter sets that give $E_{\mathrm{m}}=4.3$\,meV introduces additional discrepancies between the measured and simulated spectra at momentum positions other than $\mathbf{q} =\mathbf{q}_{\alpha}$. Thus, the observed $E_{\mathrm{m}}=4.3$\,meV at $\mathbf{q} = \mathbf{q}_{\alpha}$ suggests the presence of magnon energy renormalization beyond LSWT, which indeed could be present in the triple-${\mathbf{Q}}$ phase of \tred{\CTS{}} as described in the main text.

\newpage
\section{Magnon decay/renormalization by the Stoner continuum}
In the main text, we suggested that the pronounced magnon decay/renormalization observed on the edges of the hexagon that connects the M points in the momentum space (${\mathbf{q}}\in\{\mathbf{q}_{h}\}$) is primarily attributed to the two-magnon continuum. Nevertheless, the Stoner continuum may also partially contribute to our observations. This is because, for metallic antiferromagnets with a Fermi surface lying at ${\mathbf{q}}\in \{\mathbf{q}_{h}\}$, a gapless Stoner continuum should be present at ${\mathbf{q}} \in \{\mathbf{q}_{h}\}$. Notably, the Fermi surface of \tred{\CTS{}} is close to this condition~\cite{CTS_tripleQ_natcomm}.

\begin{figure}[h]
\includegraphics[width=1\columnwidth]{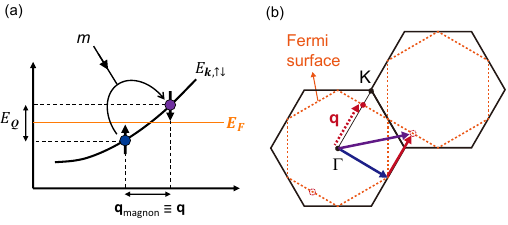} 
\caption{\label{Stoner} (a) Schematic illustration of a spin-flip electron-hole pair creation (i.e., a Stoner excitation) in a metallic antiferromagnet with spin-degenerate electron band structure. (b) The Stoner excitation process with momentum transfer $\mathbf{Q}\in\{\mathbf{q}_{h}\}$ and infinitesimally small energy transfer for a three-quarter-filled Fermi surface geometry. In this situation, the Stoner continuum becomes gapless at $\mathbf{Q}\in\{\mathbf{q}_{h}\}$.}
\end{figure}

Fig.~\ref{Stoner}(a) illustrates the decay of a magnon into a spin-flip electron-hole pair around the Fermi energy ($E_{\mathrm{F}}$), where both the electron and hole are on the same band index. Such intraband spin-flip excitations can occur in antiferromagnets that retain the spin degeneracy of electron bands (mostly due to $PT$ symmetry). Thus, for momentum vectors $\mathbf{k}$ that connect two different positions on the Fermi level, a Stoner excitation process with momentum transfer $\mathbf{k}$ can occur with infinitesimally small energy transfer. In other words, a gapless Stoner continuum is present at $\mathbf{k}$, allowing magnons with ${\mathbf{q}}=\mathbf{k}$ and any finite energy to decay. Fig.~\ref{Stoner}(b) shows that when the Fermi surface lies on ${\mathbf{q}}\in\{\mathbf{q}_{h}\}$, any momentum vectors on $\mathbf{q}_{h}$ satisfies this condition. Thus, a gapless Stoner continuum is expected along ${\mathbf{q}}\in\{\mathbf{q}_{h}\}$, potentially leading to sizable decay and renormalization of magnons at these momentum positions, such as $\mathbf{q}_{\alpha}$ in Fig.~\ref{2mag}. A more quantitative understanding of the effects of the Stoner continuum would require calculating the Stoner continuum DOS based on the full electron band structure of \tred{\CTS{}} (e.g. see~\cite{CrB2}), which is beyond the scope of this study.

However, it is important to note that the magnon decay process via the Stoner continuum could be more complex in noncollinear magnets than the simple conjecture described above. A noncollinear spin configuration complicates the spin index of electron bands in momentum space (rather than simply spin-up or spin-down, the spin index will also be noncollinear), which would affect the Stoner excitation process as it involves a full flip of a single spin~\cite{CrB2}. This complication would suppress the intraband Stoner excitation processes to some extent and potentially remove some magnon decay channels discussed in the previous paragraph.  

\clearpage


\begin{thebibliography}{86}%
\makeatletter
\providecommand \@ifxundefined [1]{%
 \@ifx{#1\undefined}
}%
\providecommand \@ifnum [1]{%
 \ifnum #1\expandafter \@firstoftwo
 \else \expandafter \@secondoftwo
 \fi
}%
\providecommand \@ifx [1]{%
 \ifx #1\expandafter \@firstoftwo
 \else \expandafter \@secondoftwo
 \fi
}%
\providecommand \natexlab [1]{#1}%
\providecommand \enquote  [1]{``#1''}%
\providecommand \bibnamefont  [1]{#1}%
\providecommand \bibfnamefont [1]{#1}%
\providecommand \citenamefont [1]{#1}%
\providecommand \href@noop [0]{\@secondoftwo}%
\providecommand \href [0]{\begingroup \@sanitize@url \@href}%
\providecommand \@href[1]{\@@startlink{#1}\@@href}%
\providecommand \@@href[1]{\endgroup#1\@@endlink}%
\providecommand \@sanitize@url [0]{\catcode `\\12\catcode `\$12\catcode `\&12\catcode `\#12\catcode `\^12\catcode `\_12\catcode `\%12\relax}%
\providecommand \@@startlink[1]{}%
\providecommand \@@endlink[0]{}%
\providecommand \url  [0]{\begingroup\@sanitize@url \@url }%
\providecommand \@url [1]{\endgroup\@href {#1}{\urlprefix }}%
\providecommand \urlprefix  [0]{URL }%
\providecommand \Eprint [0]{\href }%
\providecommand \doibase [0]{https://doi.org/}%
\providecommand \selectlanguage [0]{\@gobble}%
\providecommand \bibinfo  [0]{\@secondoftwo}%
\providecommand \bibfield  [0]{\@secondoftwo}%
\providecommand \translation [1]{[#1]}%
\providecommand \BibitemOpen [0]{}%
\providecommand \bibitemStop [0]{}%
\providecommand \bibitemNoStop [0]{.\EOS\space}%
\providecommand \EOS [0]{\spacefactor3000\relax}%
\providecommand \BibitemShut  [1]{\csname bibitem#1\endcsname}%
\let\auto@bib@innerbib\@empty
\bibitem [{\citenamefont {{\v S}mejkal}\ \emph {et~al.}(2022)\citenamefont {{\v S}mejkal}, \citenamefont {MacDonald}, \citenamefont {Sinova}, \citenamefont {Nakatsuji},\ and\ \citenamefont {Jungwirth}}]{Smejkal22}%
  \BibitemOpen
  \bibfield  {author} {\bibinfo {author} {\bibfnamefont {L.}~\bibnamefont {{\v S}mejkal}}, \bibinfo {author} {\bibfnamefont {A.~H.}\ \bibnamefont {MacDonald}}, \bibinfo {author} {\bibfnamefont {J.}~\bibnamefont {Sinova}}, \bibinfo {author} {\bibfnamefont {S.}~\bibnamefont {Nakatsuji}},\ and\ \bibinfo {author} {\bibfnamefont {T.}~\bibnamefont {Jungwirth}},\ }\bibfield  {title} {\bibinfo {title} {Anomalous {Hall} antiferromagnets},\ }\href {https://doi.org/10.1038/s41578-022-00430-3} {\bibfield  {journal} {\bibinfo  {journal} {Nature Reviews Materials}\ }\textbf {\bibinfo {volume} {7}},\ \bibinfo {pages} {482} (\bibinfo {year} {2022})}\BibitemShut {NoStop}%
\bibitem [{\citenamefont {Bonbien}\ \emph {et~al.}(2021)\citenamefont {Bonbien}, \citenamefont {Zhuo}, \citenamefont {Salimath}, \citenamefont {Ly}, \citenamefont {Abbout},\ and\ \citenamefont {Manchon}}]{Bonbien22}%
  \BibitemOpen
  \bibfield  {author} {\bibinfo {author} {\bibfnamefont {V.}~\bibnamefont {Bonbien}}, \bibinfo {author} {\bibfnamefont {F.}~\bibnamefont {Zhuo}}, \bibinfo {author} {\bibfnamefont {A.}~\bibnamefont {Salimath}}, \bibinfo {author} {\bibfnamefont {O.}~\bibnamefont {Ly}}, \bibinfo {author} {\bibfnamefont {A.}~\bibnamefont {Abbout}},\ and\ \bibinfo {author} {\bibfnamefont {A.}~\bibnamefont {Manchon}},\ }\bibfield  {title} {\bibinfo {title} {Topological aspects of antiferromagnets},\ }\href {https://doi.org/10.1088/1361-6463/ac28fa} {\bibfield  {journal} {\bibinfo  {journal} {Journal of Physics D: Applied Physics}\ }\textbf {\bibinfo {volume} {55}},\ \bibinfo {pages} {103002} (\bibinfo {year} {2021})}\BibitemShut {NoStop}%
\bibitem [{\citenamefont {Okubo}\ \emph {et~al.}(2012)\citenamefont {Okubo}, \citenamefont {Chung},\ and\ \citenamefont {Kawamura}}]{Okubo12}%
  \BibitemOpen
  \bibfield  {author} {\bibinfo {author} {\bibfnamefont {T.}~\bibnamefont {Okubo}}, \bibinfo {author} {\bibfnamefont {S.}~\bibnamefont {Chung}},\ and\ \bibinfo {author} {\bibfnamefont {H.}~\bibnamefont {Kawamura}},\ }\bibfield  {title} {\bibinfo {title} {Multiple-$q$ states and the skyrmion lattice of the triangular-lattice {Heisenberg} antiferromagnet under magnetic fields},\ }\href {https://doi.org/10.1103/PhysRevLett.108.017206} {\bibfield  {journal} {\bibinfo  {journal} {Phys. Rev. Lett.}\ }\textbf {\bibinfo {volume} {108}},\ \bibinfo {pages} {017206} (\bibinfo {year} {2012})}\BibitemShut {NoStop}%
\bibitem [{\citenamefont {Leonov}\ and\ \citenamefont {Mostovoy}(2015)}]{Leonov2015}%
  \BibitemOpen
  \bibfield  {author} {\bibinfo {author} {\bibfnamefont {A.~O.}\ \bibnamefont {Leonov}}\ and\ \bibinfo {author} {\bibfnamefont {M.}~\bibnamefont {Mostovoy}},\ }\bibfield  {title} {\bibinfo {title} {Multiply periodic states and isolated skyrmions in an anisotropic frustrated magnet},\ }\href {https://doi.org/10.1038/ncomms9275} {\bibfield  {journal} {\bibinfo  {journal} {Nature Communications}\ }\textbf {\bibinfo {volume} {6}},\ \bibinfo {pages} {8275} (\bibinfo {year} {2015})}\BibitemShut {NoStop}%
\bibitem [{\citenamefont {Hayami}\ \emph {et~al.}(2016)\citenamefont {Hayami}, \citenamefont {Lin},\ and\ \citenamefont {Batista}}]{Hayami16}%
  \BibitemOpen
  \bibfield  {author} {\bibinfo {author} {\bibfnamefont {S.}~\bibnamefont {Hayami}}, \bibinfo {author} {\bibfnamefont {S.-Z.}\ \bibnamefont {Lin}},\ and\ \bibinfo {author} {\bibfnamefont {C.~D.}\ \bibnamefont {Batista}},\ }\bibfield  {title} {\bibinfo {title} {Bubble and skyrmion crystals in frustrated magnets with easy-axis anisotropy},\ }\href {https://doi.org/10.1103/PhysRevB.93.184413} {\bibfield  {journal} {\bibinfo  {journal} {Phys. Rev. B}\ }\textbf {\bibinfo {volume} {93}},\ \bibinfo {pages} {184413} (\bibinfo {year} {2016})}\BibitemShut {NoStop}%
\bibitem [{\citenamefont {Ozawa}\ \emph {et~al.}(2016)\citenamefont {Ozawa}, \citenamefont {Hayami}, \citenamefont {Barros}, \citenamefont {Chern}, \citenamefont {Motome},\ and\ \citenamefont {Batista}}]{Ozawa16}%
  \BibitemOpen
  \bibfield  {author} {\bibinfo {author} {\bibfnamefont {R.}~\bibnamefont {Ozawa}}, \bibinfo {author} {\bibfnamefont {S.}~\bibnamefont {Hayami}}, \bibinfo {author} {\bibfnamefont {K.}~\bibnamefont {Barros}}, \bibinfo {author} {\bibfnamefont {G.-W.}\ \bibnamefont {Chern}}, \bibinfo {author} {\bibfnamefont {Y.}~\bibnamefont {Motome}},\ and\ \bibinfo {author} {\bibfnamefont {C.~D.}\ \bibnamefont {Batista}},\ }\bibfield  {title} {\bibinfo {title} {Vortex crystals with chiral stripes in itinerant magnets},\ }\href {https://doi.org/10.7566/JPSJ.85.103703} {\bibfield  {journal} {\bibinfo  {journal} {Journal of the Physical Society of Japan}\ }\textbf {\bibinfo {volume} {85}},\ \bibinfo {pages} {103703} (\bibinfo {year} {2016})},\ \Eprint {https://arxiv.org/abs/https://doi.org/10.7566/JPSJ.85.103703} {https://doi.org/10.7566/JPSJ.85.103703} \BibitemShut {NoStop}%
\bibitem [{\citenamefont {Batista}\ \emph {et~al.}(2016{\natexlab{a}})\citenamefont {Batista}, \citenamefont {Lin}, \citenamefont {Hayami},\ and\ \citenamefont {Kamiya}}]{Batista_2016}%
  \BibitemOpen
  \bibfield  {author} {\bibinfo {author} {\bibfnamefont {C.~D.}\ \bibnamefont {Batista}}, \bibinfo {author} {\bibfnamefont {S.-Z.}\ \bibnamefont {Lin}}, \bibinfo {author} {\bibfnamefont {S.}~\bibnamefont {Hayami}},\ and\ \bibinfo {author} {\bibfnamefont {Y.}~\bibnamefont {Kamiya}},\ }\bibfield  {title} {\bibinfo {title} {Frustration and chiral orderings in correlated electron systems},\ }\href {https://doi.org/10.1088/0034-4885/79/8/084504} {\bibfield  {journal} {\bibinfo  {journal} {Reports on Progress in Physics}\ }\textbf {\bibinfo {volume} {79}},\ \bibinfo {pages} {084504} (\bibinfo {year} {2016}{\natexlab{a}})}\BibitemShut {NoStop}%
\bibitem [{\citenamefont {Ozawa}\ \emph {et~al.}(2017)\citenamefont {Ozawa}, \citenamefont {Hayami},\ and\ \citenamefont {Motome}}]{Ozawa17}%
  \BibitemOpen
  \bibfield  {author} {\bibinfo {author} {\bibfnamefont {R.}~\bibnamefont {Ozawa}}, \bibinfo {author} {\bibfnamefont {S.}~\bibnamefont {Hayami}},\ and\ \bibinfo {author} {\bibfnamefont {Y.}~\bibnamefont {Motome}},\ }\bibfield  {title} {\bibinfo {title} {Zero-field skyrmions with a high topological number in itinerant magnets},\ }\href {https://doi.org/10.1103/PhysRevLett.118.147205} {\bibfield  {journal} {\bibinfo  {journal} {Phys. Rev. Lett.}\ }\textbf {\bibinfo {volume} {118}},\ \bibinfo {pages} {147205} (\bibinfo {year} {2017})}\BibitemShut {NoStop}%
\bibitem [{\citenamefont {Wang}\ \emph {et~al.}(2021)\citenamefont {Wang}, \citenamefont {Su}, \citenamefont {Lin},\ and\ \citenamefont {Batista}}]{Wang2021}%
  \BibitemOpen
  \bibfield  {author} {\bibinfo {author} {\bibfnamefont {Z.}~\bibnamefont {Wang}}, \bibinfo {author} {\bibfnamefont {Y.}~\bibnamefont {Su}}, \bibinfo {author} {\bibfnamefont {S.-Z.}\ \bibnamefont {Lin}},\ and\ \bibinfo {author} {\bibfnamefont {C.~D.}\ \bibnamefont {Batista}},\ }\bibfield  {title} {\bibinfo {title} {Meron, skyrmion, and vortex crystals in centrosymmetric tetragonal magnets},\ }\href {https://doi.org/10.1103/PhysRevB.103.104408} {\bibfield  {journal} {\bibinfo  {journal} {Phys. Rev. B}\ }\textbf {\bibinfo {volume} {103}},\ \bibinfo {pages} {104408} (\bibinfo {year} {2021})}\BibitemShut {NoStop}%
\bibitem [{\citenamefont {Hayami}\ and\ \citenamefont {Motome}(2021)}]{Motome21_review}%
  \BibitemOpen
  \bibfield  {author} {\bibinfo {author} {\bibfnamefont {S.}~\bibnamefont {Hayami}}\ and\ \bibinfo {author} {\bibfnamefont {Y.}~\bibnamefont {Motome}},\ }\bibfield  {title} {\bibinfo {title} {Topological spin crystals by itinerant frustration},\ }\href@noop {} {\bibfield  {journal} {\bibinfo  {journal} {Journal of Physics: Condensed Matter}\ }\textbf {\bibinfo {volume} {33}},\ \bibinfo {pages} {443001} (\bibinfo {year} {2021})}\BibitemShut {NoStop}%
\bibitem [{\citenamefont {Wang}\ \emph {et~al.}(2020)\citenamefont {Wang}, \citenamefont {Su}, \citenamefont {Lin},\ and\ \citenamefont {Batista}}]{Wang20}%
  \BibitemOpen
  \bibfield  {author} {\bibinfo {author} {\bibfnamefont {Z.}~\bibnamefont {Wang}}, \bibinfo {author} {\bibfnamefont {Y.}~\bibnamefont {Su}}, \bibinfo {author} {\bibfnamefont {S.-Z.}\ \bibnamefont {Lin}},\ and\ \bibinfo {author} {\bibfnamefont {C.~D.}\ \bibnamefont {Batista}},\ }\bibfield  {title} {\bibinfo {title} {Skyrmion crystal from {RKKY} interaction mediated by 2d electron gas},\ }\href {https://doi.org/10.1103/PhysRevLett.124.207201} {\bibfield  {journal} {\bibinfo  {journal} {Phys. Rev. Lett.}\ }\textbf {\bibinfo {volume} {124}},\ \bibinfo {pages} {207201} (\bibinfo {year} {2020})}\BibitemShut {NoStop}%
\bibitem [{\citenamefont {Wang}\ and\ \citenamefont {Batista}(2023)}]{wang_skyrmion}%
  \BibitemOpen
  \bibfield  {author} {\bibinfo {author} {\bibfnamefont {Z.}~\bibnamefont {Wang}}\ and\ \bibinfo {author} {\bibfnamefont {C.~D.}\ \bibnamefont {Batista}},\ }\bibfield  {title} {\bibinfo {title} {Skyrmion crystals in the triangular {Kondo} lattice model},\ }\href@noop {} {\bibfield  {journal} {\bibinfo  {journal} {SciPost Physics}\ }\textbf {\bibinfo {volume} {15}},\ \bibinfo {pages} {161} (\bibinfo {year} {2023})}\BibitemShut {NoStop}%
\bibitem [{\citenamefont {Bogdanov}\ and\ \citenamefont {Yablonskii}(1989)}]{Skyrmion_1989}%
  \BibitemOpen
  \bibfield  {author} {\bibinfo {author} {\bibfnamefont {A.~N.}\ \bibnamefont {Bogdanov}}\ and\ \bibinfo {author} {\bibfnamefont {D.}~\bibnamefont {Yablonskii}},\ }\bibfield  {title} {\bibinfo {title} {Thermodynamically stable “vortices” in magnetically ordered crystals. the mixed state of magnets},\ }\href@noop {} {\bibfield  {journal} {\bibinfo  {journal} {Zh. Eksp. Teor. Fiz}\ }\textbf {\bibinfo {volume} {95}},\ \bibinfo {pages} {178} (\bibinfo {year} {1989})}\BibitemShut {NoStop}%
\bibitem [{\citenamefont {Fert}\ \emph {et~al.}(2017)\citenamefont {Fert}, \citenamefont {Reyren},\ and\ \citenamefont {Cros}}]{skyr_review}%
  \BibitemOpen
  \bibfield  {author} {\bibinfo {author} {\bibfnamefont {A.}~\bibnamefont {Fert}}, \bibinfo {author} {\bibfnamefont {N.}~\bibnamefont {Reyren}},\ and\ \bibinfo {author} {\bibfnamefont {V.}~\bibnamefont {Cros}},\ }\bibfield  {title} {\bibinfo {title} {Magnetic skyrmions: advances in physics and potential applications},\ }\href@noop {} {\bibfield  {journal} {\bibinfo  {journal} {Nature Reviews Materials}\ }\textbf {\bibinfo {volume} {2}},\ \bibinfo {pages} {1} (\bibinfo {year} {2017})}\BibitemShut {NoStop}%
\bibitem [{\citenamefont {Romming}\ \emph {et~al.}(2013)\citenamefont {Romming}, \citenamefont {Hanneken}, \citenamefont {Menzel}, \citenamefont {Bickel}, \citenamefont {Wolter}, \citenamefont {von Bergmann}, \citenamefont {Kubetzka},\ and\ \citenamefont {Wiesendanger}}]{skyr_writing}%
  \BibitemOpen
  \bibfield  {author} {\bibinfo {author} {\bibfnamefont {N.}~\bibnamefont {Romming}}, \bibinfo {author} {\bibfnamefont {C.}~\bibnamefont {Hanneken}}, \bibinfo {author} {\bibfnamefont {M.}~\bibnamefont {Menzel}}, \bibinfo {author} {\bibfnamefont {J.~E.}\ \bibnamefont {Bickel}}, \bibinfo {author} {\bibfnamefont {B.}~\bibnamefont {Wolter}}, \bibinfo {author} {\bibfnamefont {K.}~\bibnamefont {von Bergmann}}, \bibinfo {author} {\bibfnamefont {A.}~\bibnamefont {Kubetzka}},\ and\ \bibinfo {author} {\bibfnamefont {R.}~\bibnamefont {Wiesendanger}},\ }\bibfield  {title} {\bibinfo {title} {Writing and deleting single magnetic skyrmions},\ }\href@noop {} {\bibfield  {journal} {\bibinfo  {journal} {Science}\ }\textbf {\bibinfo {volume} {341}},\ \bibinfo {pages} {636} (\bibinfo {year} {2013})}\BibitemShut {NoStop}%
\bibitem [{\citenamefont {Schulz}\ \emph {et~al.}(2012)\citenamefont {Schulz}, \citenamefont {Ritz}, \citenamefont {Bauer}, \citenamefont {Halder}, \citenamefont {Wagner}, \citenamefont {Franz}, \citenamefont {Pfleiderer}, \citenamefont {Everschor}, \citenamefont {Garst},\ and\ \citenamefont {Rosch}}]{Skyr_electrody}%
  \BibitemOpen
  \bibfield  {author} {\bibinfo {author} {\bibfnamefont {T.}~\bibnamefont {Schulz}}, \bibinfo {author} {\bibfnamefont {R.}~\bibnamefont {Ritz}}, \bibinfo {author} {\bibfnamefont {A.}~\bibnamefont {Bauer}}, \bibinfo {author} {\bibfnamefont {M.}~\bibnamefont {Halder}}, \bibinfo {author} {\bibfnamefont {M.}~\bibnamefont {Wagner}}, \bibinfo {author} {\bibfnamefont {C.}~\bibnamefont {Franz}}, \bibinfo {author} {\bibfnamefont {C.}~\bibnamefont {Pfleiderer}}, \bibinfo {author} {\bibfnamefont {K.}~\bibnamefont {Everschor}}, \bibinfo {author} {\bibfnamefont {M.}~\bibnamefont {Garst}},\ and\ \bibinfo {author} {\bibfnamefont {A.}~\bibnamefont {Rosch}},\ }\bibfield  {title} {\bibinfo {title} {Emergent electrodynamics of skyrmions in a chiral magnet},\ }\href@noop {} {\bibfield  {journal} {\bibinfo  {journal} {Nature Physics}\ }\textbf {\bibinfo {volume} {8}},\ \bibinfo {pages} {301} (\bibinfo {year} {2012})}\BibitemShut {NoStop}%
\bibitem [{\citenamefont {Fert}\ \emph {et~al.}(2013)\citenamefont {Fert}, \citenamefont {Cros},\ and\ \citenamefont {Sampaio}}]{Skyr_track}%
  \BibitemOpen
  \bibfield  {author} {\bibinfo {author} {\bibfnamefont {A.}~\bibnamefont {Fert}}, \bibinfo {author} {\bibfnamefont {V.}~\bibnamefont {Cros}},\ and\ \bibinfo {author} {\bibfnamefont {J.}~\bibnamefont {Sampaio}},\ }\bibfield  {title} {\bibinfo {title} {Skyrmions on the track},\ }\href@noop {} {\bibfield  {journal} {\bibinfo  {journal} {Nature nanotechnology}\ }\textbf {\bibinfo {volume} {8}},\ \bibinfo {pages} {152} (\bibinfo {year} {2013})}\BibitemShut {NoStop}%
\bibitem [{\citenamefont {Rousochatzakis}\ \emph {et~al.}(2016)\citenamefont {Rousochatzakis}, \citenamefont {R{\"o}ssler}, \citenamefont {Van Den~Brink},\ and\ \citenamefont {Daghofer}}]{vortex_kitaev}%
  \BibitemOpen
  \bibfield  {author} {\bibinfo {author} {\bibfnamefont {I.}~\bibnamefont {Rousochatzakis}}, \bibinfo {author} {\bibfnamefont {U.~K.}\ \bibnamefont {R{\"o}ssler}}, \bibinfo {author} {\bibfnamefont {J.}~\bibnamefont {Van Den~Brink}},\ and\ \bibinfo {author} {\bibfnamefont {M.}~\bibnamefont {Daghofer}},\ }\bibfield  {title} {\bibinfo {title} {{Kitaev} anisotropy induces mesoscopic z 2 vortex crystals in frustrated hexagonal antiferromagnets},\ }\href@noop {} {\bibfield  {journal} {\bibinfo  {journal} {Phys. Rev. B}\ }\textbf {\bibinfo {volume} {93}},\ \bibinfo {pages} {104417} (\bibinfo {year} {2016})}\BibitemShut {NoStop}%
\bibitem [{\citenamefont {Takagi}\ \emph {et~al.}(2018)\citenamefont {Takagi}, \citenamefont {White}, \citenamefont {Hayami}, \citenamefont {Arita}, \citenamefont {Honecker}, \citenamefont {R{\o}nnow}, \citenamefont {Tokura},\ and\ \citenamefont {Seki}}]{takagi2018}%
  \BibitemOpen
  \bibfield  {author} {\bibinfo {author} {\bibfnamefont {R.}~\bibnamefont {Takagi}}, \bibinfo {author} {\bibfnamefont {J.}~\bibnamefont {White}}, \bibinfo {author} {\bibfnamefont {S.}~\bibnamefont {Hayami}}, \bibinfo {author} {\bibfnamefont {R.}~\bibnamefont {Arita}}, \bibinfo {author} {\bibfnamefont {D.}~\bibnamefont {Honecker}}, \bibinfo {author} {\bibfnamefont {H.}~\bibnamefont {R{\o}nnow}}, \bibinfo {author} {\bibfnamefont {Y.}~\bibnamefont {Tokura}},\ and\ \bibinfo {author} {\bibfnamefont {S.}~\bibnamefont {Seki}},\ }\bibfield  {title} {\bibinfo {title} {Multiple-q noncollinear magnetism in an itinerant hexagonal magnet},\ }\href@noop {} {\bibfield  {journal} {\bibinfo  {journal} {Science advances}\ }\textbf {\bibinfo {volume} {4}},\ \bibinfo {pages} {eaau3402} (\bibinfo {year} {2018})}\BibitemShut {NoStop}%
\bibitem [{\citenamefont {Kurumaji}\ \emph {et~al.}(2019)\citenamefont {Kurumaji}, \citenamefont {Nakajima}, \citenamefont {Hirschberger}, \citenamefont {Kikkawa}, \citenamefont {Yamasaki}, \citenamefont {Sagayama}, \citenamefont {Nakao}, \citenamefont {Taguchi}, \citenamefont {Arima},\ and\ \citenamefont {Tokura}}]{TL_skyrmion2}%
  \BibitemOpen
  \bibfield  {author} {\bibinfo {author} {\bibfnamefont {T.}~\bibnamefont {Kurumaji}}, \bibinfo {author} {\bibfnamefont {T.}~\bibnamefont {Nakajima}}, \bibinfo {author} {\bibfnamefont {M.}~\bibnamefont {Hirschberger}}, \bibinfo {author} {\bibfnamefont {A.}~\bibnamefont {Kikkawa}}, \bibinfo {author} {\bibfnamefont {Y.}~\bibnamefont {Yamasaki}}, \bibinfo {author} {\bibfnamefont {H.}~\bibnamefont {Sagayama}}, \bibinfo {author} {\bibfnamefont {H.}~\bibnamefont {Nakao}}, \bibinfo {author} {\bibfnamefont {Y.}~\bibnamefont {Taguchi}}, \bibinfo {author} {\bibfnamefont {T.-h.}\ \bibnamefont {Arima}},\ and\ \bibinfo {author} {\bibfnamefont {Y.}~\bibnamefont {Tokura}},\ }\bibfield  {title} {\bibinfo {title} {Skyrmion lattice with a giant topological {Hall} effect in a frustrated triangular-lattice magnet},\ }\href@noop {} {\bibfield  {journal} {\bibinfo  {journal} {Science}\ }\textbf {\bibinfo {volume} {365}},\ \bibinfo {pages} {914} (\bibinfo {year} {2019})}\BibitemShut {NoStop}%
\bibitem [{\citenamefont {Mühlbauer}\ \emph {et~al.}(2009)\citenamefont {Mühlbauer}, \citenamefont {Binz}, \citenamefont {Jonietz}, \citenamefont {Pfleiderer}, \citenamefont {Rosch}, \citenamefont {Neubauer}, \citenamefont {Georgii},\ and\ \citenamefont {Böni}}]{skyrmion_diffH}%
  \BibitemOpen
  \bibfield  {author} {\bibinfo {author} {\bibfnamefont {S.}~\bibnamefont {Mühlbauer}}, \bibinfo {author} {\bibfnamefont {B.}~\bibnamefont {Binz}}, \bibinfo {author} {\bibfnamefont {F.}~\bibnamefont {Jonietz}}, \bibinfo {author} {\bibfnamefont {C.}~\bibnamefont {Pfleiderer}}, \bibinfo {author} {\bibfnamefont {A.}~\bibnamefont {Rosch}}, \bibinfo {author} {\bibfnamefont {A.}~\bibnamefont {Neubauer}}, \bibinfo {author} {\bibfnamefont {R.}~\bibnamefont {Georgii}},\ and\ \bibinfo {author} {\bibfnamefont {P.}~\bibnamefont {Böni}},\ }\bibfield  {title} {\bibinfo {title} {Skyrmion lattice in a chiral magnet},\ }\href {https://doi.org/10.1126/science.1166767} {\bibfield  {journal} {\bibinfo  {journal} {Science}\ }\textbf {\bibinfo {volume} {323}},\ \bibinfo {pages} {915} (\bibinfo {year} {2009})},\ \Eprint {https://arxiv.org/abs/https://www.science.org/doi/pdf/10.1126/science.1166767} {https://www.science.org/doi/pdf/10.1126/science.1166767} \BibitemShut {NoStop}%
\bibitem [{\citenamefont {Yu}\ \emph {et~al.}(2010)\citenamefont {Yu}, \citenamefont {Onose}, \citenamefont {Kanazawa}, \citenamefont {Park}, \citenamefont {Han}, \citenamefont {Matsui}, \citenamefont {Nagaosa},\ and\ \citenamefont {Tokura}}]{skyr_rspace}%
  \BibitemOpen
  \bibfield  {author} {\bibinfo {author} {\bibfnamefont {X.}~\bibnamefont {Yu}}, \bibinfo {author} {\bibfnamefont {Y.}~\bibnamefont {Onose}}, \bibinfo {author} {\bibfnamefont {N.}~\bibnamefont {Kanazawa}}, \bibinfo {author} {\bibfnamefont {J.~H.}\ \bibnamefont {Park}}, \bibinfo {author} {\bibfnamefont {J.}~\bibnamefont {Han}}, \bibinfo {author} {\bibfnamefont {Y.}~\bibnamefont {Matsui}}, \bibinfo {author} {\bibfnamefont {N.}~\bibnamefont {Nagaosa}},\ and\ \bibinfo {author} {\bibfnamefont {Y.}~\bibnamefont {Tokura}},\ }\bibfield  {title} {\bibinfo {title} {Real-space observation of a two-dimensional skyrmion crystal},\ }\href@noop {} {\bibfield  {journal} {\bibinfo  {journal} {Nature}\ }\textbf {\bibinfo {volume} {465}},\ \bibinfo {pages} {901} (\bibinfo {year} {2010})}\BibitemShut {NoStop}%
\bibitem [{\citenamefont {Heinze}\ \emph {et~al.}(2011)\citenamefont {Heinze}, \citenamefont {Von~Bergmann}, \citenamefont {Menzel}, \citenamefont {Brede}, \citenamefont {Kubetzka}, \citenamefont {Wiesendanger}, \citenamefont {Bihlmayer},\ and\ \citenamefont {Bl{\"u}gel}}]{Skyrmion_STM}%
  \BibitemOpen
  \bibfield  {author} {\bibinfo {author} {\bibfnamefont {S.}~\bibnamefont {Heinze}}, \bibinfo {author} {\bibfnamefont {K.}~\bibnamefont {Von~Bergmann}}, \bibinfo {author} {\bibfnamefont {M.}~\bibnamefont {Menzel}}, \bibinfo {author} {\bibfnamefont {J.}~\bibnamefont {Brede}}, \bibinfo {author} {\bibfnamefont {A.}~\bibnamefont {Kubetzka}}, \bibinfo {author} {\bibfnamefont {R.}~\bibnamefont {Wiesendanger}}, \bibinfo {author} {\bibfnamefont {G.}~\bibnamefont {Bihlmayer}},\ and\ \bibinfo {author} {\bibfnamefont {S.}~\bibnamefont {Bl{\"u}gel}},\ }\bibfield  {title} {\bibinfo {title} {Spontaneous atomic-scale magnetic skyrmion lattice in two dimensions},\ }\href@noop {} {\bibfield  {journal} {\bibinfo  {journal} {Nature Physics}\ }\textbf {\bibinfo {volume} {7}},\ \bibinfo {pages} {713} (\bibinfo {year} {2011})}\BibitemShut {NoStop}%
\bibitem [{\citenamefont {Jensen}\ and\ \citenamefont {Bak}(1981)}]{3Qdynamics1}%
  \BibitemOpen
  \bibfield  {author} {\bibinfo {author} {\bibfnamefont {J.}~\bibnamefont {Jensen}}\ and\ \bibinfo {author} {\bibfnamefont {P.}~\bibnamefont {Bak}},\ }\bibfield  {title} {\bibinfo {title} {Spin waves in triple-$\stackrel{\ensuremath{\rightarrow}}{\mathrm{q}}$ structures. application to {$\mathrm{USb}$}},\ }\href {https://doi.org/10.1103/PhysRevB.23.6180} {\bibfield  {journal} {\bibinfo  {journal} {Phys. Rev. B}\ }\textbf {\bibinfo {volume} {23}},\ \bibinfo {pages} {6180} (\bibinfo {year} {1981})}\BibitemShut {NoStop}%
\bibitem [{\citenamefont {Park}\ \emph {et~al.}(2023)\citenamefont {Park}, \citenamefont {Cho}, \citenamefont {Kim}, \citenamefont {An}, \citenamefont {Kang}, \citenamefont {Avdeev}, \citenamefont {Sibille}, \citenamefont {Iida}, \citenamefont {Kajimoto}, \citenamefont {Lee} \emph {et~al.}}]{CTS_tripleQ_natcomm}%
  \BibitemOpen
  \bibfield  {author} {\bibinfo {author} {\bibfnamefont {P.}~\bibnamefont {Park}}, \bibinfo {author} {\bibfnamefont {W.}~\bibnamefont {Cho}}, \bibinfo {author} {\bibfnamefont {C.}~\bibnamefont {Kim}}, \bibinfo {author} {\bibfnamefont {Y.}~\bibnamefont {An}}, \bibinfo {author} {\bibfnamefont {Y.-G.}\ \bibnamefont {Kang}}, \bibinfo {author} {\bibfnamefont {M.}~\bibnamefont {Avdeev}}, \bibinfo {author} {\bibfnamefont {R.}~\bibnamefont {Sibille}}, \bibinfo {author} {\bibfnamefont {K.}~\bibnamefont {Iida}}, \bibinfo {author} {\bibfnamefont {R.}~\bibnamefont {Kajimoto}}, \bibinfo {author} {\bibfnamefont {K.~H.}\ \bibnamefont {Lee}}, \emph {et~al.},\ }\bibfield  {title} {\bibinfo {title} {Tetrahedral triple-q magnetic ordering and large spontaneous {Hall} conductivity in the metallic triangular antiferromagnet {$\mathrm{Co_{1/3}TaS_{2}}$}},\ }\href@noop {} {\bibfield  {journal} {\bibinfo  {journal} {Nature Communications}\ }\textbf {\bibinfo {volume} {14}},\ \bibinfo {pages} {8346} (\bibinfo {year}
  {2023})}\BibitemShut {NoStop}%
\bibitem [{\citenamefont {Takagi}\ \emph {et~al.}(2023)\citenamefont {Takagi}, \citenamefont {Takagi}, \citenamefont {Minami}, \citenamefont {Nomoto}, \citenamefont {Ohishi}, \citenamefont {Suzuki}, \citenamefont {Yanagi}, \citenamefont {Hirayama}, \citenamefont {Khanh}, \citenamefont {Karube} \emph {et~al.}}]{CTS_tripleQ_nphys}%
  \BibitemOpen
  \bibfield  {author} {\bibinfo {author} {\bibfnamefont {H.}~\bibnamefont {Takagi}}, \bibinfo {author} {\bibfnamefont {R.}~\bibnamefont {Takagi}}, \bibinfo {author} {\bibfnamefont {S.}~\bibnamefont {Minami}}, \bibinfo {author} {\bibfnamefont {T.}~\bibnamefont {Nomoto}}, \bibinfo {author} {\bibfnamefont {K.}~\bibnamefont {Ohishi}}, \bibinfo {author} {\bibfnamefont {M.-T.}\ \bibnamefont {Suzuki}}, \bibinfo {author} {\bibfnamefont {Y.}~\bibnamefont {Yanagi}}, \bibinfo {author} {\bibfnamefont {M.}~\bibnamefont {Hirayama}}, \bibinfo {author} {\bibfnamefont {N.}~\bibnamefont {Khanh}}, \bibinfo {author} {\bibfnamefont {K.}~\bibnamefont {Karube}}, \emph {et~al.},\ }\bibfield  {title} {\bibinfo {title} {Spontaneous topological {Hall} effect induced by non-coplanar antiferromagnetic order in intercalated van der {Waals} materials},\ }\href@noop {} {\bibfield  {journal} {\bibinfo  {journal} {Nature Physics}\ }\textbf {\bibinfo {volume} {19}},\ \bibinfo {pages} {961} (\bibinfo {year} {2023})}\BibitemShut {NoStop}%
\bibitem [{\citenamefont {Dahlbom}\ \emph {et~al.}(2024)\citenamefont {Dahlbom}, \citenamefont {Brooks}, \citenamefont {Wilson}, \citenamefont {Chi}, \citenamefont {Kolesnikov}, \citenamefont {Stone}, \citenamefont {Cao}, \citenamefont {Li}, \citenamefont {Barros}, \citenamefont {Mourigal}, \citenamefont {Batista},\ and\ \citenamefont {Bai}}]{rescale_Dahlbom}%
  \BibitemOpen
  \bibfield  {author} {\bibinfo {author} {\bibfnamefont {D.}~\bibnamefont {Dahlbom}}, \bibinfo {author} {\bibfnamefont {F.~T.}\ \bibnamefont {Brooks}}, \bibinfo {author} {\bibfnamefont {M.~S.}\ \bibnamefont {Wilson}}, \bibinfo {author} {\bibfnamefont {S.}~\bibnamefont {Chi}}, \bibinfo {author} {\bibfnamefont {A.~I.}\ \bibnamefont {Kolesnikov}}, \bibinfo {author} {\bibfnamefont {M.~B.}\ \bibnamefont {Stone}}, \bibinfo {author} {\bibfnamefont {H.}~\bibnamefont {Cao}}, \bibinfo {author} {\bibfnamefont {Y.-W.}\ \bibnamefont {Li}}, \bibinfo {author} {\bibfnamefont {K.}~\bibnamefont {Barros}}, \bibinfo {author} {\bibfnamefont {M.}~\bibnamefont {Mourigal}}, \bibinfo {author} {\bibfnamefont {C.~D.}\ \bibnamefont {Batista}},\ and\ \bibinfo {author} {\bibfnamefont {X.}~\bibnamefont {Bai}},\ }\bibfield  {title} {\bibinfo {title} {Quantum-to-classical crossover in generalized spin systems: Temperature-dependent spin dynamics of {${\mathrm{FeI}}_{2}$}},\ }\href {https://doi.org/10.1103/PhysRevB.109.014427} {\bibfield
  {journal} {\bibinfo  {journal} {Phys. Rev. B}\ }\textbf {\bibinfo {volume} {109}},\ \bibinfo {pages} {014427} (\bibinfo {year} {2024})}\BibitemShut {NoStop}%
\bibitem [{\citenamefont {Kim}\ \emph {et~al.}(2023)\citenamefont {Kim}, \citenamefont {Kim}, \citenamefont {Park}, \citenamefont {Kim}, \citenamefont {Jeong}, \citenamefont {Ohira-Kawamura}, \citenamefont {Murai}, \citenamefont {Nakajima}, \citenamefont {Chernyshev}, \citenamefont {Mourigal} \emph {et~al.}}]{CoI2_nphys}%
  \BibitemOpen
  \bibfield  {author} {\bibinfo {author} {\bibfnamefont {C.}~\bibnamefont {Kim}}, \bibinfo {author} {\bibfnamefont {S.}~\bibnamefont {Kim}}, \bibinfo {author} {\bibfnamefont {P.}~\bibnamefont {Park}}, \bibinfo {author} {\bibfnamefont {T.}~\bibnamefont {Kim}}, \bibinfo {author} {\bibfnamefont {J.}~\bibnamefont {Jeong}}, \bibinfo {author} {\bibfnamefont {S.}~\bibnamefont {Ohira-Kawamura}}, \bibinfo {author} {\bibfnamefont {N.}~\bibnamefont {Murai}}, \bibinfo {author} {\bibfnamefont {K.}~\bibnamefont {Nakajima}}, \bibinfo {author} {\bibfnamefont {A.}~\bibnamefont {Chernyshev}}, \bibinfo {author} {\bibfnamefont {M.}~\bibnamefont {Mourigal}}, \emph {et~al.},\ }\bibfield  {title} {\bibinfo {title} {Bond-dependent anisotropy and magnon decay in cobalt-based kitaev triangular antiferromagnet},\ }\href@noop {} {\bibfield  {journal} {\bibinfo  {journal} {Nature Physics}\ }\textbf {\bibinfo {volume} {19}},\ \bibinfo {pages} {1624} (\bibinfo {year} {2023})}\BibitemShut {NoStop}%
\bibitem [{\citenamefont {Park}\ \emph {et~al.}(2024{\natexlab{a}})\citenamefont {Park}, \citenamefont {Ghioldi}, \citenamefont {May}, \citenamefont {Kolopus}, \citenamefont {Podlesnyak}, \citenamefont {Calder}, \citenamefont {Paddison}, \citenamefont {Trumper}, \citenamefont {Manuel}, \citenamefont {Batista} \emph {et~al.}}]{BLCTO}%
  \BibitemOpen
  \bibfield  {author} {\bibinfo {author} {\bibfnamefont {P.}~\bibnamefont {Park}}, \bibinfo {author} {\bibfnamefont {E.}~\bibnamefont {Ghioldi}}, \bibinfo {author} {\bibfnamefont {A.~F.}\ \bibnamefont {May}}, \bibinfo {author} {\bibfnamefont {J.~A.}\ \bibnamefont {Kolopus}}, \bibinfo {author} {\bibfnamefont {A.~A.}\ \bibnamefont {Podlesnyak}}, \bibinfo {author} {\bibfnamefont {S.}~\bibnamefont {Calder}}, \bibinfo {author} {\bibfnamefont {J.~A.}\ \bibnamefont {Paddison}}, \bibinfo {author} {\bibfnamefont {A.}~\bibnamefont {Trumper}}, \bibinfo {author} {\bibfnamefont {L.}~\bibnamefont {Manuel}}, \bibinfo {author} {\bibfnamefont {C.~D.}\ \bibnamefont {Batista}}, \emph {et~al.},\ }\bibfield  {title} {\bibinfo {title} {Anomalous continuum scattering and higher-order van {Hove} singularity in the strongly anisotropic {$S= 1/2$} triangular lattice antiferromagnet},\ }\href@noop {} {\bibfield  {journal} {\bibinfo  {journal} {Nature Communications}\ }\textbf {\bibinfo {volume} {15}},\ \bibinfo {pages} {7264} (\bibinfo
  {year} {2024}{\natexlab{a}})}\BibitemShut {NoStop}%
\bibitem [{\citenamefont {Park}\ \emph {et~al.}(2024{\natexlab{b}})\citenamefont {Park}, \citenamefont {Sala}, \citenamefont {Pajerowski}, \citenamefont {May}, \citenamefont {Kolopus}, \citenamefont {Dahlbom}, \citenamefont {Stone}, \citenamefont {Hal{\'a}sz},\ and\ \citenamefont {Christianson}}]{park_ZVPO}%
  \BibitemOpen
  \bibfield  {author} {\bibinfo {author} {\bibfnamefont {P.}~\bibnamefont {Park}}, \bibinfo {author} {\bibfnamefont {G.}~\bibnamefont {Sala}}, \bibinfo {author} {\bibfnamefont {D.~M.}\ \bibnamefont {Pajerowski}}, \bibinfo {author} {\bibfnamefont {A.~F.}\ \bibnamefont {May}}, \bibinfo {author} {\bibfnamefont {J.~A.}\ \bibnamefont {Kolopus}}, \bibinfo {author} {\bibfnamefont {D.}~\bibnamefont {Dahlbom}}, \bibinfo {author} {\bibfnamefont {M.~B.}\ \bibnamefont {Stone}}, \bibinfo {author} {\bibfnamefont {G.~B.}\ \bibnamefont {Hal{\'a}sz}},\ and\ \bibinfo {author} {\bibfnamefont {A.~D.}\ \bibnamefont {Christianson}},\ }\bibfield  {title} {\bibinfo {title} {Quantum and classical spin dynamics across temperature scales in the {$S=1/2$} {Heisenberg} antiferromagnet},\ }\href {https://doi.org/10.1103/PhysRevResearch.6.033184} {\bibfield  {journal} {\bibinfo  {journal} {Phys. Rev. Res.}\ }\textbf {\bibinfo {volume} {6}},\ \bibinfo {pages} {033184} (\bibinfo {year} {2024}{\natexlab{b}})}\BibitemShut {NoStop}%
\bibitem [{\citenamefont {Park}\ \emph {et~al.}(2022)\citenamefont {Park}, \citenamefont {Kang}, \citenamefont {Kim}, \citenamefont {Lee}, \citenamefont {Noh}, \citenamefont {Han},\ and\ \citenamefont {Park}}]{CTS_npj_2022}%
  \BibitemOpen
  \bibfield  {author} {\bibinfo {author} {\bibfnamefont {P.}~\bibnamefont {Park}}, \bibinfo {author} {\bibfnamefont {Y.-G.}\ \bibnamefont {Kang}}, \bibinfo {author} {\bibfnamefont {J.}~\bibnamefont {Kim}}, \bibinfo {author} {\bibfnamefont {K.~H.}\ \bibnamefont {Lee}}, \bibinfo {author} {\bibfnamefont {H.-J.}\ \bibnamefont {Noh}}, \bibinfo {author} {\bibfnamefont {M.~J.}\ \bibnamefont {Han}},\ and\ \bibinfo {author} {\bibfnamefont {J.-G.}\ \bibnamefont {Park}},\ }\bibfield  {title} {\bibinfo {title} {Field-tunable toroidal moment and anomalous {Hall} effect in noncollinear antiferromagnetic weyl semimetal {$\mathrm{Co_{1/3}TaS_{2}}$}},\ }\href@noop {} {\bibfield  {journal} {\bibinfo  {journal} {npj Quantum Materials}\ }\textbf {\bibinfo {volume} {7}},\ \bibinfo {pages} {42} (\bibinfo {year} {2022})}\BibitemShut {NoStop}%
\bibitem [{\citenamefont {Park}\ \emph {et~al.}(2024{\natexlab{c}})\citenamefont {Park}, \citenamefont {Cho}, \citenamefont {Kim}, \citenamefont {An}, \citenamefont {Avdeev}, \citenamefont {Iida}, \citenamefont {Kajimoto},\ and\ \citenamefont {Park}}]{CTS_composition_2024}%
  \BibitemOpen
  \bibfield  {author} {\bibinfo {author} {\bibfnamefont {P.}~\bibnamefont {Park}}, \bibinfo {author} {\bibfnamefont {W.}~\bibnamefont {Cho}}, \bibinfo {author} {\bibfnamefont {C.}~\bibnamefont {Kim}}, \bibinfo {author} {\bibfnamefont {Y.}~\bibnamefont {An}}, \bibinfo {author} {\bibfnamefont {M.}~\bibnamefont {Avdeev}}, \bibinfo {author} {\bibfnamefont {K.}~\bibnamefont {Iida}}, \bibinfo {author} {\bibfnamefont {R.}~\bibnamefont {Kajimoto}},\ and\ \bibinfo {author} {\bibfnamefont {J.-G.}\ \bibnamefont {Park}},\ }\bibfield  {title} {\bibinfo {title} {Composition dependence of bulk properties in the {Co}-intercalated transition metal dichalcogenide {$\mathrm{Co_{1/3}TaS_{2}}$}},\ }\href@noop {} {\bibfield  {journal} {\bibinfo  {journal} {Phys. Rev. B}\ }\textbf {\bibinfo {volume} {109}},\ \bibinfo {pages} {L060403} (\bibinfo {year} {2024}{\natexlab{c}})}\BibitemShut {NoStop}%
\bibitem [{\citenamefont {Kajimoto}\ \emph {et~al.}(2011)\citenamefont {Kajimoto}, \citenamefont {Nakamura}, \citenamefont {Inamura}, \citenamefont {Mizuno}, \citenamefont {Nakajima}, \citenamefont {Ohira-Kawamura}, \citenamefont {Yokoo}, \citenamefont {Nakatani}, \citenamefont {Maruyama}, \citenamefont {Soyama}, \citenamefont {Shibata}, \citenamefont {Suzuya}, \citenamefont {Sato}, \citenamefont {Aizawa}, \citenamefont {Arai}, \citenamefont {Wakimoto}, \citenamefont {Ishikado}, \citenamefont {Shamoto}, \citenamefont {Fujita}, \citenamefont {Hiraka}, \citenamefont {Ohoyama}, \citenamefont {Yamada},\ and\ \citenamefont {Lee}}]{4SEASONS}%
  \BibitemOpen
  \bibfield  {author} {\bibinfo {author} {\bibfnamefont {R.}~\bibnamefont {Kajimoto}}, \bibinfo {author} {\bibfnamefont {M.}~\bibnamefont {Nakamura}}, \bibinfo {author} {\bibfnamefont {Y.}~\bibnamefont {Inamura}}, \bibinfo {author} {\bibfnamefont {F.}~\bibnamefont {Mizuno}}, \bibinfo {author} {\bibfnamefont {K.}~\bibnamefont {Nakajima}}, \bibinfo {author} {\bibfnamefont {S.}~\bibnamefont {Ohira-Kawamura}}, \bibinfo {author} {\bibfnamefont {T.}~\bibnamefont {Yokoo}}, \bibinfo {author} {\bibfnamefont {T.}~\bibnamefont {Nakatani}}, \bibinfo {author} {\bibfnamefont {R.}~\bibnamefont {Maruyama}}, \bibinfo {author} {\bibfnamefont {K.}~\bibnamefont {Soyama}}, \bibinfo {author} {\bibfnamefont {K.}~\bibnamefont {Shibata}}, \bibinfo {author} {\bibfnamefont {K.}~\bibnamefont {Suzuya}}, \bibinfo {author} {\bibfnamefont {S.}~\bibnamefont {Sato}}, \bibinfo {author} {\bibfnamefont {K.}~\bibnamefont {Aizawa}}, \bibinfo {author} {\bibfnamefont {M.}~\bibnamefont {Arai}}, \bibinfo {author} {\bibfnamefont {S.}~\bibnamefont
  {Wakimoto}}, \bibinfo {author} {\bibfnamefont {M.}~\bibnamefont {Ishikado}}, \bibinfo {author} {\bibfnamefont {S.-i.}\ \bibnamefont {Shamoto}}, \bibinfo {author} {\bibfnamefont {M.}~\bibnamefont {Fujita}}, \bibinfo {author} {\bibfnamefont {H.}~\bibnamefont {Hiraka}}, \bibinfo {author} {\bibfnamefont {K.}~\bibnamefont {Ohoyama}}, \bibinfo {author} {\bibfnamefont {K.}~\bibnamefont {Yamada}},\ and\ \bibinfo {author} {\bibfnamefont {C.-H.}\ \bibnamefont {Lee}},\ }\bibfield  {title} {\bibinfo {title} {The {Fermi} chopper spectrometer {4SEASONS} at {J-PARC}},\ }\href {https://doi.org/10.1143/JPSJS.80SB.SB025} {\bibfield  {journal} {\bibinfo  {journal} {Journal of the Physical Society of Japan}\ }\textbf {\bibinfo {volume} {80}},\ \bibinfo {pages} {SB025} (\bibinfo {year} {2011})}\BibitemShut {NoStop}%
\bibitem [{\citenamefont {Nakamura}\ \emph {et~al.}(2009)\citenamefont {Nakamura}, \citenamefont {Kajimoto}, \citenamefont {Inamura}, \citenamefont {Mizuno}, \citenamefont {Fujita}, \citenamefont {Yokoo},\ and\ \citenamefont {Arai}}]{Jparc_RRM}%
  \BibitemOpen
  \bibfield  {author} {\bibinfo {author} {\bibfnamefont {M.}~\bibnamefont {Nakamura}}, \bibinfo {author} {\bibfnamefont {R.}~\bibnamefont {Kajimoto}}, \bibinfo {author} {\bibfnamefont {Y.}~\bibnamefont {Inamura}}, \bibinfo {author} {\bibfnamefont {F.}~\bibnamefont {Mizuno}}, \bibinfo {author} {\bibfnamefont {M.}~\bibnamefont {Fujita}}, \bibinfo {author} {\bibfnamefont {T.}~\bibnamefont {Yokoo}},\ and\ \bibinfo {author} {\bibfnamefont {M.}~\bibnamefont {Arai}},\ }\bibfield  {title} {\bibinfo {title} {First demonstration of novel method for inelastic neutron scattering measurement utilizing multiple incident energies},\ }\href@noop {} {\bibfield  {journal} {\bibinfo  {journal} {Journal of the Physical Society of Japan}\ }\textbf {\bibinfo {volume} {78}},\ \bibinfo {pages} {093002} (\bibinfo {year} {2009})}\BibitemShut {NoStop}%
\bibitem [{\citenamefont {Ewings}\ \emph {et~al.}(2016)\citenamefont {Ewings}, \citenamefont {Buts}, \citenamefont {Le}, \citenamefont {{van Duijn}}, \citenamefont {Bustinduy},\ and\ \citenamefont {Perring}}]{Horace}%
  \BibitemOpen
  \bibfield  {author} {\bibinfo {author} {\bibfnamefont {R.}~\bibnamefont {Ewings}}, \bibinfo {author} {\bibfnamefont {A.}~\bibnamefont {Buts}}, \bibinfo {author} {\bibfnamefont {M.}~\bibnamefont {Le}}, \bibinfo {author} {\bibfnamefont {J.}~\bibnamefont {{van Duijn}}}, \bibinfo {author} {\bibfnamefont {I.}~\bibnamefont {Bustinduy}},\ and\ \bibinfo {author} {\bibfnamefont {T.}~\bibnamefont {Perring}},\ }\bibfield  {title} {\bibinfo {title} {Horace: Software for the analysis of data from single crystal spectroscopy experiments at time-of-flight neutron instruments},\ }\href {https://doi.org/https://doi.org/10.1016/j.nima.2016.07.036} {\bibfield  {journal} {\bibinfo  {journal} {Nuclear Instruments and Methods in Physics Research Section A: Accelerators, Spectrometers, Detectors and Associated Equipment}\ }\textbf {\bibinfo {volume} {834}},\ \bibinfo {pages} {132} (\bibinfo {year} {2016})}\BibitemShut {NoStop}%
\bibitem [{\citenamefont {Inamura}\ \emph {et~al.}(2013)\citenamefont {Inamura}, \citenamefont {Nakatani}, \citenamefont {Suzuki},\ and\ \citenamefont {Otomo}}]{Utsusemi}%
  \BibitemOpen
  \bibfield  {author} {\bibinfo {author} {\bibfnamefont {Y.}~\bibnamefont {Inamura}}, \bibinfo {author} {\bibfnamefont {T.}~\bibnamefont {Nakatani}}, \bibinfo {author} {\bibfnamefont {J.}~\bibnamefont {Suzuki}},\ and\ \bibinfo {author} {\bibfnamefont {T.}~\bibnamefont {Otomo}},\ }\bibfield  {title} {\bibinfo {title} {Development status of software {“Utsusemi”} for chopper spectrometers at {MLF}, {J-PARC}},\ }\href {https://doi.org/10.7566/JPSJS.82SA.SA031} {\bibfield  {journal} {\bibinfo  {journal} {Journal of the Physical Society of Japan}\ }\textbf {\bibinfo {volume} {82}},\ \bibinfo {pages} {SA031} (\bibinfo {year} {2013})}\BibitemShut {NoStop}%
\bibitem [{\citenamefont {Toth}\ and\ \citenamefont {Lake}(2015)}]{SpinW}%
  \BibitemOpen
  \bibfield  {author} {\bibinfo {author} {\bibfnamefont {S.}~\bibnamefont {Toth}}\ and\ \bibinfo {author} {\bibfnamefont {B.}~\bibnamefont {Lake}},\ }\bibfield  {title} {\bibinfo {title} {Linear spin wave theory for single-q incommensurate magnetic structures},\ }\href@noop {} {\bibfield  {journal} {\bibinfo  {journal} {Journal of Physics: Condensed Matter}\ }\textbf {\bibinfo {volume} {27}},\ \bibinfo {pages} {166002} (\bibinfo {year} {2015})}\BibitemShut {NoStop}%
\bibitem [{Sun()}]{Sunny}%
  \BibitemOpen
  \href {https://github.com/SunnySuite/Sunny.jl} {\bibinfo {title} {Su(n)ny, spin dynamics and generalization to {SU(N)} coherent states, https://github.com/sunnysuite/sunny.jl.}}\BibitemShut {Stop}%
\bibitem [{\citenamefont {Dahlbom}\ \emph {et~al.}(2022)\citenamefont {Dahlbom}, \citenamefont {Zhang}, \citenamefont {Miles}, \citenamefont {Bai}, \citenamefont {Batista},\ and\ \citenamefont {Barros}}]{Sunny_ref1}%
  \BibitemOpen
  \bibfield  {author} {\bibinfo {author} {\bibfnamefont {D.}~\bibnamefont {Dahlbom}}, \bibinfo {author} {\bibfnamefont {H.}~\bibnamefont {Zhang}}, \bibinfo {author} {\bibfnamefont {C.}~\bibnamefont {Miles}}, \bibinfo {author} {\bibfnamefont {X.}~\bibnamefont {Bai}}, \bibinfo {author} {\bibfnamefont {C.~D.}\ \bibnamefont {Batista}},\ and\ \bibinfo {author} {\bibfnamefont {K.}~\bibnamefont {Barros}},\ }\bibfield  {title} {\bibinfo {title} {Geometric integration of classical spin dynamics via a mean-field {Schr\"odinger} equation},\ }\href {https://doi.org/10.1103/PhysRevB.106.054423} {\bibfield  {journal} {\bibinfo  {journal} {Phys. Rev. B}\ }\textbf {\bibinfo {volume} {106}},\ \bibinfo {pages} {054423} (\bibinfo {year} {2022})}\BibitemShut {NoStop}%
\bibitem [{\citenamefont {Dahlbom}\ \emph {et~al.}(2023)\citenamefont {Dahlbom}, \citenamefont {Zhang}, \citenamefont {Laraib}, \citenamefont {Pajerowski}, \citenamefont {Barros},\ and\ \citenamefont {Batista}}]{dahlbom_renormalized}%
  \BibitemOpen
  \bibfield  {author} {\bibinfo {author} {\bibfnamefont {D.}~\bibnamefont {Dahlbom}}, \bibinfo {author} {\bibfnamefont {H.}~\bibnamefont {Zhang}}, \bibinfo {author} {\bibfnamefont {Z.}~\bibnamefont {Laraib}}, \bibinfo {author} {\bibfnamefont {D.~M.}\ \bibnamefont {Pajerowski}}, \bibinfo {author} {\bibfnamefont {K.}~\bibnamefont {Barros}},\ and\ \bibinfo {author} {\bibfnamefont {C.}~\bibnamefont {Batista}},\ }\bibfield  {title} {\bibinfo {title} {Renormalized classical theory of quantum magnets},\ }\href@noop {} {\bibfield  {journal} {\bibinfo  {journal} {arXiv preprint arXiv:2304.03874}\ } (\bibinfo {year} {2023})}\BibitemShut {NoStop}%
\bibitem [{\citenamefont {Parkin}\ and\ \citenamefont {Friend}(1980{\natexlab{a}})}]{Parkin_80_v1}%
  \BibitemOpen
  \bibfield  {author} {\bibinfo {author} {\bibfnamefont {S.~S.~P.}\ \bibnamefont {Parkin}}\ and\ \bibinfo {author} {\bibfnamefont {R.~H.}\ \bibnamefont {Friend}},\ }\bibfield  {title} {\bibinfo {title} {3d transition-metal intercalates of the niobium and tantalum dichalcogenides. i. magnetic properties},\ }\href {https://doi.org/10.1080/13642818008245370} {\bibfield  {journal} {\bibinfo  {journal} {Philosophical Magazine B}\ }\textbf {\bibinfo {volume} {41}},\ \bibinfo {pages} {65} (\bibinfo {year} {1980}{\natexlab{a}})}\BibitemShut {NoStop}%
\bibitem [{\citenamefont {Martin}\ and\ \citenamefont {Batista}(2008)}]{Batista_3q_08}%
  \BibitemOpen
  \bibfield  {author} {\bibinfo {author} {\bibfnamefont {I.}~\bibnamefont {Martin}}\ and\ \bibinfo {author} {\bibfnamefont {C.~D.}\ \bibnamefont {Batista}},\ }\bibfield  {title} {\bibinfo {title} {Itinerant electron-driven chiral magnetic ordering and spontaneous quantum {Hall} effect in triangular lattice models},\ }\href {https://doi.org/10.1103/PhysRevLett.101.156402} {\bibfield  {journal} {\bibinfo  {journal} {Phys. Rev. Lett.}\ }\textbf {\bibinfo {volume} {101}},\ \bibinfo {pages} {156402} (\bibinfo {year} {2008})}\BibitemShut {NoStop}%
\bibitem [{\citenamefont {Huang}\ and\ \citenamefont {Chien}(2012)}]{FeGe_THE}%
  \BibitemOpen
  \bibfield  {author} {\bibinfo {author} {\bibfnamefont {S.}~\bibnamefont {Huang}}\ and\ \bibinfo {author} {\bibfnamefont {C.}~\bibnamefont {Chien}},\ }\bibfield  {title} {\bibinfo {title} {Extended skyrmion phase in epitaxial {FeGe} (111) thin films},\ }\href@noop {} {\bibfield  {journal} {\bibinfo  {journal} {Phys. Rev. Lett.}\ }\textbf {\bibinfo {volume} {108}},\ \bibinfo {pages} {267201} (\bibinfo {year} {2012})}\BibitemShut {NoStop}%
\bibitem [{\citenamefont {Neubauer}\ \emph {et~al.}(2009)\citenamefont {Neubauer}, \citenamefont {Pfleiderer}, \citenamefont {Binz}, \citenamefont {Rosch}, \citenamefont {Ritz}, \citenamefont {Niklowitz},\ and\ \citenamefont {B{\"o}ni}}]{MnSi_THE}%
  \BibitemOpen
  \bibfield  {author} {\bibinfo {author} {\bibfnamefont {A.}~\bibnamefont {Neubauer}}, \bibinfo {author} {\bibfnamefont {C.}~\bibnamefont {Pfleiderer}}, \bibinfo {author} {\bibfnamefont {B.}~\bibnamefont {Binz}}, \bibinfo {author} {\bibfnamefont {A.}~\bibnamefont {Rosch}}, \bibinfo {author} {\bibfnamefont {R.}~\bibnamefont {Ritz}}, \bibinfo {author} {\bibfnamefont {P.}~\bibnamefont {Niklowitz}},\ and\ \bibinfo {author} {\bibfnamefont {P.}~\bibnamefont {B{\"o}ni}},\ }\bibfield  {title} {\bibinfo {title} {Topological {Hall} effect in the {A} phase of {MnSi}},\ }\href@noop {} {\bibfield  {journal} {\bibinfo  {journal} {Phys. Rev. Lett.}\ }\textbf {\bibinfo {volume} {102}},\ \bibinfo {pages} {186602} (\bibinfo {year} {2009})}\BibitemShut {NoStop}%
\bibitem [{\citenamefont {Villain}\ \emph {et~al.}(1980)\citenamefont {Villain}, \citenamefont {Bidaux}, \citenamefont {Carton},\ and\ \citenamefont {Conte}}]{od_by_disod1}%
  \BibitemOpen
  \bibfield  {author} {\bibinfo {author} {\bibfnamefont {J.}~\bibnamefont {Villain}}, \bibinfo {author} {\bibfnamefont {R.}~\bibnamefont {Bidaux}}, \bibinfo {author} {\bibfnamefont {J.-P.}\ \bibnamefont {Carton}},\ and\ \bibinfo {author} {\bibfnamefont {R.}~\bibnamefont {Conte}},\ }\bibfield  {title} {\bibinfo {title} {Order as an effect of disorder},\ }\href@noop {} {\bibfield  {journal} {\bibinfo  {journal} {Journal de Physique}\ }\textbf {\bibinfo {volume} {41}},\ \bibinfo {pages} {1263} (\bibinfo {year} {1980})}\BibitemShut {NoStop}%
\bibitem [{\citenamefont {Henley}(1989)}]{od_by_disod2}%
  \BibitemOpen
  \bibfield  {author} {\bibinfo {author} {\bibfnamefont {C.~L.}\ \bibnamefont {Henley}},\ }\bibfield  {title} {\bibinfo {title} {Ordering due to disorder in a frustrated vector antiferromagnet},\ }\href@noop {} {\bibfield  {journal} {\bibinfo  {journal} {Phys. Rev. Lett.}\ }\textbf {\bibinfo {volume} {62}},\ \bibinfo {pages} {2056} (\bibinfo {year} {1989})}\BibitemShut {NoStop}%
\bibitem [{\citenamefont {Sharma}\ \emph {et~al.}(2023)\citenamefont {Sharma}, \citenamefont {Wang},\ and\ \citenamefont {Batista}}]{4spin_general}%
  \BibitemOpen
  \bibfield  {author} {\bibinfo {author} {\bibfnamefont {V.}~\bibnamefont {Sharma}}, \bibinfo {author} {\bibfnamefont {Z.}~\bibnamefont {Wang}},\ and\ \bibinfo {author} {\bibfnamefont {C.~D.}\ \bibnamefont {Batista}},\ }\bibfield  {title} {\bibinfo {title} {Machine learning assisted derivation of minimal low-energy models for metallic magnets},\ }\href@noop {} {\bibfield  {journal} {\bibinfo  {journal} {npj Computational Materials}\ }\textbf {\bibinfo {volume} {9}},\ \bibinfo {pages} {192} (\bibinfo {year} {2023})}\BibitemShut {NoStop}%
\bibitem [{\citenamefont {Diallo}\ \emph {et~al.}(2009)\citenamefont {Diallo}, \citenamefont {Antropov}, \citenamefont {Perring}, \citenamefont {Broholm}, \citenamefont {Pulikkotil}, \citenamefont {Ni}, \citenamefont {Bud’ko}, \citenamefont {Canfield}, \citenamefont {Kreyssig}, \citenamefont {Goldman} \emph {et~al.}}]{AFM_metal_1}%
  \BibitemOpen
  \bibfield  {author} {\bibinfo {author} {\bibfnamefont {S.}~\bibnamefont {Diallo}}, \bibinfo {author} {\bibfnamefont {V.}~\bibnamefont {Antropov}}, \bibinfo {author} {\bibfnamefont {T.}~\bibnamefont {Perring}}, \bibinfo {author} {\bibfnamefont {C.}~\bibnamefont {Broholm}}, \bibinfo {author} {\bibfnamefont {J.}~\bibnamefont {Pulikkotil}}, \bibinfo {author} {\bibfnamefont {N.}~\bibnamefont {Ni}}, \bibinfo {author} {\bibfnamefont {S.}~\bibnamefont {Bud’ko}}, \bibinfo {author} {\bibfnamefont {P.}~\bibnamefont {Canfield}}, \bibinfo {author} {\bibfnamefont {A.}~\bibnamefont {Kreyssig}}, \bibinfo {author} {\bibfnamefont {A.}~\bibnamefont {Goldman}}, \emph {et~al.},\ }\bibfield  {title} {\bibinfo {title} {Itinerant magnetic excitations in antiferromagnetic {CaFe$_{2}$As$_{2}$}},\ }\href@noop {} {\bibfield  {journal} {\bibinfo  {journal} {Phys. Rev. Lett.}\ }\textbf {\bibinfo {volume} {102}},\ \bibinfo {pages} {187206} (\bibinfo {year} {2009})}\BibitemShut {NoStop}%
\bibitem [{\citenamefont {Ibuka}\ \emph {et~al.}(2017)\citenamefont {Ibuka}, \citenamefont {Itoh}, \citenamefont {Yokoo},\ and\ \citenamefont {Endoh}}]{AFM_metal_2}%
  \BibitemOpen
  \bibfield  {author} {\bibinfo {author} {\bibfnamefont {S.}~\bibnamefont {Ibuka}}, \bibinfo {author} {\bibfnamefont {S.}~\bibnamefont {Itoh}}, \bibinfo {author} {\bibfnamefont {T.}~\bibnamefont {Yokoo}},\ and\ \bibinfo {author} {\bibfnamefont {Y.}~\bibnamefont {Endoh}},\ }\bibfield  {title} {\bibinfo {title} {Damped spin-wave excitations in the itinerant antiferromagnet $\gamma$-{Fe$_{0.7}$Mn$_{0.3}$}},\ }\href@noop {} {\bibfield  {journal} {\bibinfo  {journal} {Phys. Rev. B}\ }\textbf {\bibinfo {volume} {95}},\ \bibinfo {pages} {224406} (\bibinfo {year} {2017})}\BibitemShut {NoStop}%
\bibitem [{\citenamefont {Adams}\ \emph {et~al.}(2000)\citenamefont {Adams}, \citenamefont {Mason}, \citenamefont {Fawcett}, \citenamefont {Menshikov}, \citenamefont {Frost}, \citenamefont {Forsyth}, \citenamefont {Perring},\ and\ \citenamefont {Holden}}]{AFM_metal_3}%
  \BibitemOpen
  \bibfield  {author} {\bibinfo {author} {\bibfnamefont {C.}~\bibnamefont {Adams}}, \bibinfo {author} {\bibfnamefont {T.}~\bibnamefont {Mason}}, \bibinfo {author} {\bibfnamefont {E.}~\bibnamefont {Fawcett}}, \bibinfo {author} {\bibfnamefont {A.}~\bibnamefont {Menshikov}}, \bibinfo {author} {\bibfnamefont {C.}~\bibnamefont {Frost}}, \bibinfo {author} {\bibfnamefont {J.}~\bibnamefont {Forsyth}}, \bibinfo {author} {\bibfnamefont {T.}~\bibnamefont {Perring}},\ and\ \bibinfo {author} {\bibfnamefont {T.}~\bibnamefont {Holden}},\ }\bibfield  {title} {\bibinfo {title} {High-energy magnetic excitations and anomalous spin-wave damping in {FeGe$_{2}$}},\ }\href@noop {} {\bibfield  {journal} {\bibinfo  {journal} {Journal of Physics: Condensed Matter}\ }\textbf {\bibinfo {volume} {12}},\ \bibinfo {pages} {8487} (\bibinfo {year} {2000})}\BibitemShut {NoStop}%
\bibitem [{\citenamefont {Zhao}\ \emph {et~al.}(2009)\citenamefont {Zhao}, \citenamefont {Adroja}, \citenamefont {Yao}, \citenamefont {Bewley}, \citenamefont {Li}, \citenamefont {Wang}, \citenamefont {Wu}, \citenamefont {Chen}, \citenamefont {Hu},\ and\ \citenamefont {Dai}}]{CaFe2As2_nphys}%
  \BibitemOpen
  \bibfield  {author} {\bibinfo {author} {\bibfnamefont {J.}~\bibnamefont {Zhao}}, \bibinfo {author} {\bibfnamefont {D.}~\bibnamefont {Adroja}}, \bibinfo {author} {\bibfnamefont {D.-X.}\ \bibnamefont {Yao}}, \bibinfo {author} {\bibfnamefont {R.}~\bibnamefont {Bewley}}, \bibinfo {author} {\bibfnamefont {S.}~\bibnamefont {Li}}, \bibinfo {author} {\bibfnamefont {X.}~\bibnamefont {Wang}}, \bibinfo {author} {\bibfnamefont {G.}~\bibnamefont {Wu}}, \bibinfo {author} {\bibfnamefont {X.}~\bibnamefont {Chen}}, \bibinfo {author} {\bibfnamefont {J.}~\bibnamefont {Hu}},\ and\ \bibinfo {author} {\bibfnamefont {P.}~\bibnamefont {Dai}},\ }\bibfield  {title} {\bibinfo {title} {Spin waves and magnetic exchange interactions in {CaFe$_{2}$As$_{2}$}},\ }\href@noop {} {\bibfield  {journal} {\bibinfo  {journal} {Nature Physics}\ }\textbf {\bibinfo {volume} {5}},\ \bibinfo {pages} {555} (\bibinfo {year} {2009})}\BibitemShut {NoStop}%
\bibitem [{\citenamefont {Do}\ \emph {et~al.}(2022)\citenamefont {Do}, \citenamefont {Kaneko}, \citenamefont {Kajimoto}, \citenamefont {Kamazawa}, \citenamefont {Stone}, \citenamefont {Lin}, \citenamefont {Itoh}, \citenamefont {Masuda}, \citenamefont {Samolyuk}, \citenamefont {Dagotto}, \citenamefont {Meier}, \citenamefont {Sales}, \citenamefont {Miao},\ and\ \citenamefont {Christianson}}]{FeSn_INS}%
  \BibitemOpen
  \bibfield  {author} {\bibinfo {author} {\bibfnamefont {S.-H.}\ \bibnamefont {Do}}, \bibinfo {author} {\bibfnamefont {K.}~\bibnamefont {Kaneko}}, \bibinfo {author} {\bibfnamefont {R.}~\bibnamefont {Kajimoto}}, \bibinfo {author} {\bibfnamefont {K.}~\bibnamefont {Kamazawa}}, \bibinfo {author} {\bibfnamefont {M.~B.}\ \bibnamefont {Stone}}, \bibinfo {author} {\bibfnamefont {J.~Y.~Y.}\ \bibnamefont {Lin}}, \bibinfo {author} {\bibfnamefont {S.}~\bibnamefont {Itoh}}, \bibinfo {author} {\bibfnamefont {T.}~\bibnamefont {Masuda}}, \bibinfo {author} {\bibfnamefont {G.~D.}\ \bibnamefont {Samolyuk}}, \bibinfo {author} {\bibfnamefont {E.}~\bibnamefont {Dagotto}}, \bibinfo {author} {\bibfnamefont {W.~R.}\ \bibnamefont {Meier}}, \bibinfo {author} {\bibfnamefont {B.~C.}\ \bibnamefont {Sales}}, \bibinfo {author} {\bibfnamefont {H.}~\bibnamefont {Miao}},\ and\ \bibinfo {author} {\bibfnamefont {A.~D.}\ \bibnamefont {Christianson}},\ }\bibfield  {title} {\bibinfo {title} {Damped dirac magnon in the metallic kagome
  antiferromagnet {$\mathrm{FeSn}$}},\ }\href {https://doi.org/10.1103/PhysRevB.105.L180403} {\bibfield  {journal} {\bibinfo  {journal} {Phys. Rev. B}\ }\textbf {\bibinfo {volume} {105}},\ \bibinfo {pages} {L180403} (\bibinfo {year} {2022})}\BibitemShut {NoStop}%
\bibitem [{\citenamefont {Park}\ \emph {et~al.}(2020)\citenamefont {Park}, \citenamefont {Park}, \citenamefont {Kim}, \citenamefont {Kousaka}, \citenamefont {Lee}, \citenamefont {Perring}, \citenamefont {Jeong}, \citenamefont {Stuhr}, \citenamefont {Akimitsu}, \citenamefont {Kenzelmann},\ and\ \citenamefont {Park}}]{CrB2}%
  \BibitemOpen
  \bibfield  {author} {\bibinfo {author} {\bibfnamefont {P.}~\bibnamefont {Park}}, \bibinfo {author} {\bibfnamefont {K.}~\bibnamefont {Park}}, \bibinfo {author} {\bibfnamefont {T.}~\bibnamefont {Kim}}, \bibinfo {author} {\bibfnamefont {Y.}~\bibnamefont {Kousaka}}, \bibinfo {author} {\bibfnamefont {K.~H.}\ \bibnamefont {Lee}}, \bibinfo {author} {\bibfnamefont {T.~G.}\ \bibnamefont {Perring}}, \bibinfo {author} {\bibfnamefont {J.}~\bibnamefont {Jeong}}, \bibinfo {author} {\bibfnamefont {U.}~\bibnamefont {Stuhr}}, \bibinfo {author} {\bibfnamefont {J.}~\bibnamefont {Akimitsu}}, \bibinfo {author} {\bibfnamefont {M.}~\bibnamefont {Kenzelmann}},\ and\ \bibinfo {author} {\bibfnamefont {J.-G.}\ \bibnamefont {Park}},\ }\bibfield  {title} {\bibinfo {title} {Momentum-dependent magnon lifetime in the metallic noncollinear triangular antiferromagnet {${\mathrm{CrB}}_{2}$}},\ }\href {https://doi.org/10.1103/PhysRevLett.125.027202} {\bibfield  {journal} {\bibinfo  {journal} {Phys. Rev. Lett.}\ }\textbf {\bibinfo {volume}
  {125}},\ \bibinfo {pages} {027202} (\bibinfo {year} {2020})}\BibitemShut {NoStop}%
\bibitem [{\citenamefont {Moriya}(2012)}]{moriya}%
  \BibitemOpen
  \bibfield  {author} {\bibinfo {author} {\bibfnamefont {T.}~\bibnamefont {Moriya}},\ }\href@noop {} {\emph {\bibinfo {title} {Spin fluctuations in itinerant electron magnetism}}},\ Vol.~\bibinfo {volume} {56}\ (\bibinfo  {publisher} {Springer Science \& Business Media},\ \bibinfo {year} {2012})\BibitemShut {NoStop}%
\bibitem [{\citenamefont {Chernyshev}\ and\ \citenamefont {Zhitomirsky}(2006)}]{Noncollinear_SWT1}%
  \BibitemOpen
  \bibfield  {author} {\bibinfo {author} {\bibfnamefont {A.}~\bibnamefont {Chernyshev}}\ and\ \bibinfo {author} {\bibfnamefont {M.}~\bibnamefont {Zhitomirsky}},\ }\bibfield  {title} {\bibinfo {title} {Magnon decay in noncollinear quantum antiferromagnets},\ }\href@noop {} {\bibfield  {journal} {\bibinfo  {journal} {Phys. Rev. Lett.}\ }\textbf {\bibinfo {volume} {97}},\ \bibinfo {pages} {207202} (\bibinfo {year} {2006})}\BibitemShut {NoStop}%
\bibitem [{\citenamefont {Chernyshev}\ and\ \citenamefont {Zhitomirsky}(2009)}]{Noncollinear_SWT2}%
  \BibitemOpen
  \bibfield  {author} {\bibinfo {author} {\bibfnamefont {A.~L.}\ \bibnamefont {Chernyshev}}\ and\ \bibinfo {author} {\bibfnamefont {M.~E.}\ \bibnamefont {Zhitomirsky}},\ }\bibfield  {title} {\bibinfo {title} {Spin waves in a triangular lattice antiferromagnet: Decays, spectrum renormalization, and singularities},\ }\href {https://doi.org/10.1103/PhysRevB.79.144416} {\bibfield  {journal} {\bibinfo  {journal} {Phys. Rev. B}\ }\textbf {\bibinfo {volume} {79}},\ \bibinfo {pages} {144416} (\bibinfo {year} {2009})}\BibitemShut {NoStop}%
\bibitem [{\citenamefont {Zhitomirsky}\ and\ \citenamefont {Chernyshev}(2013)}]{RMP_Mdecay}%
  \BibitemOpen
  \bibfield  {author} {\bibinfo {author} {\bibfnamefont {M.~E.}\ \bibnamefont {Zhitomirsky}}\ and\ \bibinfo {author} {\bibfnamefont {A.~L.}\ \bibnamefont {Chernyshev}},\ }\bibfield  {title} {\bibinfo {title} {Colloquium: Spontaneous magnon decays},\ }\href {https://doi.org/10.1103/RevModPhys.85.219} {\bibfield  {journal} {\bibinfo  {journal} {Rev. Mod. Phys.}\ }\textbf {\bibinfo {volume} {85}},\ \bibinfo {pages} {219} (\bibinfo {year} {2013})}\BibitemShut {NoStop}%
\bibitem [{\citenamefont {Batista}\ \emph {et~al.}(2016{\natexlab{b}})\citenamefont {Batista}, \citenamefont {Lin}, \citenamefont {Hayami},\ and\ \citenamefont {Kamiya}}]{Batista_2016_review}%
  \BibitemOpen
  \bibfield  {author} {\bibinfo {author} {\bibfnamefont {C.~D.}\ \bibnamefont {Batista}}, \bibinfo {author} {\bibfnamefont {S.-Z.}\ \bibnamefont {Lin}}, \bibinfo {author} {\bibfnamefont {S.}~\bibnamefont {Hayami}},\ and\ \bibinfo {author} {\bibfnamefont {Y.}~\bibnamefont {Kamiya}},\ }\bibfield  {title} {\bibinfo {title} {Frustration and chiral orderings in correlated electron systems},\ }\href@noop {} {\bibfield  {journal} {\bibinfo  {journal} {Reports on Progress in Physics}\ }\textbf {\bibinfo {volume} {79}},\ \bibinfo {pages} {084504} (\bibinfo {year} {2016}{\natexlab{b}})}\BibitemShut {NoStop}%
\bibitem [{\citenamefont {Mourigal}\ \emph {et~al.}(2013)\citenamefont {Mourigal}, \citenamefont {Fuhrman}, \citenamefont {Chernyshev},\ and\ \citenamefont {Zhitomirsky}}]{TLAF_NLSWT}%
  \BibitemOpen
  \bibfield  {author} {\bibinfo {author} {\bibfnamefont {M.}~\bibnamefont {Mourigal}}, \bibinfo {author} {\bibfnamefont {W.}~\bibnamefont {Fuhrman}}, \bibinfo {author} {\bibfnamefont {A.}~\bibnamefont {Chernyshev}},\ and\ \bibinfo {author} {\bibfnamefont {M.}~\bibnamefont {Zhitomirsky}},\ }\bibfield  {title} {\bibinfo {title} {Dynamical structure factor of the triangular-lattice antiferromagnet},\ }\href@noop {} {\bibfield  {journal} {\bibinfo  {journal} {Phys. Rev. B—Condensed Matter and Materials Physics}\ }\textbf {\bibinfo {volume} {88}},\ \bibinfo {pages} {094407} (\bibinfo {year} {2013})}\BibitemShut {NoStop}%
\bibitem [{\citenamefont {Luo}\ \emph {et~al.}(2020)\citenamefont {Luo}, \citenamefont {Marcus}, \citenamefont {Trump}, \citenamefont {Kindervater}, \citenamefont {Stone}, \citenamefont {Rodriguez-Rivera}, \citenamefont {Qiu}, \citenamefont {McQueen}, \citenamefont {Tchernyshyov},\ and\ \citenamefont {Broholm}}]{tmdos_ref2}%
  \BibitemOpen
  \bibfield  {author} {\bibinfo {author} {\bibfnamefont {Y.}~\bibnamefont {Luo}}, \bibinfo {author} {\bibfnamefont {G.}~\bibnamefont {Marcus}}, \bibinfo {author} {\bibfnamefont {B.}~\bibnamefont {Trump}}, \bibinfo {author} {\bibfnamefont {J.}~\bibnamefont {Kindervater}}, \bibinfo {author} {\bibfnamefont {M.}~\bibnamefont {Stone}}, \bibinfo {author} {\bibfnamefont {J.}~\bibnamefont {Rodriguez-Rivera}}, \bibinfo {author} {\bibfnamefont {Y.}~\bibnamefont {Qiu}}, \bibinfo {author} {\bibfnamefont {T.}~\bibnamefont {McQueen}}, \bibinfo {author} {\bibfnamefont {O.}~\bibnamefont {Tchernyshyov}},\ and\ \bibinfo {author} {\bibfnamefont {C.}~\bibnamefont {Broholm}},\ }\bibfield  {title} {\bibinfo {title} {Low-energy magnons in the chiral ferrimagnet {$\mathrm{Cu_{2}OSeO_{3}}$}: A coarse-grained approach},\ }\href@noop {} {\bibfield  {journal} {\bibinfo  {journal} {Phys. Rev. B}\ }\textbf {\bibinfo {volume} {101}},\ \bibinfo {pages} {144411} (\bibinfo {year} {2020})}\BibitemShut {NoStop}%
\bibitem [{\citenamefont {Chubukov}\ \emph {et~al.}(1994)\citenamefont {Chubukov}, \citenamefont {Sachdev},\ and\ \citenamefont {Senthil}}]{Chubukov1994}%
  \BibitemOpen
  \bibfield  {author} {\bibinfo {author} {\bibfnamefont {A.}~\bibnamefont {Chubukov}}, \bibinfo {author} {\bibfnamefont {S.}~\bibnamefont {Sachdev}},\ and\ \bibinfo {author} {\bibfnamefont {T.}~\bibnamefont {Senthil}},\ }\bibfield  {title} {\bibinfo {title} {Large-{$S$} expansion for quantum antiferromagnets on a triangular lattice},\ }\href@noop {} {\bibfield  {journal} {\bibinfo  {journal} {Journal of Physics: Condensed Matter}\ }\textbf {\bibinfo {volume} {6}},\ \bibinfo {pages} {8891} (\bibinfo {year} {1994})}\BibitemShut {NoStop}%
\bibitem [{\citenamefont {Parkin}\ \emph {et~al.}(1983)\citenamefont {Parkin}, \citenamefont {Marseglia},\ and\ \citenamefont {Brown}}]{Parkin_1983_magstr}%
  \BibitemOpen
  \bibfield  {author} {\bibinfo {author} {\bibfnamefont {S.~S.~P.}\ \bibnamefont {Parkin}}, \bibinfo {author} {\bibfnamefont {E.~A.}\ \bibnamefont {Marseglia}},\ and\ \bibinfo {author} {\bibfnamefont {P.~J.}\ \bibnamefont {Brown}},\ }\bibfield  {title} {\bibinfo {title} {Magnetic structure of {$\mathrm{Co_{1/3}NbS_{2}}$} and {$\mathrm{Co_{1/3}TaS_{2}}$}},\ }\href {https://doi.org/10.1088/0022-3719/16/14/016} {\bibfield  {journal} {\bibinfo  {journal} {Journal of Physics C: Solid State Physics}\ }\textbf {\bibinfo {volume} {16}},\ \bibinfo {pages} {2765} (\bibinfo {year} {1983})}\BibitemShut {NoStop}%
\bibitem [{\citenamefont {Ghimire}\ \emph {et~al.}(2018)\citenamefont {Ghimire}, \citenamefont {Botana}, \citenamefont {Jiang}, \citenamefont {Zhang}, \citenamefont {Chen},\ and\ \citenamefont {Mitchell}}]{CNS_ncomm}%
  \BibitemOpen
  \bibfield  {author} {\bibinfo {author} {\bibfnamefont {N.~J.}\ \bibnamefont {Ghimire}}, \bibinfo {author} {\bibfnamefont {A.}~\bibnamefont {Botana}}, \bibinfo {author} {\bibfnamefont {J.}~\bibnamefont {Jiang}}, \bibinfo {author} {\bibfnamefont {J.}~\bibnamefont {Zhang}}, \bibinfo {author} {\bibfnamefont {Y.-S.}\ \bibnamefont {Chen}},\ and\ \bibinfo {author} {\bibfnamefont {J.}~\bibnamefont {Mitchell}},\ }\bibfield  {title} {\bibinfo {title} {Large anomalous {Hall} effect in the chiral-lattice antiferromagnet {CoNb$_{3}$S$_{6}$}},\ }\href@noop {} {\bibfield  {journal} {\bibinfo  {journal} {Nature communications}\ }\textbf {\bibinfo {volume} {9}},\ \bibinfo {pages} {3280} (\bibinfo {year} {2018})}\BibitemShut {NoStop}%
\bibitem [{\citenamefont {Zager}\ \emph {et~al.}(2023)\citenamefont {Zager}, \citenamefont {Fan}, \citenamefont {Steadman},\ and\ \citenamefont {Plumb}}]{CNS_RXS}%
  \BibitemOpen
  \bibfield  {author} {\bibinfo {author} {\bibfnamefont {B.}~\bibnamefont {Zager}}, \bibinfo {author} {\bibfnamefont {R.}~\bibnamefont {Fan}}, \bibinfo {author} {\bibfnamefont {P.}~\bibnamefont {Steadman}},\ and\ \bibinfo {author} {\bibfnamefont {K.}~\bibnamefont {Plumb}},\ }\bibfield  {title} {\bibinfo {title} {Double-{$Q$} spin chirality stripes in the anomalous {Hall} antiferromagnet {$\mathrm{CoNb}_{3}\mathrm{S}_{6}$}},\ }\href@noop {} {\bibfield  {journal} {\bibinfo  {journal} {arXiv preprint arXiv:2307.03776}\ } (\bibinfo {year} {2023})}\BibitemShut {NoStop}%
\bibitem [{\citenamefont {Lu}\ \emph {et~al.}(2022)\citenamefont {Lu}, \citenamefont {Murzabekova}, \citenamefont {Shim}, \citenamefont {Park}, \citenamefont {Kim}, \citenamefont {Kish}, \citenamefont {Wu}, \citenamefont {DeBeer-Schmitt}, \citenamefont {Aczel}, \citenamefont {Schleife} \emph {et~al.}}]{Co1/3NbS2_ND}%
  \BibitemOpen
  \bibfield  {author} {\bibinfo {author} {\bibfnamefont {K.}~\bibnamefont {Lu}}, \bibinfo {author} {\bibfnamefont {A.}~\bibnamefont {Murzabekova}}, \bibinfo {author} {\bibfnamefont {S.}~\bibnamefont {Shim}}, \bibinfo {author} {\bibfnamefont {J.}~\bibnamefont {Park}}, \bibinfo {author} {\bibfnamefont {S.}~\bibnamefont {Kim}}, \bibinfo {author} {\bibfnamefont {L.}~\bibnamefont {Kish}}, \bibinfo {author} {\bibfnamefont {Y.}~\bibnamefont {Wu}}, \bibinfo {author} {\bibfnamefont {L.}~\bibnamefont {DeBeer-Schmitt}}, \bibinfo {author} {\bibfnamefont {A.}~\bibnamefont {Aczel}}, \bibinfo {author} {\bibfnamefont {A.}~\bibnamefont {Schleife}}, \emph {et~al.},\ }\bibfield  {title} {\bibinfo {title} {Understanding the anomalous {Hall} effect in {$\mathrm{Co_{1/3}NbS_{2}}$} from crystal and magnetic structures},\ }\href@noop {} {\bibfield  {journal} {\bibinfo  {journal} {arXiv preprint arXiv:2212.14762}\ } (\bibinfo {year} {2022})}\BibitemShut {NoStop}%
\bibitem [{\citenamefont {Lin}\ \emph {et~al.}(2021)\citenamefont {Lin}, \citenamefont {Jeong}, \citenamefont {Kim}, \citenamefont {Wang}, \citenamefont {Huang}, \citenamefont {Masuda}, \citenamefont {Asai}, \citenamefont {Itoh}, \citenamefont {G{\"u}nther}, \citenamefont {Russina} \emph {et~al.}}]{NCTO_field}%
  \BibitemOpen
  \bibfield  {author} {\bibinfo {author} {\bibfnamefont {G.}~\bibnamefont {Lin}}, \bibinfo {author} {\bibfnamefont {J.}~\bibnamefont {Jeong}}, \bibinfo {author} {\bibfnamefont {C.}~\bibnamefont {Kim}}, \bibinfo {author} {\bibfnamefont {Y.}~\bibnamefont {Wang}}, \bibinfo {author} {\bibfnamefont {Q.}~\bibnamefont {Huang}}, \bibinfo {author} {\bibfnamefont {T.}~\bibnamefont {Masuda}}, \bibinfo {author} {\bibfnamefont {S.}~\bibnamefont {Asai}}, \bibinfo {author} {\bibfnamefont {S.}~\bibnamefont {Itoh}}, \bibinfo {author} {\bibfnamefont {G.}~\bibnamefont {G{\"u}nther}}, \bibinfo {author} {\bibfnamefont {M.}~\bibnamefont {Russina}}, \emph {et~al.},\ }\bibfield  {title} {\bibinfo {title} {Field-induced quantum spin disordered state in spin-1/2 honeycomb magnet {$\mathrm{Na_{2}Co_{2}TeO_{6}}$}},\ }\href@noop {} {\bibfield  {journal} {\bibinfo  {journal} {Nature communications}\ }\textbf {\bibinfo {volume} {12}},\ \bibinfo {pages} {5559} (\bibinfo {year} {2021})}\BibitemShut {NoStop}%
\bibitem [{\citenamefont {Songvilay}\ \emph {et~al.}(2020)\citenamefont {Songvilay}, \citenamefont {Robert}, \citenamefont {Petit}, \citenamefont {Rodriguez-Rivera}, \citenamefont {Ratcliff}, \citenamefont {Damay}, \citenamefont {Bal{\'e}dent}, \citenamefont {Jim{\'e}nez-Ruiz}, \citenamefont {Lejay}, \citenamefont {Pachoud} \emph {et~al.}}]{NCTO_INS1}%
  \BibitemOpen
  \bibfield  {author} {\bibinfo {author} {\bibfnamefont {M.}~\bibnamefont {Songvilay}}, \bibinfo {author} {\bibfnamefont {J.}~\bibnamefont {Robert}}, \bibinfo {author} {\bibfnamefont {S.}~\bibnamefont {Petit}}, \bibinfo {author} {\bibfnamefont {J.}~\bibnamefont {Rodriguez-Rivera}}, \bibinfo {author} {\bibfnamefont {W.}~\bibnamefont {Ratcliff}}, \bibinfo {author} {\bibfnamefont {F.}~\bibnamefont {Damay}}, \bibinfo {author} {\bibfnamefont {V.}~\bibnamefont {Bal{\'e}dent}}, \bibinfo {author} {\bibfnamefont {M.}~\bibnamefont {Jim{\'e}nez-Ruiz}}, \bibinfo {author} {\bibfnamefont {P.}~\bibnamefont {Lejay}}, \bibinfo {author} {\bibfnamefont {E.}~\bibnamefont {Pachoud}}, \emph {et~al.},\ }\bibfield  {title} {\bibinfo {title} {{Kitaev} interactions in the co honeycomb antiferromagnets {$\mathrm{Na_{3}Co_{2}SbO_{6}}$} and {$\mathrm{Na_{2}Co_{2}TeO_{6}}$}},\ }\href@noop {} {\bibfield  {journal} {\bibinfo  {journal} {Phys. Rev. B}\ }\textbf {\bibinfo {volume} {102}},\ \bibinfo {pages} {224429} (\bibinfo {year}
  {2020})}\BibitemShut {NoStop}%
\bibitem [{\citenamefont {Kim}\ \emph {et~al.}(2021)\citenamefont {Kim}, \citenamefont {Jeong}, \citenamefont {Lin}, \citenamefont {Park}, \citenamefont {Masuda}, \citenamefont {Asai}, \citenamefont {Itoh}, \citenamefont {Kim}, \citenamefont {Zhou}, \citenamefont {Ma} \emph {et~al.}}]{NCTO_INS2}%
  \BibitemOpen
  \bibfield  {author} {\bibinfo {author} {\bibfnamefont {C.}~\bibnamefont {Kim}}, \bibinfo {author} {\bibfnamefont {J.}~\bibnamefont {Jeong}}, \bibinfo {author} {\bibfnamefont {G.}~\bibnamefont {Lin}}, \bibinfo {author} {\bibfnamefont {P.}~\bibnamefont {Park}}, \bibinfo {author} {\bibfnamefont {T.}~\bibnamefont {Masuda}}, \bibinfo {author} {\bibfnamefont {S.}~\bibnamefont {Asai}}, \bibinfo {author} {\bibfnamefont {S.}~\bibnamefont {Itoh}}, \bibinfo {author} {\bibfnamefont {H.-S.}\ \bibnamefont {Kim}}, \bibinfo {author} {\bibfnamefont {H.}~\bibnamefont {Zhou}}, \bibinfo {author} {\bibfnamefont {J.}~\bibnamefont {Ma}}, \emph {et~al.},\ }\bibfield  {title} {\bibinfo {title} {Antiferromagnetic {Kitaev} interaction in {$J_{eff}= 1/2$} cobalt honeycomb materials {$\mathrm{Na_{3}Co_{2}SbO_{6}}$} and {$\mathrm{Na_{2}Co_{2}TeO_{6}}$}},\ }\href@noop {} {\bibfield  {journal} {\bibinfo  {journal} {Journal of Physics: Condensed Matter}\ }\textbf {\bibinfo {volume} {34}},\ \bibinfo {pages} {045802} (\bibinfo {year}
  {2021})}\BibitemShut {NoStop}%
\bibitem [{\citenamefont {Jiao}\ \emph {et~al.}(2024)\citenamefont {Jiao}, \citenamefont {Li}, \citenamefont {Lin}, \citenamefont {Shu}, \citenamefont {Xu}, \citenamefont {Zaharko}, \citenamefont {Shiroka}, \citenamefont {Hong}, \citenamefont {Kolesnikov}, \citenamefont {Deng} \emph {et~al.}}]{Ma_NCTO_1q}%
  \BibitemOpen
  \bibfield  {author} {\bibinfo {author} {\bibfnamefont {J.}~\bibnamefont {Jiao}}, \bibinfo {author} {\bibfnamefont {X.}~\bibnamefont {Li}}, \bibinfo {author} {\bibfnamefont {G.}~\bibnamefont {Lin}}, \bibinfo {author} {\bibfnamefont {M.}~\bibnamefont {Shu}}, \bibinfo {author} {\bibfnamefont {W.}~\bibnamefont {Xu}}, \bibinfo {author} {\bibfnamefont {O.}~\bibnamefont {Zaharko}}, \bibinfo {author} {\bibfnamefont {T.}~\bibnamefont {Shiroka}}, \bibinfo {author} {\bibfnamefont {T.}~\bibnamefont {Hong}}, \bibinfo {author} {\bibfnamefont {A.~I.}\ \bibnamefont {Kolesnikov}}, \bibinfo {author} {\bibfnamefont {G.}~\bibnamefont {Deng}}, \emph {et~al.},\ }\bibfield  {title} {\bibinfo {title} {Static magnetic order with strong quantum fluctuations in spin-1/2 honeycomb magnet {$\mathrm{Na_{2}Co_{2}TeO_{6}}$}},\ }\href@noop {} {\bibfield  {journal} {\bibinfo  {journal} {Communications Materials}\ }\textbf {\bibinfo {volume} {5}},\ \bibinfo {pages} {159} (\bibinfo {year} {2024})}\BibitemShut {NoStop}%
\bibitem [{\citenamefont {Lefran\ifmmode~\mbox{\c{c}}\else \c{c}\fi{}ois}\ \emph {et~al.}(2016)\citenamefont {Lefran\ifmmode~\mbox{\c{c}}\else \c{c}\fi{}ois}, \citenamefont {Songvilay}, \citenamefont {Robert}, \citenamefont {Nataf}, \citenamefont {Jordan}, \citenamefont {Chaix}, \citenamefont {Colin}, \citenamefont {Lejay}, \citenamefont {Hadj-Azzem}, \citenamefont {Ballou},\ and\ \citenamefont {Simonet}}]{NCTO_1q_2}%
  \BibitemOpen
  \bibfield  {author} {\bibinfo {author} {\bibfnamefont {E.}~\bibnamefont {Lefran\ifmmode~\mbox{\c{c}}\else \c{c}\fi{}ois}}, \bibinfo {author} {\bibfnamefont {M.}~\bibnamefont {Songvilay}}, \bibinfo {author} {\bibfnamefont {J.}~\bibnamefont {Robert}}, \bibinfo {author} {\bibfnamefont {G.}~\bibnamefont {Nataf}}, \bibinfo {author} {\bibfnamefont {E.}~\bibnamefont {Jordan}}, \bibinfo {author} {\bibfnamefont {L.}~\bibnamefont {Chaix}}, \bibinfo {author} {\bibfnamefont {C.~V.}\ \bibnamefont {Colin}}, \bibinfo {author} {\bibfnamefont {P.}~\bibnamefont {Lejay}}, \bibinfo {author} {\bibfnamefont {A.}~\bibnamefont {Hadj-Azzem}}, \bibinfo {author} {\bibfnamefont {R.}~\bibnamefont {Ballou}},\ and\ \bibinfo {author} {\bibfnamefont {V.}~\bibnamefont {Simonet}},\ }\bibfield  {title} {\bibinfo {title} {Magnetic properties of the honeycomb oxide {${\mathrm{Na}}_{2}{\mathrm{Co}}_{2}{\mathrm{TeO}}_{6}$}},\ }\href {https://doi.org/10.1103/PhysRevB.94.214416} {\bibfield  {journal} {\bibinfo  {journal} {Phys. Rev. B}\ }\textbf
  {\bibinfo {volume} {94}},\ \bibinfo {pages} {214416} (\bibinfo {year} {2016})}\BibitemShut {NoStop}%
\bibitem [{\citenamefont {Bera}\ \emph {et~al.}(2017)\citenamefont {Bera}, \citenamefont {Yusuf}, \citenamefont {Kumar},\ and\ \citenamefont {Ritter}}]{NCTO_1q_3}%
  \BibitemOpen
  \bibfield  {author} {\bibinfo {author} {\bibfnamefont {A.~K.}\ \bibnamefont {Bera}}, \bibinfo {author} {\bibfnamefont {S.~M.}\ \bibnamefont {Yusuf}}, \bibinfo {author} {\bibfnamefont {A.}~\bibnamefont {Kumar}},\ and\ \bibinfo {author} {\bibfnamefont {C.}~\bibnamefont {Ritter}},\ }\bibfield  {title} {\bibinfo {title} {Zigzag antiferromagnetic ground state with anisotropic correlation lengths in the quasi-two-dimensional honeycomb lattice compound {$\mathrm{N}{\mathrm{a}}_{2}\mathrm{C}{\mathrm{o}}_{2}\mathrm{Te}{\mathrm{O}}_{6}$}},\ }\href {https://doi.org/10.1103/PhysRevB.95.094424} {\bibfield  {journal} {\bibinfo  {journal} {Phys. Rev. B}\ }\textbf {\bibinfo {volume} {95}},\ \bibinfo {pages} {094424} (\bibinfo {year} {2017})}\BibitemShut {NoStop}%
\bibitem [{\citenamefont {Chen}\ \emph {et~al.}(2021)\citenamefont {Chen}, \citenamefont {Li}, \citenamefont {Hu}, \citenamefont {Hu}, \citenamefont {Yue}, \citenamefont {Sutarto}, \citenamefont {He}, \citenamefont {Iida}, \citenamefont {Kamazawa}, \citenamefont {Yu} \emph {et~al.}}]{Li_NCTO_3q_1}%
  \BibitemOpen
  \bibfield  {author} {\bibinfo {author} {\bibfnamefont {W.}~\bibnamefont {Chen}}, \bibinfo {author} {\bibfnamefont {X.}~\bibnamefont {Li}}, \bibinfo {author} {\bibfnamefont {Z.}~\bibnamefont {Hu}}, \bibinfo {author} {\bibfnamefont {Z.}~\bibnamefont {Hu}}, \bibinfo {author} {\bibfnamefont {L.}~\bibnamefont {Yue}}, \bibinfo {author} {\bibfnamefont {R.}~\bibnamefont {Sutarto}}, \bibinfo {author} {\bibfnamefont {F.}~\bibnamefont {He}}, \bibinfo {author} {\bibfnamefont {K.}~\bibnamefont {Iida}}, \bibinfo {author} {\bibfnamefont {K.}~\bibnamefont {Kamazawa}}, \bibinfo {author} {\bibfnamefont {W.}~\bibnamefont {Yu}}, \emph {et~al.},\ }\bibfield  {title} {\bibinfo {title} {Spin-orbit phase behavior of {$\mathrm{Na_{2}Co_{2}TeO_{6}}$} at low temperatures},\ }\href@noop {} {\bibfield  {journal} {\bibinfo  {journal} {Phys. Rev. B}\ }\textbf {\bibinfo {volume} {103}},\ \bibinfo {pages} {L180404} (\bibinfo {year} {2021})}\BibitemShut {NoStop}%
\bibitem [{\citenamefont {Kr\"uger}\ \emph {et~al.}(2023)\citenamefont {Kr\"uger}, \citenamefont {Chen}, \citenamefont {Jin}, \citenamefont {Li},\ and\ \citenamefont {Janssen}}]{Li_NCTO_3q2}%
  \BibitemOpen
  \bibfield  {author} {\bibinfo {author} {\bibfnamefont {W.~G.~F.}\ \bibnamefont {Kr\"uger}}, \bibinfo {author} {\bibfnamefont {W.}~\bibnamefont {Chen}}, \bibinfo {author} {\bibfnamefont {X.}~\bibnamefont {Jin}}, \bibinfo {author} {\bibfnamefont {Y.}~\bibnamefont {Li}},\ and\ \bibinfo {author} {\bibfnamefont {L.}~\bibnamefont {Janssen}},\ }\bibfield  {title} {\bibinfo {title} {Triple-q order in {${\mathrm{Na}}_{2}{\mathrm{Co}}_{2}{\mathrm{TeO}}_{6}$} from proximity to hidden-{SU(2)}-symmetric point},\ }\href {https://doi.org/10.1103/PhysRevLett.131.146702} {\bibfield  {journal} {\bibinfo  {journal} {Phys. Rev. Lett.}\ }\textbf {\bibinfo {volume} {131}},\ \bibinfo {pages} {146702} (\bibinfo {year} {2023})}\BibitemShut {NoStop}%
\bibitem [{\citenamefont {Yao}\ \emph {et~al.}(2023)\citenamefont {Yao}, \citenamefont {Zhao}, \citenamefont {Qiu}, \citenamefont {Balz}, \citenamefont {Stewart}, \citenamefont {Lynn},\ and\ \citenamefont {Li}}]{Li_NCTO_3q_3}%
  \BibitemOpen
  \bibfield  {author} {\bibinfo {author} {\bibfnamefont {W.}~\bibnamefont {Yao}}, \bibinfo {author} {\bibfnamefont {Y.}~\bibnamefont {Zhao}}, \bibinfo {author} {\bibfnamefont {Y.}~\bibnamefont {Qiu}}, \bibinfo {author} {\bibfnamefont {C.}~\bibnamefont {Balz}}, \bibinfo {author} {\bibfnamefont {J.~R.}\ \bibnamefont {Stewart}}, \bibinfo {author} {\bibfnamefont {J.~W.}\ \bibnamefont {Lynn}},\ and\ \bibinfo {author} {\bibfnamefont {Y.}~\bibnamefont {Li}},\ }\bibfield  {title} {\bibinfo {title} {Magnetic ground state of the {Kitaev} {$\mathrm{Na_{2}Co_{2}TeO_{6}}$} spin liquid candidate},\ }\href@noop {} {\bibfield  {journal} {\bibinfo  {journal} {Physical Review Research}\ }\textbf {\bibinfo {volume} {5}},\ \bibinfo {pages} {L022045} (\bibinfo {year} {2023})}\BibitemShut {NoStop}%
\bibitem [{\citenamefont {Husremovic}\ \emph {et~al.}(2022)\citenamefont {Husremovic}, \citenamefont {Groschner}, \citenamefont {Inzani}, \citenamefont {Craig}, \citenamefont {Bustillo}, \citenamefont {Ercius}, \citenamefont {Kazmierczak}, \citenamefont {Syndikus}, \citenamefont {Van~Winkle}, \citenamefont {Aloni} \emph {et~al.}}]{TTS_chem_inter}%
  \BibitemOpen
  \bibfield  {author} {\bibinfo {author} {\bibfnamefont {S.}~\bibnamefont {Husremovic}}, \bibinfo {author} {\bibfnamefont {C.~K.}\ \bibnamefont {Groschner}}, \bibinfo {author} {\bibfnamefont {K.}~\bibnamefont {Inzani}}, \bibinfo {author} {\bibfnamefont {I.~M.}\ \bibnamefont {Craig}}, \bibinfo {author} {\bibfnamefont {K.~C.}\ \bibnamefont {Bustillo}}, \bibinfo {author} {\bibfnamefont {P.}~\bibnamefont {Ercius}}, \bibinfo {author} {\bibfnamefont {N.~P.}\ \bibnamefont {Kazmierczak}}, \bibinfo {author} {\bibfnamefont {J.}~\bibnamefont {Syndikus}}, \bibinfo {author} {\bibfnamefont {M.}~\bibnamefont {Van~Winkle}}, \bibinfo {author} {\bibfnamefont {S.}~\bibnamefont {Aloni}}, \emph {et~al.},\ }\bibfield  {title} {\bibinfo {title} {Hard ferromagnetism down to the thinnest limit of iron-intercalated tantalum disulfide},\ }\href@noop {} {\bibfield  {journal} {\bibinfo  {journal} {Journal of the American Chemical Society}\ }\textbf {\bibinfo {volume} {144}},\ \bibinfo {pages} {12167} (\bibinfo {year} {2022})}\BibitemShut
  {NoStop}%
\bibitem [{\citenamefont {He}\ \emph {et~al.}(2024)\citenamefont {He}, \citenamefont {Si}, \citenamefont {Xu}, \citenamefont {Wang}, \citenamefont {Jin}, \citenamefont {Yang}, \citenamefont {Wei}, \citenamefont {Meng}, \citenamefont {Zhai}, \citenamefont {Zhang} \emph {et~al.}}]{TTS_2D_limit}%
  \BibitemOpen
  \bibfield  {author} {\bibinfo {author} {\bibfnamefont {Q.}~\bibnamefont {He}}, \bibinfo {author} {\bibfnamefont {K.}~\bibnamefont {Si}}, \bibinfo {author} {\bibfnamefont {Z.}~\bibnamefont {Xu}}, \bibinfo {author} {\bibfnamefont {X.}~\bibnamefont {Wang}}, \bibinfo {author} {\bibfnamefont {C.}~\bibnamefont {Jin}}, \bibinfo {author} {\bibfnamefont {Y.}~\bibnamefont {Yang}}, \bibinfo {author} {\bibfnamefont {J.}~\bibnamefont {Wei}}, \bibinfo {author} {\bibfnamefont {L.}~\bibnamefont {Meng}}, \bibinfo {author} {\bibfnamefont {P.}~\bibnamefont {Zhai}}, \bibinfo {author} {\bibfnamefont {P.}~\bibnamefont {Zhang}}, \emph {et~al.},\ }\bibfield  {title} {\bibinfo {title} {Direct synthesis of controllable ultrathin heteroatoms-intercalated 2d layered materials},\ }\href@noop {} {\bibfield  {journal} {\bibinfo  {journal} {Nature Communications}\ }\textbf {\bibinfo {volume} {15}},\ \bibinfo {pages} {6320} (\bibinfo {year} {2024})}\BibitemShut {NoStop}%
\bibitem [{\citenamefont {Husremovi{\'c}}\ \emph {et~al.}(2025)\citenamefont {Husremovi{\'c}}, \citenamefont {Gonzalez}, \citenamefont {Goodge}, \citenamefont {Xie}, \citenamefont {Kong}, \citenamefont {Zhang}, \citenamefont {Ryu}, \citenamefont {Ribet}, \citenamefont {Fender}, \citenamefont {Bustillo} \emph {et~al.}}]{TTS_chem_inter2}%
  \BibitemOpen
  \bibfield  {author} {\bibinfo {author} {\bibfnamefont {S.}~\bibnamefont {Husremovi{\'c}}}, \bibinfo {author} {\bibfnamefont {O.}~\bibnamefont {Gonzalez}}, \bibinfo {author} {\bibfnamefont {B.~H.}\ \bibnamefont {Goodge}}, \bibinfo {author} {\bibfnamefont {L.~S.}\ \bibnamefont {Xie}}, \bibinfo {author} {\bibfnamefont {Z.}~\bibnamefont {Kong}}, \bibinfo {author} {\bibfnamefont {W.}~\bibnamefont {Zhang}}, \bibinfo {author} {\bibfnamefont {S.~H.}\ \bibnamefont {Ryu}}, \bibinfo {author} {\bibfnamefont {S.~M.}\ \bibnamefont {Ribet}}, \bibinfo {author} {\bibfnamefont {S.~S.}\ \bibnamefont {Fender}}, \bibinfo {author} {\bibfnamefont {K.~C.}\ \bibnamefont {Bustillo}}, \emph {et~al.},\ }\bibfield  {title} {\bibinfo {title} {Tailored topotactic chemistry unlocks heterostructures of magnetic intercalation compounds},\ }\href@noop {} {\bibfield  {journal} {\bibinfo  {journal} {Nature Communications}\ }\textbf {\bibinfo {volume} {16}},\ \bibinfo {pages} {1208} (\bibinfo {year} {2025})}\BibitemShut {NoStop}%
\bibitem [{\citenamefont {Burch}\ \emph {et~al.}(2018)\citenamefont {Burch}, \citenamefont {Mandrus},\ and\ \citenamefont {Park}}]{Nature_review}%
  \BibitemOpen
  \bibfield  {author} {\bibinfo {author} {\bibfnamefont {K.~S.}\ \bibnamefont {Burch}}, \bibinfo {author} {\bibfnamefont {D.}~\bibnamefont {Mandrus}},\ and\ \bibinfo {author} {\bibfnamefont {J.-G.}\ \bibnamefont {Park}},\ }\bibfield  {title} {\bibinfo {title} {Magnetism in two-dimensional van der {Waals} materials},\ }\href@noop {} {\bibfield  {journal} {\bibinfo  {journal} {Nature}\ }\textbf {\bibinfo {volume} {563}},\ \bibinfo {pages} {47} (\bibinfo {year} {2018})}\BibitemShut {NoStop}%
\bibitem [{\citenamefont {Parkin}\ and\ \citenamefont {Friend}(1980{\natexlab{b}})}]{Parkin_80_v2}%
  \BibitemOpen
  \bibfield  {author} {\bibinfo {author} {\bibfnamefont {S.~S.~P.}\ \bibnamefont {Parkin}}\ and\ \bibinfo {author} {\bibfnamefont {R.~H.}\ \bibnamefont {Friend}},\ }\bibfield  {title} {\bibinfo {title} {3d transition-metal intercalates of the niobium and tantalum dichalcogenides. ii. transport properties},\ }\href {https://doi.org/10.1080/13642818008245371} {\bibfield  {journal} {\bibinfo  {journal} {Philosophical Magazine B}\ }\textbf {\bibinfo {volume} {41}},\ \bibinfo {pages} {95} (\bibinfo {year} {1980}{\natexlab{b}})}\BibitemShut {NoStop}%
\bibitem [{\citenamefont {Park}\ \emph {et~al.}(2018)\citenamefont {Park}, \citenamefont {Oh}, \citenamefont {Uhl{\'\i}{\v{r}}ov{\'a}}, \citenamefont {Jackson}, \citenamefont {De{\'a}k}, \citenamefont {Szunyogh}, \citenamefont {Lee}, \citenamefont {Cho}, \citenamefont {Kim}, \citenamefont {Walker} \emph {et~al.}}]{Mn3Sn_INS}%
  \BibitemOpen
  \bibfield  {author} {\bibinfo {author} {\bibfnamefont {P.}~\bibnamefont {Park}}, \bibinfo {author} {\bibfnamefont {J.}~\bibnamefont {Oh}}, \bibinfo {author} {\bibfnamefont {K.}~\bibnamefont {Uhl{\'\i}{\v{r}}ov{\'a}}}, \bibinfo {author} {\bibfnamefont {J.}~\bibnamefont {Jackson}}, \bibinfo {author} {\bibfnamefont {A.}~\bibnamefont {De{\'a}k}}, \bibinfo {author} {\bibfnamefont {L.}~\bibnamefont {Szunyogh}}, \bibinfo {author} {\bibfnamefont {K.~H.}\ \bibnamefont {Lee}}, \bibinfo {author} {\bibfnamefont {H.}~\bibnamefont {Cho}}, \bibinfo {author} {\bibfnamefont {H.-L.}\ \bibnamefont {Kim}}, \bibinfo {author} {\bibfnamefont {H.~C.}\ \bibnamefont {Walker}}, \emph {et~al.},\ }\bibfield  {title} {\bibinfo {title} {Magnetic excitations in non-collinear antiferromagnetic {Weyl} semimetal {$\mathrm{Mn_{3}Sn}$}},\ }\href@noop {} {\bibfield  {journal} {\bibinfo  {journal} {npj Quantum Materials}\ }\textbf {\bibinfo {volume} {3}},\ \bibinfo {pages} {63} (\bibinfo {year} {2018})}\BibitemShut {NoStop}%
\bibitem [{\citenamefont {Maksimov}\ \emph {et~al.}(2019)\citenamefont {Maksimov}, \citenamefont {Zhu}, \citenamefont {White},\ and\ \citenamefont {Chernyshev}}]{TLAF_generalized}%
  \BibitemOpen
  \bibfield  {author} {\bibinfo {author} {\bibfnamefont {P.}~\bibnamefont {Maksimov}}, \bibinfo {author} {\bibfnamefont {Z.}~\bibnamefont {Zhu}}, \bibinfo {author} {\bibfnamefont {S.~R.}\ \bibnamefont {White}},\ and\ \bibinfo {author} {\bibfnamefont {A.}~\bibnamefont {Chernyshev}},\ }\bibfield  {title} {\bibinfo {title} {Anisotropic-exchange magnets on a triangular lattice: spin waves, accidental degeneracies, and dual spin liquids},\ }\href@noop {} {\bibfield  {journal} {\bibinfo  {journal} {Physical Review X}\ }\textbf {\bibinfo {volume} {9}},\ \bibinfo {pages} {021017} (\bibinfo {year} {2019})}\BibitemShut {NoStop}%
\bibitem [{\citenamefont {Brochu}\ \emph {et~al.}(2010)\citenamefont {Brochu}, \citenamefont {Cora},\ and\ \citenamefont {De~Freitas}}]{Ref_Bopt}%
  \BibitemOpen
  \bibfield  {author} {\bibinfo {author} {\bibfnamefont {E.}~\bibnamefont {Brochu}}, \bibinfo {author} {\bibfnamefont {V.~M.}\ \bibnamefont {Cora}},\ and\ \bibinfo {author} {\bibfnamefont {N.}~\bibnamefont {De~Freitas}},\ }\bibfield  {title} {\bibinfo {title} {A tutorial on bayesian optimization of expensive cost functions, with application to active user modeling and hierarchical reinforcement learning},\ }\href@noop {} {\bibfield  {journal} {\bibinfo  {journal} {arXiv preprint arXiv:1012.2599}\ } (\bibinfo {year} {2010})}\BibitemShut {NoStop}%
\bibitem [{\citenamefont {Pedregosa}(2011)}]{Bopt_python}%
  \BibitemOpen
  \bibfield  {author} {\bibinfo {author} {\bibfnamefont {F.}~\bibnamefont {Pedregosa}},\ }\bibfield  {title} {\bibinfo {title} {Scikit-learn: Machine learning in python fabian},\ }\href@noop {} {\bibfield  {journal} {\bibinfo  {journal} {Journal of machine learning research}\ }\textbf {\bibinfo {volume} {12}},\ \bibinfo {pages} {2825} (\bibinfo {year} {2011})}\BibitemShut {NoStop}%
\bibitem [{\citenamefont {{Sakib Matin}}()}]{Bopt_Matin}%
  \BibitemOpen
  \bibfield  {author} {\bibinfo {author} {\bibnamefont {{Sakib Matin}}},\ }\href@noop {} {}\bibinfo {note} {An opinionated Julia wrapper for Bayesian Optimization. Available online: \url{https://github.com/sakibmatin/BayesOptim.jl}}\BibitemShut {NoStop}%
\bibitem [{\citenamefont {Kim}\ \emph {et~al.}(2019)\citenamefont {Kim}, \citenamefont {Lim}, \citenamefont {Lee}, \citenamefont {Lee}, \citenamefont {Kim}, \citenamefont {Park}, \citenamefont {Jeon}, \citenamefont {Park}, \citenamefont {Park},\ and\ \citenamefont {Cheong}}]{NiPS3_ncomm}%
  \BibitemOpen
  \bibfield  {author} {\bibinfo {author} {\bibfnamefont {K.}~\bibnamefont {Kim}}, \bibinfo {author} {\bibfnamefont {S.~Y.}\ \bibnamefont {Lim}}, \bibinfo {author} {\bibfnamefont {J.-U.}\ \bibnamefont {Lee}}, \bibinfo {author} {\bibfnamefont {S.}~\bibnamefont {Lee}}, \bibinfo {author} {\bibfnamefont {T.~Y.}\ \bibnamefont {Kim}}, \bibinfo {author} {\bibfnamefont {K.}~\bibnamefont {Park}}, \bibinfo {author} {\bibfnamefont {G.~S.}\ \bibnamefont {Jeon}}, \bibinfo {author} {\bibfnamefont {C.-H.}\ \bibnamefont {Park}}, \bibinfo {author} {\bibfnamefont {J.-G.}\ \bibnamefont {Park}},\ and\ \bibinfo {author} {\bibfnamefont {H.}~\bibnamefont {Cheong}},\ }\bibfield  {title} {\bibinfo {title} {Suppression of magnetic ordering in xxz-type antiferromagnetic monolayer {$\mathrm{NiPS_{3}}$}},\ }\href@noop {} {\bibfield  {journal} {\bibinfo  {journal} {Nature communications}\ }\textbf {\bibinfo {volume} {10}},\ \bibinfo {pages} {345} (\bibinfo {year} {2019})}\BibitemShut {NoStop}%
\bibitem [{\citenamefont {Son}\ \emph {et~al.}(2021)\citenamefont {Son}, \citenamefont {Son}, \citenamefont {Park}, \citenamefont {Kim}, \citenamefont {Tao}, \citenamefont {Oh}, \citenamefont {Lee}, \citenamefont {Lee}, \citenamefont {Kim}, \citenamefont {Zhang} \emph {et~al.}}]{CrPS4}%
  \BibitemOpen
  \bibfield  {author} {\bibinfo {author} {\bibfnamefont {J.}~\bibnamefont {Son}}, \bibinfo {author} {\bibfnamefont {S.}~\bibnamefont {Son}}, \bibinfo {author} {\bibfnamefont {P.}~\bibnamefont {Park}}, \bibinfo {author} {\bibfnamefont {M.}~\bibnamefont {Kim}}, \bibinfo {author} {\bibfnamefont {Z.}~\bibnamefont {Tao}}, \bibinfo {author} {\bibfnamefont {J.}~\bibnamefont {Oh}}, \bibinfo {author} {\bibfnamefont {T.}~\bibnamefont {Lee}}, \bibinfo {author} {\bibfnamefont {S.}~\bibnamefont {Lee}}, \bibinfo {author} {\bibfnamefont {J.}~\bibnamefont {Kim}}, \bibinfo {author} {\bibfnamefont {K.}~\bibnamefont {Zhang}}, \emph {et~al.},\ }\bibfield  {title} {\bibinfo {title} {Air-stable and layer-dependent ferromagnetism in atomically thin van der {Waals} {$\mathrm{CrPS_{4}}$}},\ }\href@noop {} {\bibfield  {journal} {\bibinfo  {journal} {ACS nano}\ }\textbf {\bibinfo {volume} {15}},\ \bibinfo {pages} {16904} (\bibinfo {year} {2021})}\BibitemShut {NoStop}%
\end{thebibliography}
\end{document}